\documentclass{aa}  
\usepackage{graphicx}
\usepackage{txfonts}
\usepackage{natbib}
\bibpunct{(}{)}{;}{a}{}{,}   

\begin{document}
   \title{How old are the stars in the halo of NGC 5128 (Centaurus A)?
}

%
   
\author{M.~Rejkuba\inst{1}
	\and W.~E.~Harris\inst{2}
	\and L.~Greggio\inst{3}
	\and G.~L.~H.~Harris\inst{4}
}

   \offprints{M. Rejkuba}

   \institute{ESO, Karl-Schwarzschild-Strasse
           2, D-85748 Garching, Germany\\
              \email{mrejkuba@eso.org}
	   \and
	      Department of Physics and Astronomy, McMaster University, Hamilton ON
L8S 4M1, Canada\\
	      \email{harris@physics.mcmaster.ca}  
         \and
             INAF, Osservatorio Astronomico di Padova, Vicolo dell'Osservatorio 5,
35122 Padova, Italy\\
             \email{greggio@pd.astro.it}
	  \and 
	     Department of Physics and Astronomy, University of Waterloo, Waterloo ON N2L 3G1, 
	     Canada\\
	      \email{glharris@astro.uwaterloo.ca} 
             }

   \date{Received 25 August 2010 / Accepted 12 November 2010}
\titlerunning{Star formation history of NGC~5128}

 
  \abstract
   {NGC 5128 (Centaurus A) is, at the distance of just 3.8~Mpc, the nearest easily observable
giant elliptical galaxy. Therefore it is the best target to investigate the 
early star formation history of an elliptical galaxy. }
   {Our aims are to establish when the oldest stars formed in NGC~5128, and
whether this galaxy formed stars over a long period.}
   {We compare simulated colour-magnitude diagrams with the deep ACS/HST
photometry. The simulations assume in input
either the observed metallicity distribution function, based on the colour 
distribution of the upper red giant branch stars, or the closed box chemical
enrichment. Simulations are constructed for single age bursts using BASTI
evolutionary isochrones;  more complex star formation histories are
constructed as well by combining several individual simulations. Comparisons with 
data are made by fitting the whole colour-magnitude diagram as well as the
the luminosity functions in $V$ and $I$ band. In addition we inspect
carefully the red clump and asymptotic giant branch bump luminosities and number
counts, since these features  are the primary constraints on the ages of the observed stars.}
   {We find that that the observed colour-magnitude diagram can be reproduced
satisfactorily only by simulations that have the bulk of the stars with ages in excess
of $\sim 10$~Gyr, and that the alpha-enhanced models fit the data much better than
the solar scaled ones. Data are not consistent with extended star formation over
more than $3-4$~Gyr. Two burst models, with 70-80\% of the stars formed
$12\pm1$~Gyr ago and with 20-30\% younger contribution with $2-4$~Gyr old stars
provide the best agreement with the data.  The old component spans the whole
metallicity range of the models ($Z=0.0001 - 0.04$), while for the young component 
the best fitting models indicate higher minimum metallicity ($\sim 1/10 - 1/4$ Z$_\odot$).
}
  {The bulk of the halo stars in NGC~5128 must have formed at redshift $z \ga 2$ 
and the chemical enrichment was very fast, reaching solar or even 
twice-solar metallicity already for the $\sim 11-12$~Gyr old population. 
The minor young component, adding $\sim 20-30$\%  of the stars to the halo,  and contributing
less than 10\% of the mass, 
may have resulted from a later star formation event $\sim 2-4$~Gyr ago.}

   \keywords{Galaxies: elliptical and lenticular, cD --
             Galaxies: Individual: NGC~5128 --
             Galaxies: stellar content --
             Galaxies: star formation
               }

   \maketitle
%

\section{Introduction}

NGC 5128 (Centaurus A) is by far the nearest easily observable giant E
galaxy and the centrally dominant object in the Centaurus group
of galaxies \citep{karachentsev05}.  At D=3.8 Mpc \citep{harris+10},
it is more than 2 magnitudes closer than the ellipticals in the
Leo group and 3 magnitudes closer than the Virgo cluster.  As such,
it offers an unparalleled opportunity for studying the nature of
stellar populations in a large elliptical. Particularly interesting is
the old-halo component whose basic properties (age distribution, 
metallicity distribution, star formation history) are
difficult to measure in detail for  galaxies beyond the Local Group.

In Paper I \citep{rejkuba+05}, we presented HST ACS/WFC photometry of
the stars in an outer-halo field of NGC 5128.  The photometric limits
of the data, resulting from 12 full-orbit exposures in each of
the F606W and F814W filters, 
were deep enough to reveal both the old red-giant branch (RGB)
to its reddest, most metal-rich extent, and the core-helium-burning ``red clump'' or
horizontal branch (RC or HB) stars.  The main purpose of our deep ACS/WFC photometric
program was to make a first attempt at \emph{directly} measuring the
earliest star formation history in this keystone galaxy, because the
HB is the most luminous  stellar component that can unambiguously
reveal very old populations.  Previous work with the HST/WFPC2 camera which probed the
brightest $\sim 2$ magnitudes of the RGB at other locations in
the mid- and outer-halo \citep{harris+99,harris+harris00,harris+harris02} indicated
that the stellar population in these fields is dominated by normal, old, moderately
metal-rich red-giant stars, with extremely few if any ``young'' ($\tau \lesssim 5$ Gyr)
evolved stars. 
However, the well known age/metallicity degeneracy that strongly
affects the old RGB stars prevented any more precise statements about
the age distribution. Several other resolved stellar population studies of
halo stars in this galaxy have claimed the presence of an up to $\sim 15$\%
intermediate-age stellar component \citep{soria+96,marleau+00,rejkuba+03},
similar to what is deduced for more distant field elliptical galaxies
\citep{thomas+05}. 

It is an unfortunate historical accident that
NGC 5128 is still frequently thought of as a ``peculiar'' galaxy.
This view (dating back more than half a century, when little was known
about the range of normal-galaxy properties compared with the present time)
is based on the obvious presence of components such as
the central dust lane and accompanying recent star formation 
\citep{graham79,moellenhoff81,quillen+93,minniti+04_Sersic13,ferrarese+07}, 
the central supermassive black hole \citep{krajnovic+07,cappellari+09, neumayer10} 
and the jets at various scales most easily visible in radio and X-ray 
wavelengths \citep{kraft+02,hardcastle+03,goodger+10}, 
as well as other markers of activity that
lie in the inner $\sim 5$ kpc of the bulge \citep{neumayer+07}.  Further out, faint shells can
be seen that are presumably the remnants of a long-ago satellite accretion \citep{malin+83},
as well as faint filaments of ionized gas and young stars along the northern radio 
and X-ray jet \citep{graham98,mould+00,rejkuba+01,rejkuba+02},
a young blue arc of star formation \citep{peng+02}, and 
diffuse radio lobes that extend out hundreds of kiloparsecs \citep{morganti+99,feain+09}. For 
extensive reviews we refer to \citet{ebneter+balick83} and \citet{israel98}.
This range of properties often prompts the response that anything learned about
the old stellar population of NGC 5128 will be ``anomalous'' and thus
not applicable to other giant ellipticals.

Our view is that such attitudes are far too 
dismissive and should long since have 
been put aside \citep[see][for a comprehensive discussion]{harris10}.  
We now know that many large E galaxies  have
subcomponents of various kinds which trace ongoing, sporadic accretion events 
\citep[such as central black holes and jets, dust lanes, modest amounts of young star
formation, and so on; see e.g.][]{vandokkum05}. In these respects NGC 5128 can no longer be said 
to stand out among other similarly massive ellipticals anywhere else. 
Its active features and evidence for an accretion/merger history have
unusual prominence in the literature \emph{simply because it is the
closest and brightest example}, providing an unexcelled stage on which
these processes can be studied in unique detail.
Exactly this point of view has been in the literature for a remarkably long
time \citep[e.g.][]{graham79,ebneter+balick83,israel98,harris10}, 
but has still not reached the wide recognition 
it deserves.   The intensive work on these active components of Centaurus A during
the 1970's and 1980's \citep[thoroughly reviewed in ][]{ebneter+balick83,israel98}
was, unfortunately, not paralleled during the same period
by a comparable amount of work on its underlying stellar populations.  
The paradoxical result was that by about 1990 we knew a good deal more about this galaxy's 
peculiarities than its normalities.

We call these active subcomponents peculiarities, because they make up quite a small
fraction of the total mass of the galaxy.  The mass in recently formed stars in the 
north-eastern halo is several times $10^6$~$M_\odot$ \citep{rejkuba+04}, 
and the amount of neutral HI and H$_2$ gas \citep{schiminovich+94, charmandaris+00} 
corresponds to few times $10^7$~$M_\odot$.  The mass of the black hole in the centre 
of the galaxy is $(5.5 \pm 3.0) \times 10^7$~$M_\odot$ \citep{cappellari+09}.
The amount of HI associated with the central dust lane is $3.5 \times 10^8 M_{\odot}$, and 
there is about the same amount of $H_2$ \citep{charmandaris+00}.
At $\lesssim 10^9 M_{\odot}$ \citep{charmandaris+00}, the total gas mass is 
much less than the total dynamically determined mass of the galaxy
of  $(1.3 \pm 0.5) \times 10^{12}$~$M_\odot$ within a galactocentric radius of 
45 kpc \citep{woodley06, woodley+07, woodley+10b}, equivalent to a total stellar mass
$\sim 10^{11} M_{\odot}$.

Since the 1990s, largely thanks to the high-resolution imaging 
capabilities of the Hubble Space Telescope (HST) and 
improvements in  
modern ground based spectroscopic and imaging instrumentation,
we have gathered evidence indicating that the main body of
the galaxy is, in fact, a rather conventional 
giant elliptical.\footnote{\citet{graham79} 30 years 
ago said explicitly that ``The present observations
reinforce the view that NGC 5128 is a giant elliptical galaxy in which is
embedded an inclined and rotating disk composed partly of gas ... [resulting
from] addition of gaseous material to a basically normal elliptical galaxy.''
\citet{ebneter+balick83} arrived at the same view, one which is quite plausible today:
``Cen A has a probably undeserved reputation for being one of the most
peculiar galaxies in the sky ... it is not significantly different from
either other dusty elliptical[s] or other active galaxies.  Most of
Cen A's major features are probably the result of the collision and merger
of a small spiral galaxy with a giant elliptical.''}
On large scales, the light distribution has long been known to follow a
standard $r^{1/4}$ profile \citep{vandenberg76,dufour+79}.
Further observations of the stellar populations 
since that time have continued to support its underlying normality.
Direct spectroscopic measurements of the ages and
compositions of its globular clusters throughout the bulge and halo
\citep{peng+04GCS,beasley+08,woodley+10a} show that their age distribution has a
clearly wider range than is the case for
 the Milky Way clusters.  But the great
majority of them are older than $\tau \gtrsim 8$ Gy, with a small
fraction that may have arisen in later formation events.  This pattern
is very much like what is seen in a number of other ellipticals \citep[e.g.][]{puzia+05}.
In addition, the low-metallicity clusters that traditionally mark
the earliest star formation epoch $ 11 - 13$ Gyr ago in large galaxies
are strongly present in NGC 5128.
The kinematics and dynamics of the halo as measured through its
globular clusters \citep{woodley+07, woodley+10b} and planetary nebulae \citep{peng+04PN}
also do not present anomalies compared with other gE systems.
Lastly, as is mentioned above, the halo field stars as sampled
so far show a predominant uniformly old population with a wide range
of metallicities. 

In summary, the existing data indicate that we may be able to learn
a great deal about the old stellar populations in giant E galaxies
by an intensive study of NGC 5128.

To date, there is only a handful of luminous galaxies in which studies of resolved 
halo stars have been carried out. A recent such investigation of the M81 
halo \citep{durrell+10} provides a summary of the results for both the spiral galaxy 
halos: the Milky Way \citep[e.g.][]{ryan+norris91,carollo+07,ivezic+08,juric+08},  M31 
 \citep[e.g.]{mould+kristian86,durrell+01,ferguson+02,brown+03,kalirai+06,chapman+06,ibata+07,mcconnachie+09}, and NGC~891 \citep{rejkuba+09},  as well as for elliptical galaxies NGC~5128, NGC~3377  \citep{harris+07a}, and NGC~3379 \citep{harris+07b}. Few other luminous galaxy halos beyond the Local Group have been also resolved, for example  M87 \citep{bird+10} and Sombrero \citep{mould+spitler10}, but their colour-magnitude diagrams (CMDs) are not deep enough for detailed population studies. 
 
The metallicity distributions of these large galaxies display a wide diversity. However, taking into account the different locations sampled, as well as the presence of rather ubiquitous substructures  in the stellar density and metallicity distributions, the emerging picture from these studies seems to point to the fact that large galaxies host a relatively more metal-rich inner halo component and a metal-poor outer component, which starts to dominate beyond $\sim 10-12 \mathrm{R}_{eff}$ \citep{harris+07b}. 

The {\it ages} of halo stars are even less well known than their metallicity distributions. In the Milky Way there are halo stars that are as old as the oldest globular clusters, but the overall age distribution of the halo stars is uncertain. Currently only for Local Group galaxies can the age, and the detailed star formation history, be derived based on observations that reach as deep as the oldest main sequence turn-off. The mean age of  M31 halo fields studied by \citet{brown+08} is between 9.7-11~Gyr. For more distant galaxies the mean age can be obtained from the fits to the age and metallicity sensitive luminosity features such as red clump, asymptotic giant branch bump and red giant branch bump.  \citet{rejkuba+05} derived luminosity weighted mean age of $8^{+3}_{-3.5}$~Gyr for NGC~5128 halo, and \citet{durrell+10} obtained a mean age of M81 halo stars of $9 \pm 1$~Gyr. In all three galaxies (M31, M81 and NGC~5128) there are stars younger than $\la 8$~Gyr, but the bulk of the population is old.  In M31 the intermediate-age component contributes to about 30\% of the halo mass \citep{brown+08}. For galaxies beyond the Local Group this is an open question. Here we address this question for NGC~5128.

\section{The data and goals for this study}

In Paper I, we presented a full description of the data and photometry
in this field $\sim$33' south of the galaxy center and
then used interpolation within the grid of evolved low-mass stellar models 
of \citet{vandenberg+00},  with $(V-I)$ colours calibrated against fiducial 
Milky Way globular clusters, to derive an empirical metallicity distribution.  
Furthermore we compared the observed luminosity function (LF) with 
theoretical LFs for single age populations convolved with the observed 
metallicity distribution function (MDF). These theoretical LFs
were constructed using the  BASTI stellar evolutionary 
tracks \citep{pietrinferni+04}. Comparing them with the observed LF, 
and in particular with the luminosities of the RC/HB and the AGB 
(asymptotic giant branch) bump allowed us to estimate the \emph{mean age} 
of the NGC 5128 halo stars to be $8^{+3}_{-3.5}$~Gyr.
Although both, RC and AGB bump, features point to an old mean age, we
found discrepancies between these average ages
both between the RC and AGB bump positions, and between the
$V$ and $I$ luminosity for a given feature.
These offsets suggest to us that a single age stellar population, 
albeit with a wide metallicity spread, is inadequate to represent the data.
In any case, these rough indicators cannot replace a
more complete analysis of the entire CMD through
simulations with built-in age and metallicity distributions, that
allow for investigation of more complex star formation histories.  
The purpose of this
paper is to take the next step into these higher-level CMD comparisons.

Our dataset consists of 55,000 stars drawn from a location 38 kpc in
projected distance from the center of NGC 5128, brighter than the 50\% 
completeness limiting magnitudes $m_I=28.8$ and $m_V=29.7$ ($M_I = +0.7$).  
For the purposes of this study, we can
describe this sample as both \emph{unique} and \emph{limited} in the
context of other giant ellipticals:
\begin{description}
\item{a)} The CMD we have is \emph{unique} because it reaches deeper in
luminosity than for any other E galaxy beyond the Local Group.  More to the
point, it is the only gE in which we can capture direct, star-by-star photometry
of both the RGB and the red clump, and thus have 
direct leverage on the age distribution of the
oldest component of the parent galaxy.  This  is one of the most important
factors making NGC 5128 a unique resource for stellar population 
studies.\footnote{The next nearest large E galaxy is Maffei~1, but it is
impossible to explore its stellar populations at similar detail due to very
high extinction and its location behind the Galactic disk. The next giant E
is NGC 3379 in the Leo group at 11 Mpc.  HST imaging deep enough to resolve
the RC halo stars in $V,I$ and comparable with our NGC 5128 data 
is not completely out of the question, but would
require more than 350 orbits to complete just one field. The many  
attractive Virgo target E's at 16 Mpc would take considerably longer.}

\item{b)} It is also sharply \emph{limited} because the most important
age-sensitive features of the CMD that we would in principle very much
like to study (the turnoff and subgiant populations for the oldest component) are well beyond
reach and will remain so for many years \citep{olsen+03}.\footnote{The HST/ACS
camera could theoretically reach the old-halo turnoff of NGC 5128 with exposures
of about 3000 orbits in $V$ and $I$ combined, a prohibitively expensive prospect.}
In short, there is little prospect for improving soon the depth of our probe into
the CMD for gE halo stars beyond what we already have in hand, so it is clearly
worth developing the most complete analysis of it that we can.
\end{description}

\section{Modeling description}

\subsection{Synthetic CMD simulator}

The basic approach we use for gauging the star formation history of
the old halo of NGC 5128 is to construct synthetic CMDs from a library of
stellar models, and then to vary the input age and metallicity ranges
until we achieve a close match to the data.   
This general technique has become increasingly well developed 
since its first conception about 20 years ago \citep{tosi+89}, 
with several mathematical and statistical approaches that are now 
thoroughly described in the literature.
Useful descriptions of these methods are given at length in, for example,
\citet{gallart+96}, \citet{hernandez+99}, \citet{harris+zaritsky01}, 
\citet{dolphin02}, \citet{aparicio+gallart04}, \citet{aparicio+hidalgo09} and 
\citet{tolstoy+09}, and we will not discuss these in detail here.

Our approach is very much as is done in the codes described in those papers. 
Here we build many model CMDs each
populated with approximately the same total numbers of RGB, HB, and AGB
stars as in the data, by drawing from a library of evolved stellar
models covering a wide range of metallicities and masses.  
In each model specific assumptions are made about the star formation rate
(implicitly, the relative numbers of stars in each age bin), the IMF, 
and the metallicity distribution (including age-metallicity correlations).  
The synthetic CMD is then numerically broadened by the
measurement scatter and is cut off by the photometric incompleteness
functions as derived from the observed CMD.  

The CMD simulator is based on the code developed by \citet{greggio+98}, 
which has been adapted to simulations of single age populations with a wide
range of metallicities. Its input parameters are described 
in \citet{zoccali+03}, and the details of the Monte Carlo extractions
and interpolation of simulated stars on the stellar evolutionary grids
are described in \citet{rejkuba+04}. We summarize here only 
the main points and highlight those details that differ 
from the simulations of the Milky Way bulge \citep{zoccali+03}. 

The observational CMD shows a very wide RGB, readily interpreted
as trace of a wide metallicity distribution. Therefore we first 
calculated a large number of synthetic CMDs for single age 
populations. These simulations were then used as building blocks 
to construct more complex models. Unless specified otherwise, 
the adopted MDF was the one derived from the interpolation of 
colours of upper RGB stars. Here we stress another important 
difference with respect to our approach in Paper~I
concerning the consistency of the MDF. 
The metallicity distribution in Paper~I was derived based 
on the tracks from \citet{vandenberg+00}, which
were scaled as described in \citet{harris+harris00}, to match 
the Galactic globular clusters. Here for consistency we
re-derive the MDF using the same isochrones as used in the simulations. 

The synthetic CMDs are constructed by interpolating between the 
isochrones of BASTI models (referred to also as Teramo models)  
with both solar scaled  \citep{pietrinferni+04} and 
alpha-enhanced  \citep{pietrinferni+06} metal mixtures. 
These isochrones include the full set of evolutionary stages 
from main sequence (MS) to AGB that we need for this analysis.
The alpha-enhanced isochrones have included 
the thermally-pulsing AGB (TP-AGB) phase \citep{cordier+07}. 
The 
adopted $\alpha$-enhancement in the models is described in
detail in \citet{pietrinferni+06}. The overall average enhancement
for alpha-enhanced models is $[\alpha/\mathrm{Fe}] \sim 0.4$, consistent
with observations of the Galactic halo population.

\begin{figure}
\centering
\resizebox{\hsize}{!}{
\includegraphics[angle=0]{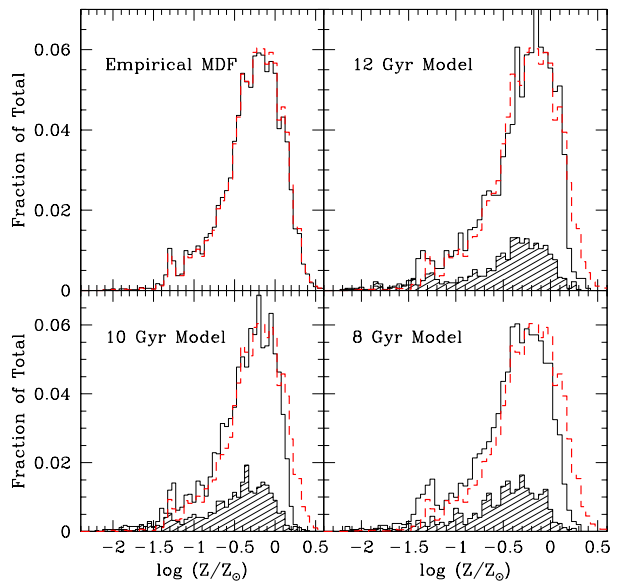}
}
\caption[]{Upper left panel: Differential metallicity distribution 
function (MDF),  normalized by the total number of 
stars,  derived from the observed CMD through the interpolation
on the grid of Teramo alpha-enhanced isochrones is shown with solid (black) line. 
The observed MDF including the correction for the AGB bias is overplotted as
dashed red histogram. The other three panels show the 
MDFs from single burst simulations, where simulations had in input 
the observed (upper left black histogram) MDF and single age bursts of 8, 10, and 12 Gyr. 
The shaded histograms show the contribution of 
AGB stars as a function of metallicity, while the open histograms show
the contribution of only the RGB stars for each
simulation. The MDFs of these simulated data are constructed 
in exactly the same way as for the
observations, by interpolating over the isochrones in CMDs after the photometric 
errors were added to the simulated stars, and can therefore be compared directly
with the observed AGB bias corrected MDF, which is overplotted with red dashed histogram.
}
\label{fig:MDF_AGBcontrib}
\end{figure}

The bolometric corrections in this work are taken from 
\citet{girardi+02}, while in Paper~I we used the tables 
of \citet{zoccali+03}, based on the empirically determined 
BCs \citep{montegriffo+98}.  The main difference 
between the two is in regard to the treatment of the red giants
with colours $V-I \ga 3.6$, equivalent to temperatures below 
$\sim 3250$~K, for which the \citet{montegriffo+98} bolometric corrections 
are several magnitudes larger (in absolute value) 
than those of \citet{girardi+02}. For giants with temperatures 
between $\sim 3800-3250$~K ($1.8 \la V-I \la 3.6$) the difference 
between the two scales is up to 0.5 mag; in this temperature
range the \citet{montegriffo+98} bolometric corrections are actually smaller
in absolute value, 
while for the hotter giants the two scales match very well. 
For the derivation of the empirical MDF from the observations 
the bolometric corrections were calibrated on Galactic globular clusters, 
as described in \citet{harris+harris00}.  

The MDF we used in input for the simulations has been carefully derived. 
In particular we introduce a correction to the empirical first-order MDF,
to remove what we will call the \emph{AGB bias}.  The grid of
isochrones we use to determine the empirical MDF consists of
the evolutionary tracks for the RGB stars, i.e.\ those along the
first ascent of the giant branch.  However, in any real sample of stars,
AGB stars (second ascent of the giant branch) are also present, and
these are slightly bluer than the RGB at a given metallicity and
luminosity.  Thus the empirical MDF derived from all the stars
will end up slightly  biased to
lower metallicity than it should.  In addition, because our data contain
a wide range of metallicities, the AGB and RGB populations are heavily
overlapped and we cannot remove the AGB bias just by cutting off the
bluest end.  However, the \emph{simulated} CMDs contain all the information
we need about the evolutionary stages of the stars in any given region
of the CMD, so we use these to find out what fraction
of the total population belongs to the AGB, and how they are distributed
in metallicity.

We first create the synthetic CMDs for single age populations with 8, 10 and
12~Gyr isochrones and the input ``empirical MDF'' derived from the 
interpolation of colours of all stars with magnitudes between $-3.6 < M_{bol} < -1.4$ 
in the observed CMD. In these synthetic CMDs we selected only the first 
ascent giants (RGB stars) and re-derived the new MDF in the same way 
as for the observations: interpolating the RGB stars colours with $-3.6 < M_{bol}< -1.4$
on the same set of models. This new MDF is the so-called AGB bias corrected MDF.
In the same way we also derive the MDF for only AGB stars for each simulation. 
In Fig.~\ref{fig:MDF_AGBcontrib} we compare the differential MDF histogram
derived from observations (upper left) with MDFs from three single age 
simulations run with empirical observed MDF in input 
and ages of 8, 10, and 12~Gyr. In all cases MDFs were constructed with 
stars in the range $-3.6<M_{bol}<-1.4$ and are normalized by the total number of stars. 
The shaded histograms show the MDFs of AGB stars, while the open histograms 
show the MDFs of only the RGB stars in the given simulation. For comparison
in each diagram we also overplot the empirical observed MDF including the 
AGB bias correction as a red dashed
histogram, renormalized to the same total number of stars.
As can be seen from the upper left panel of Fig.~\ref{fig:MDF_AGBcontrib},
the AGB bias correction turns out to have quite a small systematic effect on
our empirical MDF, with the main correction being the reduction of the 
already-small metal-poor tail.

Evaluation of our models in the $8 - 12$ Gyr range shows that over the
luminosity range $-3.6 < M_{bol} < -1.4$ 
that we use to determine the MDF, the RGB stars contribute 
76\% of the total population, with the rest being the AGB contaminants.
The actual ratio depends weakly on metallicity itself, but is well
represented by a simple interpolation curve that we obtain from the simulations,
\begin{equation}
\frac{N_{RGB}}{N_{total}}  =  0.098{\rm [Z/H]} + 0.816 \, .
\end{equation}
This curve accurately represents the ratio to within $\pm 0.05$ at any
metallicity and any age within the stated range; the scatter about
this mean line is produced mainly by the bin-to-bin Poisson fluctuations 
in the sample sizes that we are working with.
Thus at the extreme metal-poor end the AGB stars make up almost 40\% of the
population, falling to $\sim20$\% at solar metallicity and above.
Because the metal-poor bins in our observations contribute only small numbers
of stars, the global average of $\mathrm{N}_{RGB}/\mathrm{N}_{total} = 0.78$ is 
dominated by the heavily populated metal-richer bins.

Fig.~\ref{fig:MDF_AGBcontrib} also shows the classic age-metallicity degeneracy.
The 12~Gyr model matches the empirical MDF used in input closely, as expected 
because 12~Gyr isochrones are used to determine the MDF.  
For successively younger ages (most noticeable for the 8~Gyr model), 
the deduced MDF maintains the same shape but shifts
slowly to more metal-poor values, at the rate of about 0.1 dex per 3-4 Gyr.
In Paper I, we found the same amount of shift when using the Victoria isochrones
\citep{vandenberg+00}.  In short, from the RGB stars alone, we cannot
tell the difference between a 12~Gyr population with the input MDF, and
a population that is a few Gyr younger and intrinsically more metal-rich by
0.1-0.2 dex in the mean.  The most important way we have to break this
degeneracy is to use the colour distribution and luminosity function of
the RC, as we show later.

The observed empirical MDF, and the resulting MDF corrected for the AGB bias,
are given in Table~\ref{tab:MDF_AGBcorr}. In the rest of the paper, when 
we refer to ``input observed'' cumulative MDF, we mean this bias 
corrected MDF (third column of Table~\ref{tab:MDF_AGBcorr}).

\begin{table}
\caption{Observed metallicity distribution function for stars in the halo field
of NGC~5128. The second column lists the total number of stars in each metallicity
bin, while the third column gives the number of first ascent red giants in each metallicity bin,
after the correction for the AGB contribution.}
 \begin{tabular}{rrrrrr}
  $\log(Z/Z_\odot)$ & N & N$_{\mathrm{RGB}}$&  $\log(Z/Z_\odot)$ & N & N$_{\mathrm{RGB}}$ \\
\hline
 -2.05  &   0    & 0   & -0.65  & 101  &  76.0 \\
 -1.95   &  3    & 1.9   & -0.55  & 122  &  93.0\\
 -1.85   &  1    & 0.6   & -0.45  & 188  & 145.1\\
 -1.75   &  1    & 0.6   & -0.35  & 194  & 151.6\\
 -1.65   &  2    & 1.3   & -0.25  & 247  & 195.5\\
 -1.55   &  3    & 2.0  & -0.15  & 237  & 189.9\\
 -1.45   & 10   & 6.7  & -0.05  & 225  & 182.5 \\
 -1.35   & 37   & 25.3 &  0.05  & 218  & 179.0\\
 -1.25   & 23   & 16.0 &  0.15  & 111  &  92.2\\
 -1.15   & 38   & 26.7 &  0.25   & 35  &  29.4 \\
 -1.05   & 48   & 34.2 &  0.35   &  8   &  6.8 \\
 -0.95   & 46   & 33.3 &  0.45   &  1   &  0.9\\
 -0.85   & 59   & 43.2 &  0.55   &  3   &  2.6\\
 -0.75   & 97   & 72.0 &   0.65   &  0  &   0\\
\hline
 \end{tabular}
\label{tab:MDF_AGBcorr}
\end{table}

As in Paper~I we adopt  E(B-V)=0.11, the 
\citet{cardelli+89} extinction law, and the distance modulus of
$(m-M)_0=27.92$ \citep{rejkuba04,harris+10}. All simulations are derived with a 
single slope IMF with $\alpha=2.35$ \citep{salpeter55}, 
and masses $0.5<M/M_\odot<3$. Furthermore the simulations 
include the correct photometric errors and
completeness as derived from the artificial star experiments in 
Paper~1. These parameters are kept constant throughout. What we
change are the evolutionary background (solar scaled, alpha enhanced 
models), the age, and the metallicity distribution.

\begin{figure}
\centering
\resizebox{\hsize}{!}{
\includegraphics[angle=0]{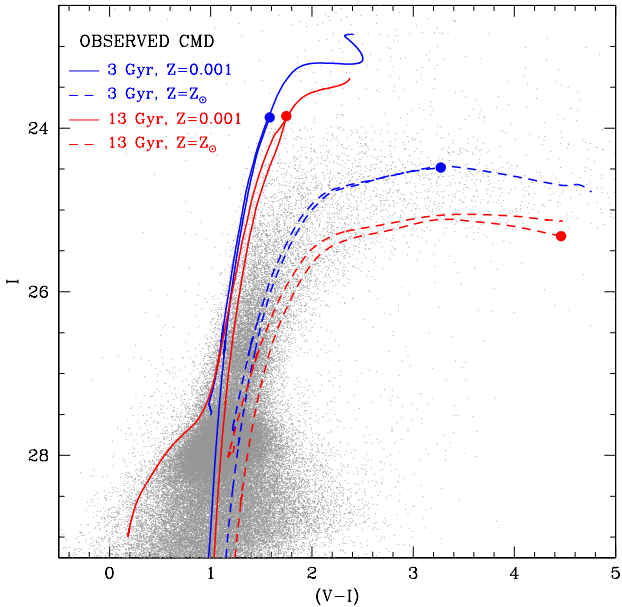}
}
\caption[]{Observed CMD with 3 (blue) and 13~Gyr (red) isochrones overplotted.
Dashed lines are used for solar (Z=0.0198) isochrones, while solid lines show metal-poor
(Z=0.001) isochrones. 
}
\label{fig:cmd_isochrones}
\end{figure}

The final set of single age simulations that we use to compare with the
observations includes:
\begin{enumerate}
\item single age burst models with input observed MDF 
and ages ranging from 2 to 13 Gyr, with a step size of 0.5 Gyr;
These simulations were run until the number of stars in a
box on the upper RGB ($24.5<I<26$ and $1.0<V-I<2.3$) matched
the observed $N_{box}=3131$ stars (Sect.~\ref{sect:1ageMDFrgb}).
\item single age burst models with a MDF following the closed box 
model with a range of yields, and minimum/maximum
metallicities. As for the previous case the number of stars
in the same box on the upper RGB was used as the condition to
stop the simulation (Sect.~\ref{sect:closedbox}).
\item single age burst models with a flat MDF selecting different 
metallicity ranges. These simulations had as a stopping condition
50,000 output stars (stars that passed all observational tests).
\end{enumerate}
The simulations with either the input observed MDF or a closed box MDF 
were compared directly with the observed CMD. 
The set of isochrones used to generate simulated CMDs
includes the following metallicities:
Z=0.0001, 0.0003, 0.0006, 0.001, 0.002, 0.004, 0.008, 0.01, 0.0198 (solar), 0.03,
and 0.04. 

Going beyond these single-age models, we next used these
as input to construct more complex (multi-burst) star 
formation histories. In particular single age flat MDF simulations 
were used to explore some complex star formation histories with 
specific age-metallicity relation. 
We did not include the simulations having single metallicity and 
age distribution, because the observed RGB is far
too wide to be reproduced in this way. 
The colour distribution on the bright RGB is far more sensitive to
metallicity than it is to age, compared with other parts of the observed CMD. 
Therefore we use it as our primary metallicity indicator.

As will be seen in the next sections,
the basic direction in which our conclusions are heading is that
the NGC 5128 halo stars are \emph{predominantly old} and with
a \emph{very wide metallicity range}.  We illustrate the essential
idea in Figure \ref{fig:cmd_isochrones}, where the
observed CMD is overplotted with isochrones 
for low Z (Z=0.001) populations of two ages that cover the range
we will be interested in (3 and 13 Gyr), and also
solar metallicity ($Z_{\odot}=0.0198$) isochrones of the same two ages.
A wide range of ages at any single metallicity cannot accommodate
the wide observed range of colours; the low metallicity isochrones are too blue for 
any age and the high metallicity isochrones are too red.
Moreover, the young, 3 Gyr old isochrones
overshoot the upper envelope of the observed RGB by about 1 mag: such a young
component, if present, must involve a very modest amount of stellar mass
to match the lack of stars brighter than the RGB tip.
In addition, NGC 5128 must contain a large component that is both old and
very metal-rich, because many RGB stars are redder than the young (3 Gyr) solar isochrone. 

Finally, the model is matched
to the observed CMD by dividing the colour/magnitude plane into a grid,
comparing the number of real and model stars in each cell of the
grid, and calculating the $\chi^2$ goodness of fit between the two.
Other diagnostics, such as luminosity function fits and
ratios of stars in different evolutionary phases, are used as well
to decide on the best result. We first describe in detail our choice of 
full CMD fitting, and then give some details about other diagnostic
diagrams, before launching into the discussion of the results of our experiment.

Appendices A and B (published only in the electronic form) 
list all single age simulations we explored as well as double burst simulations
that were compared with the observations.  
All the single age simulations that are made using
solar scaled isochrones have names starting with "sol*" and those that have
alpha enhanced models are named "aen*". All combined simulations 
have names starting with  "cmb*".

\subsection{Approach to matching the full CMD}

In previous studies, opinions have varied about how best to perform
the matchup between the observed and simulated 
CMDs.  For example, \citet{harris+zaritsky01, aparicio+hidalgo09} and others have used a 
$\chi^2-$minimization criterion across the grid to 
find the best range of input-model
parameter space.  On the other hand, \citet{dolphin02} strongly advocates
the use of a cumulative likelihood ratio, arguing that the number
counts within the grid cells intrinsically 
follow the Poisson distribution rather
than the Gaussian statistical rules that are implicit in 
a $\chi^2$ calculation.  We would agree with Dolphin's precepts in the
limit where the number $n$ of objects per cell is small, e.g. $n \lesssim 10$.
Such a limit would apply for cases in which many of the regions in the CMD
that are of key interest are very sparsely populated, even when the
total population $N$ is large and the 
cell size is comparable with the size of the measurement errors in
magnitude and colour.  However, that situation is not the case for our
data since, for our grid
definition, $n$ is always larger than 200.
In the limit of large $n$ the Poisson distribution converges
smoothly to the Gaussian one, and the practical difference 
between the $\chi^2$ statistic
and the likelihood ratio is moot 
\citep[see also][for direct comparisons and a similar 
conclusion]{brown+06,tolstoy+09}.

Another initial choice for the numerical setup is 
how to lay out the grid cells.  
We have experimented with a uniform grid (all cells 
the same size in both coordinates)
and also with an ``adaptive grid'' \citep{harris+zaritsky01,aparicio+hidalgo09} 
where the cells are smaller in areas of higher
stellar density, and where the stellar evolutionary models are more accurate and precise.  
Based on similar experiments Harris \& Zaritsky,  find that the best-fit
solutions are relatively insensitive to the particular grid structure. We have
found the same basic effect.  For the final runs 
we adopt an adaptive grid (see below) in which
the number $n$ per cell remains very roughly constant, though it is still fine
enough to track the most important changes in the stellar distribution with age and metallicity.
\citet{dolphin02} suggests that the bin size should be comparable to the
size of the smallest features in the CMD to which we want to be sensitive. 
In practice, however, this
criterion can be compromised by the photometric measurement scatter
and incompleteness, which (especially at the faint end) 
set a lower limit to the cell size that will be physically meaningful.
Smaller differences in the stellar distribution will be blurred out even if they
resulted from important differences in the age/metallicity history that we might
have hoped to measure.
Therefore, in our grid the \emph{minimum} cell size is 
similar to the observational uncertainties
in magnitude and colour.



\begin{table*}
\caption[]{Adopted grid for the computation of the $\chi^2$ for the full CMD fit. 
Magnitudes and colours are given in the observed plane, without extinction correction.
Column 6 lists the weight, column 7 gives the number of stars in the observed CMD.
}
\label{tab:grid}
\begin{tabular}{rrrrrrrl}
\hline \hline
\multicolumn{1}{c}{(1)} & \multicolumn{1}{c}{(2)} & 
\multicolumn{1}{c}{(3)}  & \multicolumn{1}{c}{(4)}&
\multicolumn{1}{c}{(5)}  & \multicolumn{1}{c}{(6)}& 
\multicolumn{1}{c}{(7)}  & \multicolumn{1}{c}{(8)}\\
\multicolumn{1}{c}{\#} & \multicolumn{1}{c}{$I_1$} &
\multicolumn{1}{c}{$I_2$}&\multicolumn{1}{c}{$(V-I)_1$} &
\multicolumn{1}{c}{$(V-I)_2$} & \multicolumn{1}{c}{$w$} & 
\multicolumn{1}{c}{$N_{obs}$} & \multicolumn{1}{c}{Comment}\\
\hline
  1 &  28.40 &  28.80 &  -0.25 &   0.55 & 0.3 &   781 & at 50\% completeness edge; $\sigma_I$=0.25-0.3, $\sigma_V=0.15-0.2$\\
  2 &  28.40 &  28.80 &   0.55 &   0.90 & 0.4 &  2027 & partly below 50\% line; $\sigma_I=0.25-0.3$, $\sigma_V=0.2-0.25$                                    \\
  3 &  28.40 &  28.80 &   0.90 &   1.35 & 0.2 &  1508 & partly below 50\% line; $\sigma_I=0.25-0.3$, $\sigma_V=0.2-0.25$                                    \\
  4 &  28.05 &  28.40 &   0.15 &   0.70 & 0.8 &  1145 & $\sigma_I=0.2-0.25$, $\sigma_V=0.2$                                                                   \\
  5 &  28.05 &  28.40 &   0.70 &   0.90 & 0.8 &  2111 & $\sigma_I=0.2-0.25$, $\sigma_V=0.2$                                                                   \\
  6 &  28.05 &  28.40 &   0.90 &   1.10 & 0.6 &  2894 & close to 50\% line; $\sigma_I=0.2-0.25$, $\sigma_V=0.25$                                              \\
  7 &  28.05 &  28.40 &   1.10 &   1.35 & 0.5 &  2854 & close to 50\% line; $\sigma_I=0.2-0.25$, $\sigma_V=0.25$                                         \\
  8 &  28.05 &  28.40 &   1.35 &   1.65 & 0.4 &   639 & partly below 50\% line; $\sigma_I=0.2-0.25$, $\sigma_V=0.3$                                           \\
  9 &  27.70 &  28.05 &   0.30 &   0.70 & 0.4 &   418 & some scattered/foreground stars?                                                              \\
 10 &  27.70 &  28.05 &   0.70 &   0.90 & 1.0 &  2212 & $2 \sigma$ off in color from the RC                                                              \\
 11 &  27.70 &  28.05 &   0.90 &   1.01 & 1.0 &  2750 &     RED CLUMP MAXIMUM $\sigma_I=0.18$, $\sigma_{(V-I)}=0.2$                                                 \\
 12 &  27.70 &  28.05 &   1.01 &   1.24 & 1.0 &  7351 &     RED CLUMP MAXIMUM $\sigma_I=0.18$, $\sigma_{(V-I)}=0.2$                                                 \\
 13 &  27.70 &  28.05 &   1.24 &   1.36 & 1.0 &  2775 &     RED CLUMP MAXIMUM $\sigma_I=0.18$, $\sigma_{(V-I)}=0.2$                                                 \\
 14 &  27.70 &  28.05 &   1.36 &   1.59 & 1.0 &  2399 & $2 \sigma$ in color off from the RC                                                              \\
 15 &  27.70 &  28.05 &   1.59 &   1.90 & 0.4 &   444 & at the 50\% limit                                                                              \\
 16 &  27.40 &  27.70 &   0.40 &   0.93 & 0.8 &   633 & $\sigma_I=0.15$, $\sigma_V=0.1$                                                                       \\
 17 &  27.40 &  27.70 &   0.93 &   1.13 & 1.0 &  2039 & $1-3 \sigma$ above the RED CLUMP                                                                 \\
 18 &  27.40 &  27.70 &   1.13 &   1.28 & 1.0 &  2077 & $1-3 \sigma$ above the RED CLUMP                                                                 \\
 19 &  27.40 &  27.70 &   1.28 &   1.48 & 1.0 &  1624 & $1-3 \sigma$ above the RED CLUMP                                                                 \\
 20 &  27.40 &  27.70 &   1.48 &   2.00 & 0.8 &   631 & $\sigma_I=0.15$, $\sigma_V=0.2$  (lower weight due to larger V error)                                 \\
 21 &  27.20 &  27.40 &   0.55 &   1.02 & 0.9 &   299 &                                                                                                   \\
 22 &  27.20 &  27.40 &   1.02 &   1.18 & 1.0 &   627 &                                                                                                   \\
 23 &  27.20 &  27.40 &   1.18 &   1.33 & 1.0 &   747 & $\sigma_I=0.1$, $\sigma_V=0.1$                                                                        \\
 24 &  27.20 &  27.40 &   1.33 &   1.48 & 1.0 &   414 &                                                                                                   \\
 25 &  27.20 &  27.40 &   1.48 &   1.95 & 0.9 &   255 &                                                                                                   \\
 26 &  27.00 &  27.20 &   0.60 &   1.10 & 1.0 &   291 &                                                                                                   \\
 27 &  27.00 &  27.20 &   1.10 &   1.25 & 1.0 &   489 &                                                                                                   \\
 28 &  27.00 &  27.20 &   1.25 &   1.40 & 1.0 &   487 &                                                                                                   \\
 29 &  27.00 &  27.20 &   1.40 &   1.92 & 1.0 &   392 &                                                                                                   \\
 30 &  26.85 &  27.00 &   0.70 &   1.15 & 0.9 &   258 &                                                                                                   \\
 31 &  26.85 &  27.00 &   1.15 &   1.30 & 1.0 &   458 &                                                                                                   \\
 32 &  26.85 &  27.00 &   1.30 &   1.45 & 1.0 &   458 &                                                                                                   \\
 33 &  26.85 &  27.00 &   1.45 &   1.90 & 0.9 &   262 &                                                                                                   \\
 34 &  26.70 &  26.85 &   0.75 &   1.18 & 1.0 &   268 &                                                                                                   \\
 35 &  26.70 &  26.85 &   1.18 &   1.26 & 1.0 &   241 & within $1.5 \sigma$ from AGB bump                                                               \\
 36 &  26.70 &  26.85 &   1.26 &   1.35 & 1.0 &   333 & AGB BUMP PEAK $\sigma_I, \sigma_{(V-I)} = 0.085$                                           \\
 37 &  26.70 &  26.85 &   1.35 &   1.46 & 1.0 &   336 & within $1.5 \sigma$ from AGB bump                                                               \\
 38 &  26.70 &  26.85 &   1.46 &   1.90 & 1.0 &   327 &                                                                                                   \\
 39 &  26.55 &  26.70 &   0.85 &   1.25 & 1.0 &   337 &                                                                                                   \\
 40 &  26.55 &  26.70 &   1.25 &   1.43 & 1.0 &   508 &      $2 \sigma$ from AGB bump in magnitude                                                           \\
 41 &  26.55 &  26.70 &   1.43 &   1.92 & 1.0 &   367 &                                                                                                   \\
 42 &  26.30 &  26.55 &   0.90 &   1.25 & 1.0 &   305 &                                                                                                   \\
 43 &  26.30 &  26.55 &   1.25 &   1.40 & 1.0 &   522 &                                                                                                   \\
 44 &  26.30 &  26.55 &   1.40 &   1.55 & 1.0 &   433 &                                                                                                   \\
 45 &  26.30 &  26.55 &   1.55 &   2.00 & 1.0 &   306 &                                                                                                   \\
 46 &  26.00 &  26.30 &   0.95 &   1.30 & 1.0 &   276 &                                                                                                   \\
 47 &  26.00 &  26.30 &   1.30 &   1.45 & 1.0 &   432 &                                                                                                   \\
 48 &  26.00 &  26.30 &   1.45 &   1.60 & 1.0 &   383 &                                                                                                   \\
 49 &  26.00 &  26.30 &   1.60 &   2.10 & 1.0 &   375 &                                                                                                   \\
 50 &  25.60 &  26.00 &   1.05 &   1.40 & 1.0 &   309 &                                                                                                   \\
 51 &  25.60 &  26.00 &   1.40 &   1.55 & 1.0 &   387 &                                                                                                   \\
 52 &  25.60 &  26.00 &   1.55 &   1.80 & 1.0 &   420 &                                                                                                   \\
 53 &  25.60 &  26.00 &   1.80 &   2.50 & 1.0 &   314 &                                                                                                   \\
 54 &  25.00 &  25.60 &   1.10 &   1.50 & 1.0 &   325 &                                                                                                   \\
 55 &  25.00 &  25.60 &   1.50 &   1.75 & 1.0 &   413 &                                                                                                   \\
 56 &  25.00 &  25.60 &   1.75 &   2.10 & 1.0 &   359 &                                                                                                   \\
 57 &  25.00 &  25.60 &   2.10 &   3.00 & 1.0 &   338 &                                                                                                   \\
 58 &  23.80 &  25.00 &   1.25 &   1.85 & 1.0 &   338 &                                                                                                   \\
 59 &  23.90 &  25.00 &   1.85 &   2.40 & 1.0 &   406 &                                                                                                   \\
 60 &  24.00 &  25.00 &   2.40 &   3.00 & 1.0 &   300 &                                                                                                   \\
 61 &  24.25 &  25.60 &   3.00 &   4.50 & 0.8 &   380 &  lower weight due to uncertain bolometric corrections for cool RGB stars   \\
\hline                                                                                                                                                    
\end{tabular}                                                                                                                                             
\end{table*}                                                                                                                                               

Beyond these numerical criteria, our approach to matching the model
 and observed CMDs is more strongly driven by 
the astrophysical limitations of our data  
than by statistical formalism.  Most previous studies of this type
\citep[][among others]{aparicio+97,dolphin02,harris+zaritsky04,brown+06,mcquinn+09,vanhollebeke+09}, including such targets as the Galactic
bulge, the Magellanic Cloud field-star populations, the M31 outer disk,
and nearby dwarf galaxies, employ CMD data that cover a 
wide luminosity range from the tip of the RGB down to below the turnoff point
of the classic old population, giving the
strongest possible leverage on the age distribution independently of
metallicity.  These target fields also typically include stars over wide
ranges of both age and metallicity, with very significant young
components.  

By contrast, the NGC 5128 halo stars 
cover a relatively small range in age (with only a small and perhaps
negligible fraction younger than $\sim 5$ Gy) but a very large range in 
metallicity (see Paper I).  In addition,
we have only the luminosity range of the HB and above to work with.  
This more restricted range in the evolutionary stages of the stars can still yield solutions
for the age distribution that are \emph{accurate} (that is,
they return systematically correct age ranges), though they are very definitely
less \emph{precise} (that is, with larger uncertainties) than if the turnoff
region were included; see, for example, the simulation tests in
\citet{dolphin02}, particularly his Fig.~7 and accompanying text. 

The size of our dataset of $\sim 70,000$ stars 
(less than 56,000 are above 50\% completeness 
limit and only these are compared to the models) is smaller than 
samples studied for example by \citet{harris+zaritsky01,harris+zaritsky04},  
who observed $4 \times 10^6$ LMC field stars and $6 \times 10^6$ stars in the SMC, as well as 
the sample of  \citet{vanhollebeke+09}, who had $6 \times 10^6$ stars in their Galactic-bulge study, and of \citet{brown+06}, who had $\sim 10^6$ stars in their study of the M31 disk and halo. 
It resembles more closely instead the sample sizes of 
the Fornax dSph \citep{coleman+dejong08},  the nearby starburst dwarfs
studied by \citet{mcquinn+09}, and the M81 outer-disk and dwarf satellite
studies of \citet{weisz+08} and \citet{williams+09}. We note however that 
the targets from the above mentioned studies exhibit either well sampled main sequences or in some cases very obvious young components, hence providing evidence of a wide total age mixtures unlike our pure old-halo sample.  

Thus, within the limitations of
 the present data we can ultimately determine only
some appropriate \emph{ranges} for the age distribution and the star formation
history.  The model fits to be discussed below are definitely
 capable of ruling out 
large sections of the total parameter space.  But the classic 
age/metallicity/alpha-enhancement
degeneracies that affect the high-luminosity 
regions of the CMD for extremely old
stellar populations leave us unable to isolate a single ``best'' solution.
Nevertheless, some clear conclusions emerge about such
important features as the minimum age spread
and the relative proportions of stars in different age ranges, that go
well beyond our initial study in Paper I.

\subsection{Diagnostics}

\begin{figure}
\centering
\resizebox{\hsize}{!}{
\includegraphics[angle=0]{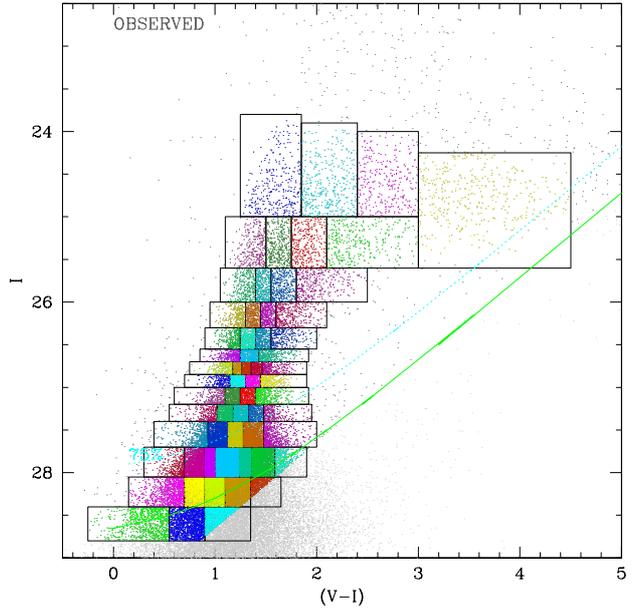}}
\caption[]{Observed CMD with overplotted ``adaptive grid'' based on which
 $\chi^2$ of the CMD fit is computed.
}
\label{fig:gridcmd}
\end{figure}

Before going into the results of the modeling, we describe here 
the diagnostics used to evaluate the goodness of the
fit between the various models and the observations. 

\begin{figure}
\centering
\resizebox{\hsize}{!}{
\includegraphics[angle=0]{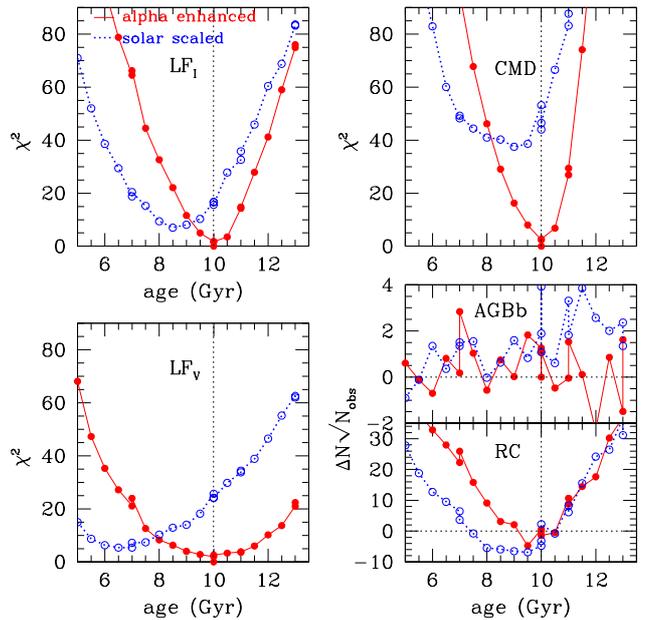}}
\caption[]{Comparing a single burst simulation for 10 Gyr old burst with 
alpha enhanced stars against all single age simulations as a test of the diagnostics. 
On the left are $\chi^2$ diagnostics based on the I-band LF (upper left) and V-band
LF (lower left). The upper right panel shows the diagnostic power of the full CMD
fitting, while the lower right panel compares the number of red clump (RC) and
AGB bump stars as a function of age. $\Delta N$ is the difference between the number
of stars in the ``input" (in this case simulation aen016) and the number of stars in the
same boxes in the simulated CMD, normalized to the Poissonian noise of the number
of ``observed" (input) stars.  
Simulations constructed with $\alpha$-enhanced isochrones are shown with
red filled dots, and solar scaled models are plotted with blue open symbols.
}
\label{fig:diagnostics_chi2}
\end{figure}

\begin{enumerate}
\item  \emph{Comparison of the luminosity functions in  V and I:}
 
We check the positions of maxima of the RC and AGB
bump features, the width of these features, and the $\chi^2$ fit 
over the whole LF. This comparison is done only for
magnitudes above the 50\% completeness limits: $V<29.65$, and $I<28.8$.

\item \emph{Comparison with overall $\chi^2$ of the CMD:}

The $\chi^2$ is calculated with the 
following formula:
\begin{equation}
\chi^2 = \frac{\Sigma_i (N_{obs,i}-N_{sim,i})^2 * w_i}{N_{obs,i}}
\end{equation}
where $N_{obs,i}$ is the number of observed stars in the $i-th$ box, 
$N_{sim,i}$ is the number of simulated stars in the $i-th$ box, and $w_i$ are
the weights (normalized to have the sum of 1). The reduced  
$\chi_\nu^2$ is given by the ratio of $\chi^2$ to the number of boxes. 

The boxes are shown in Figure~\ref{fig:gridcmd}, and are listed in Table~\ref{tab:grid}. 
They have been chosen based on
features in the observed CMD, combined with the photometric accuracy,
completeness and uncertainty in the bolometric corrections. In particular, we
selected larger box sizes in the faint part of the CMD to accommodate for the
scattering of stars due to photometric errors. The boxes in the middle of the
CMD where the number of stars is large are small in colour-magnitude space,
 but get wider at the edges of
the RGB, in order to maintain the statistics. In the upper part of the diagram
the boxes are very large to sample at least 250 stars, and the reddest box is
wide enough to cope with the changing shape of the simulated extent of the cool
red giant branch possibly due to inaccuracies in the colour-temperature
transformations. 

We also tested our results using the regular grid with smaller boxes over the
CMD, but excluding the reddest part of the RGB ($V-I>2.4$). While the values
of the reduced $\chi_\nu^2$ turn out to be sensitive to the grid size and number
of cells, the indication as to the best fitting models was robust.

To check the sensitivity of the $\chi^2$ tests, we run several simulations with
different random seeds in input and 
then compared those with other simulations with
the same set of parameters. This then allows us 
to estimate an error on $\chi^2$ values 
due to simple Poissonian statistics (since the sample of the observed/simulated 
stars is limited). The systematics can be 
assessed better by comparing simulations
vs.\ simulations while changing one of the parameters
(see discussion below and Figure~\ref{fig:diagnostics_chi2}).

\item \emph{Comparison with stellar counts on the RC and AGB bump:}

This diagnostic is based on selected boxes along the RGB that target
features sensitive to age and metallicity distribution, like the RC and the AGB bump.
The stellar counts for the RC are assumed to be the sum of all stars in the boxes 
11, 12 and 13 indicated as RC maximum boxes in Table~\ref{tab:grid} and 
covering the range $27.7<I<28.05$ and $0.9<V-I<1.36$. 
The total number of stars in the observed CMD within these boxes is
12876. The stellar counts for the AGB bump feature are constructed by
summing all stars within boxes 31, 32, 35, 36, 37 and 40 (Table~\ref{tab:grid}) that
cover the range between $26.55<I<26.7$ and $1.15<V-I<1.46$ which is within 
$2\sigma$ of the AGB bump magnitude and colour peak. The total number
of stars in the observed CMD within these boxes is 2334. 

\end{enumerate}

To gauge the sensitivity of our diagnostics on the age
we show in Figure~\ref{fig:diagnostics_chi2} the results of a comparison 
between 
a set of single age simulations using a template simulation (aen016) 
constructed with alpha-enhanced 10 Gyr old models and the input observed MDF.
The four panels refer to different diagnostics, namely the $\chi^2$ fit
of the luminosity functions in the $V$ and $I$ bands, and of the overall
CMD, and the number of stars in the Red Clump and AGB bump.
The various models assume the same input MDF as the template simulation and the
models shown in red (filled dots) assume alpha enhanced isochrones with various ages.
The $\chi^2$ value of the fit for the red model with an age of 10~Gyr 
with respect to the template simulation is 1, as expected when comparing
two simulations with all input parameters the same, except the difference in 
the initial random number seed. The blue dotted curves (and open circles), instead,
refer to models based on the solar scaled isochrones of the BASTI set,
and are meant to explore what happens if we use solar scaled tracks to 
interpret alpha enhanced stars. The pronounced minima in the $\chi^2$ curves show
that (in this very basic comparison) the diagnostic on the age is good.
Different $\chi^2$ diagnostics show that full CMD fitting and the I-band
LF are the most sensitive.
When using solar scaled isochrones, the LFs yield an age systematically too
young, but the $\chi^2$ for the overall CMD clearly indicates the need
for alpha-enhanced tracks to match the template simulation. The number of
red-clump stars is also a fair age indicator, with the appropriate (alpha
enhanced) isochrones, though we note that the sensitivity of this indicator is 
somewhat lower than for the full LF and CMD fits. 

The population size in 
the AGB bump, instead, is not very sensitive to age. This is certainly in part
due to the smaller overall number of stars in the AGB boxes, combined with the
internal photometric scatter. Therefore the total number of stars is a less  
useful diagnostic, than is the position (luminosity and colour) of the AGB bump. 

The real sensitivity of our diagnostics will certainly be worse than what is
described so far, since observations differ from the template simulation in
many respects, e.g. in the age spread. In the next section we compare 
the data and simulations.

\section{Results}

In this section we take the approach of exploring the range of 
acceptable ages and age distributions for the observed CMD. First 
we compare with the single-age simulations and show that a single-age 
burst cannot fully reproduce the observations. 
Next, we show that some two burst simulations fit the observations equally
well in terms of overall CMD fit, but significantly better when comparing
the luminosity functions.   
Finally, we explore some simple solutions with multi-age and multi-enrichment 
components. In principle by adding additional free parameters
(percentage of stars of a given age and metallicity with respect to 
the total population), it should be possible to find some ``best fit'' 
model(s).  In practice the $\chi^2$ values for the full CMD fit
do not go below $\sim 50$ likely due to the age-metallicity degeneracy in our 
observational dataset and the possibly inadequate combination of abundance
ratios in the isochrones (see Figure~\ref{fig:diagnostics_chi2}). 
As stated earlier, the fact that the CMD does not reach the much more 
age-sensitive main sequence turnoff region limits the interpretation.
We show only some selected plausible star formation histories that 
provide as good a fit to the 
observations as at least the best fitting double burst model does. 

In addition to the comparison of the observed CMD with synthetic CMDs made with
input observed MDF, we also explore alternative, physically motivated
closed box chemical enrichment models.

\subsection{Single age models with observed MDF}
\label{sect:1ageMDFrgb}

\begin{figure}
\centering
\resizebox{\hsize}{!}{
\includegraphics[angle=0]{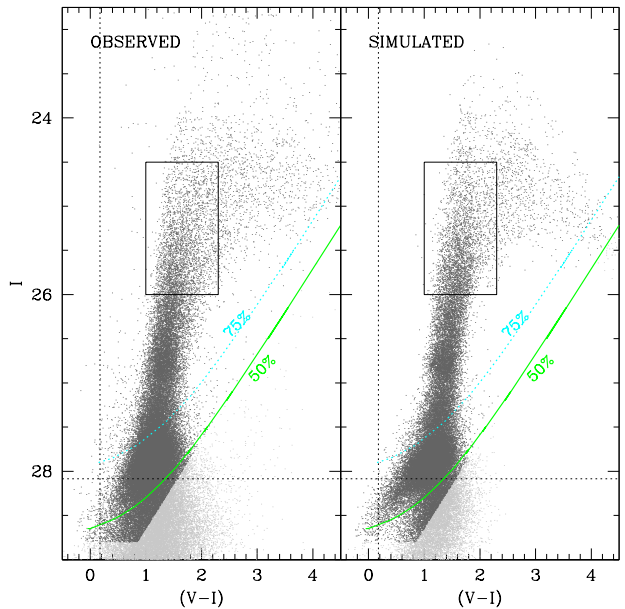}
}
\caption[]{Left panel shows the observed and the right panel the simulated CMD
for a single age burst population of 11 Gyr with the input observed MDF. The
simulation uses alpha enhanced stellar evolutionary models.
}
\label{fig:single_11Gyr}
\end{figure}

In the right panel of Figure~\ref{fig:single_11Gyr} we show the simulated CMD for 
a single age burst population of 11 Gyr with the input observed MDF made 
using the alpha enhanced isochrones. The number of stars in this simulation (and
other single age + MDF simulations) was constrained to match the star counts in
the box shown in the figure ($24.5<I<26$, $1.0<V-I<2.3$). This box contains 3131
stars and was chosen from the part of the CMD that is least sensitive to age, 
with the best photometric accuracy, and smaller
uncertainty in the colour-temperature transformations.

\begin{figure}
\centering
\resizebox{\hsize}{!}{
\includegraphics[angle=0]{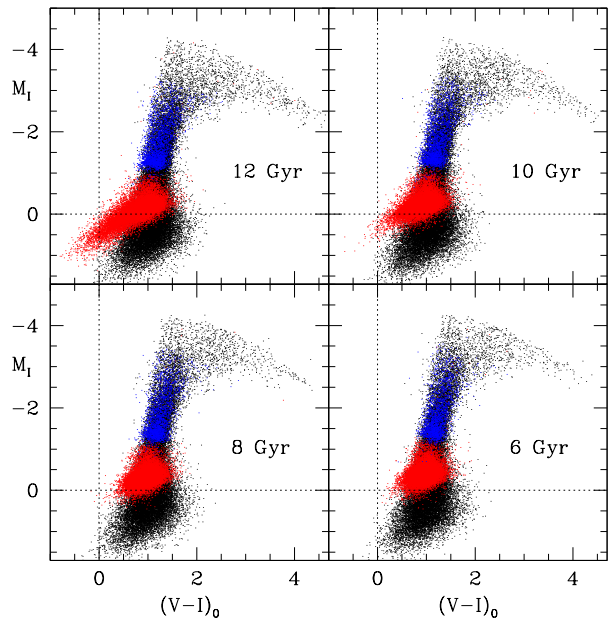}
}
\caption[]{Simulations of CMDs with input observed MDF and single age burst using
alpha enhanced stellar evolutionary models.
}
\label{fig:single_burst}
\end{figure}

In the left panel of Figure~\ref{fig:single_11Gyr} is the observed CMD. 
The overall appearance of the simulated CMD is fairly similar to
the data; however some discrepancies are evident. 
Note the difference in the upper envelope slope for the RGB
 between models and observation: the simulations have a
much stronger bend towards lower luminosity for the reddest red giants 
than is seen in the observed CMD. Possibly this is due to an overestimate
of the bolometric correction (in absolute value) to the I-band magnitudes at low temperatures. 
We also notice that the RGB is narrower in the simulated diagram, 
likely indicating that a single age is not a good fit to the
data. In the faintest part of the diagram the simulation shows more 
distinctly the core helium burning population, which in the data is more 
extended in colour, perhaps due to a combination of photometric errors and 
some age distribution effect. Finally, the red extension of the faintest 
portion of the diagram, below the 50\% completeness line, is affected 
more by the uncertainty in photometric accuracy and completeness. 
Therefore we do not include that part (shown in light gray) in our fitting. 

The widths of the RC and AGB bump features are smaller in the simulated CMD, 
also indicating that there may be a need 
to consider an extended star formation history.

In Figure~\ref{fig:single_burst} we show the 
sensitivity of the CMD to age.  As expected, changes in the age
make the most obvious differences in the RC feature;
In this figure we colour-code the simulated stars according to their 
evolutionary status. In black we show first-ascent giants (RGB) stars, red shows
core helium burners (RC) and in blue are the 
shell helium burners (AGB stars).
It is clear that in the models with oldest 
ages ($> 10$~Gyr) the blue tail
of the core helium burning stars extends to much 
fainter magnitudes and bluer colours
than in our data. The youngest model on the other hand has a brighter RC
than does our observed CMD.
These simulations, although simplistic, already
suggest to us that the bulk of the stars in the NGC~5128 halo 
formed $\sim 10$~Gyr ago. 

In the observed CMD within the RC region, the wide range of metallicities
and the photometric measurement scatter make it impossible to say 
whether any particular star belongs to the RC or the RGB.
Thus the RC colour difference with respect to the RGB colour at the same
magnitude unfortunately cannot be used as an age discriminator
(Hatzidimitriou 1991).

\begin{figure}
\centering
\resizebox{\hsize}{!}{
\includegraphics[angle=0]{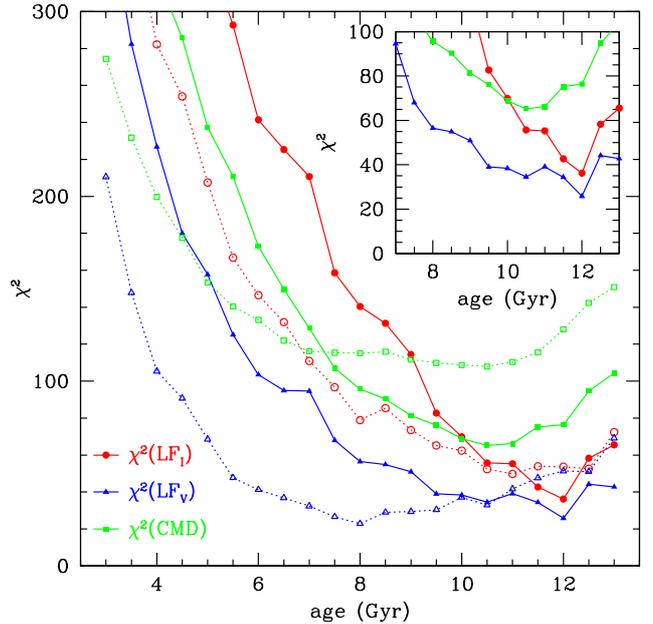}}
\caption[]{The $\chi^2$ diagnostic for luminosity function and full CMD fitting 
for observations with respect to all single age simulations. Comparison 
with the solar scaled single age simulations is shown with dotted lines 
(open symbols), and that with the alpha enhanced models  with solid lines 
(solid symbols).
}
\label{fig:diagnosticplotchi2}
\end{figure}

Figure~\ref{fig:diagnosticplotchi2} shows the $\chi^2$ 
diagnostic plots for observations with respect to all
single age simulations. Each dot on this graph represents one
single-burst model.  These indicate that the dominant 
stellar population in the observed CMD is $11 \pm 1$~Gyr old, and it 
appears to be more consistent with alpha enhanced abundance ratios.
The solar scaled models produce CMDs that differ significantly more
from the observed CMD, because they lack the stars along the blue 
edge of the RGB and RC. Notice that this applies also to the models 
constructed with the closed box metallicity distribution which has 
much more substantial population of low metallicity stars (Sect.~\ref{sect:closedbox}).

\begin{table}
\caption[]{The three best fitting models for each diagnostic are listed for 
the single age simulations with input observed MDF.
Diagnostics are $\chi^2$ for the full CMD, $\chi^2$ of the LF fit for I, and
V-bands, and in the last two columns 
$\Delta N / \sqrt{N_{obs}}= \mathrm{N}_{obs}-\mathrm{N}_{sim}/ \sqrt{N_{obs}}$ 
for the RC and AGBb boxes in our grid.  
}
\label{tab:bestfitmodels1mdf}
\begin{tabular}{crrrrrr}
\hline
\multicolumn{7}{c}{Single age models with input observed MDF} \\
\hline\hline
\multicolumn{1}{c}{simulation} & \multicolumn{1}{c}{age} &
\multicolumn{1}{c}{$\chi^2$}  & \multicolumn{1}{c}{$\chi^2$} &
\multicolumn{1}{c}{$\chi^2$}  & 
\multicolumn{1}{c}{$\frac{\Delta N}{\sqrt{N_{obs}}}$} &
\multicolumn{1}{c}{$\frac{\Delta N}{\sqrt{N_{obs}}}$} \\
\multicolumn{1}{c}{ID}& \multicolumn{1}{c}{(Gyr)}&
\multicolumn{1}{c}{CMD} & \multicolumn{1}{c}{LF$_I$} &
\multicolumn{1}{c}{LF$_V$} & \multicolumn{1}{c}{RC} &
\multicolumn{1}{c}{AGBb} \\
\hline
\multicolumn{7}{l}{CMD:} \\
 aen025 & 10.5 &   65.2 &   55.7 &   34.5 &  -10.6 &    7.0 \\ 
 aen018 & 11.0 &   65.6 &   53.6 &   40.5 &   -0.9 &    8.8 \\ 
 aen024 & 11.0 &   66.2 &   55.3 &   39.1 &    1.1 &    7.4 \\ 
\multicolumn{7}{l}{LF$_I$:}\\
 aen022 & 12.0 &   76.4 &   36.2 &   25.8 &    8.4 &    5.3 \\ 
 aen023 & 11.5 &   75.3 &   42.7 &   34.4 &    5.1 &    7.5 \\ 
 sol018 & 11.0 &  107.2 &   48.4 &   36.7 &   -3.7 &    9.1 \\ 
\multicolumn{7}{l}{LF$_V$:}\\
 sol030 &  8.0 &  115.1 &   78.9 &   22.9 &  -15.8 &    7.4 \\ 
 aen022 & 12.0 &   76.4 &   36.2 &   25.8 &    8.4 &    5.3 \\ 
 sol031 &  7.5 &  115.4 &   96.7 &   26.6 &  -10.9 &    8.9 \\ 
\multicolumn{7}{l}{RC:}\\
 sol033 &  6.5 &  122.0 &  131.9 &   36.8 &   -0.1 &    7.8 \\ 
 aen030 &  8.0 &   95.7 &  140.4 &   56.5 &   -0.5 &    6.9 \\ 
 aen018 & 11.0 &   65.6 &   53.6 &   40.5 &   -0.9 &    8.8 \\ 
\multicolumn{7}{l}{AGBb:}\\
 aen022 & 12.0 &   76.4 &   36.2 &   25.8 &    8.4 &    5.3 \\ 
 aen020 & 13.0 &  104.4 &   65.5 &   42.8 &   28.6 &    6.1 \\ 
 sol040 &  3.0 &  274.4 &  426.2 &  210.5 &   48.1 &    6.1 \\ 
\hline
\end{tabular}
\end{table}

Looking at individual diagnostics, the luminosity function $\chi^2$ 
fits for both I- and V-band point to an average single age of $12$~Gyr, 
while the overall CMD fit has a wider minimum at 10.5-11~Gyr. 
Table~\ref{tab:chisq1age} (published in the electronic version) 
reports the  $\chi^2$  values for all diagnostics, while the three best 
models for each individual diagnostic are reported in Table~\ref{tab:bestfitmodels1mdf}.
From the inspection of this table it is evident that the counts 
in the RC region are consistent with $11$~Gyr model as
well. We note however that this indicator is 
consistent also with the younger age of 8 Gyr, as found in Paper I.  
In these single age simulations AGB bump boxes are systematically less
populated with respect to the observed CMD. 

\begin{figure*}
\centering
\resizebox{\hsize}{!}{
\includegraphics[angle=0]{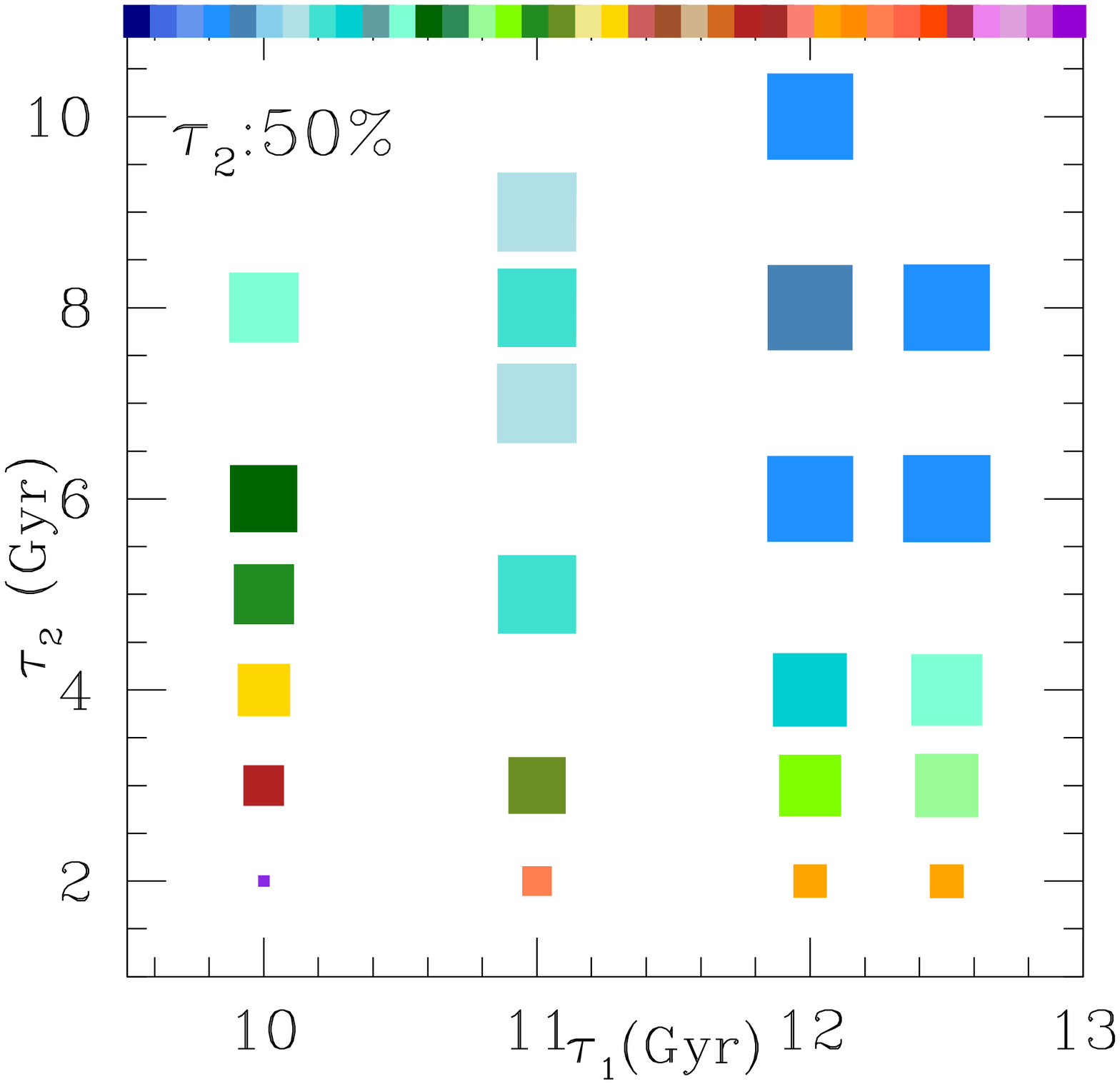}
\includegraphics[angle=0]{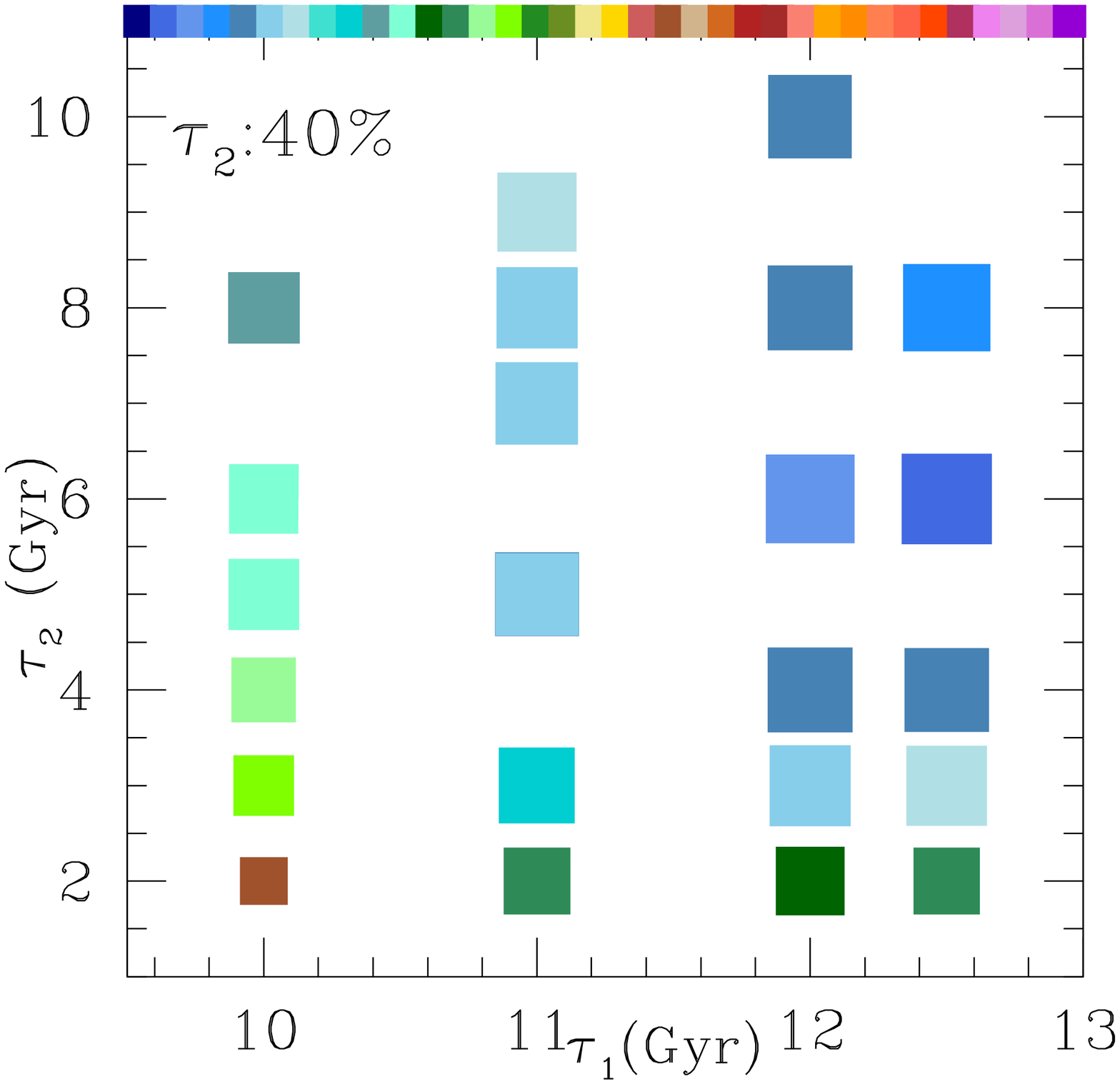}
\includegraphics[angle=0]{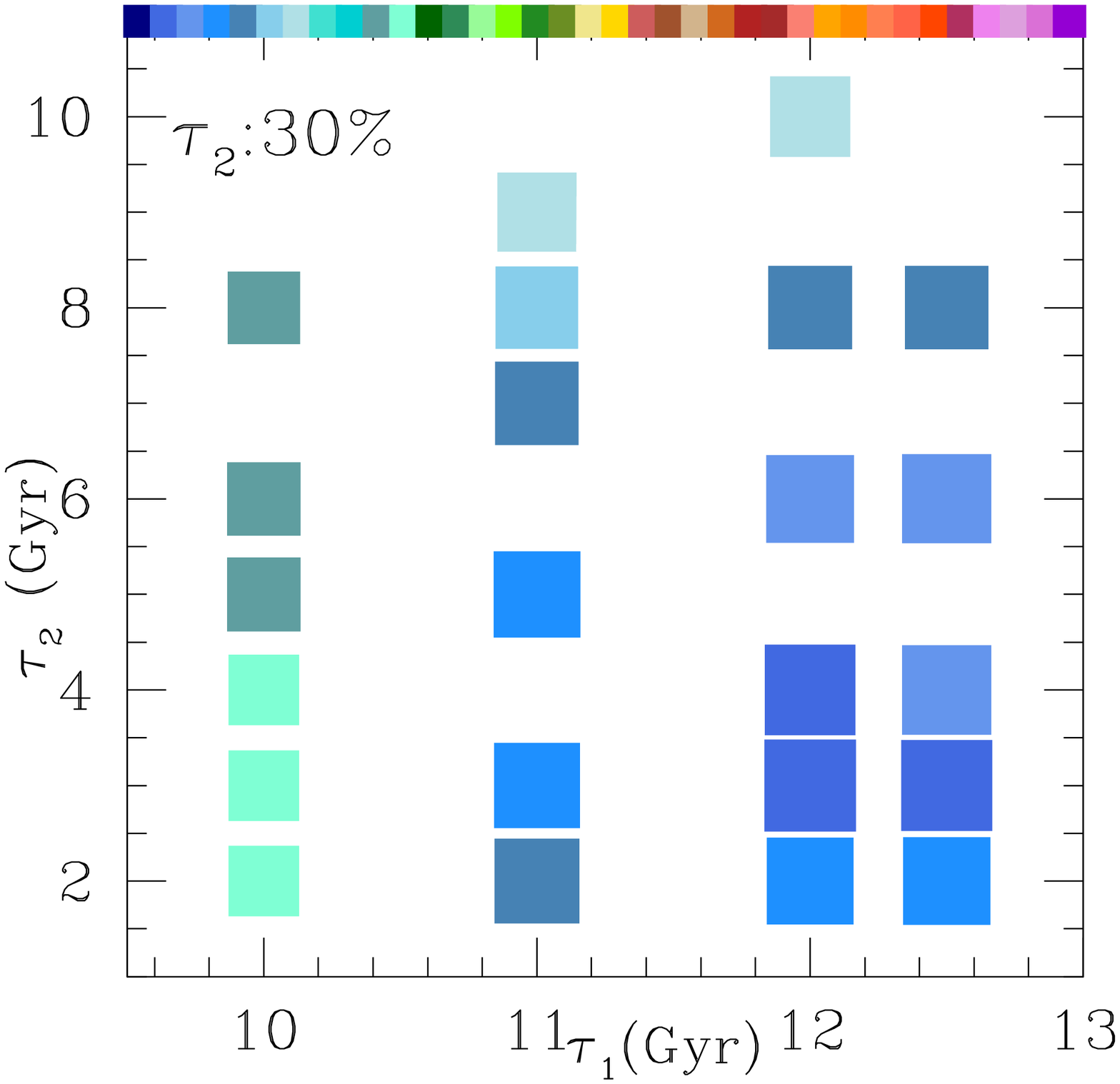}
\includegraphics[angle=0]{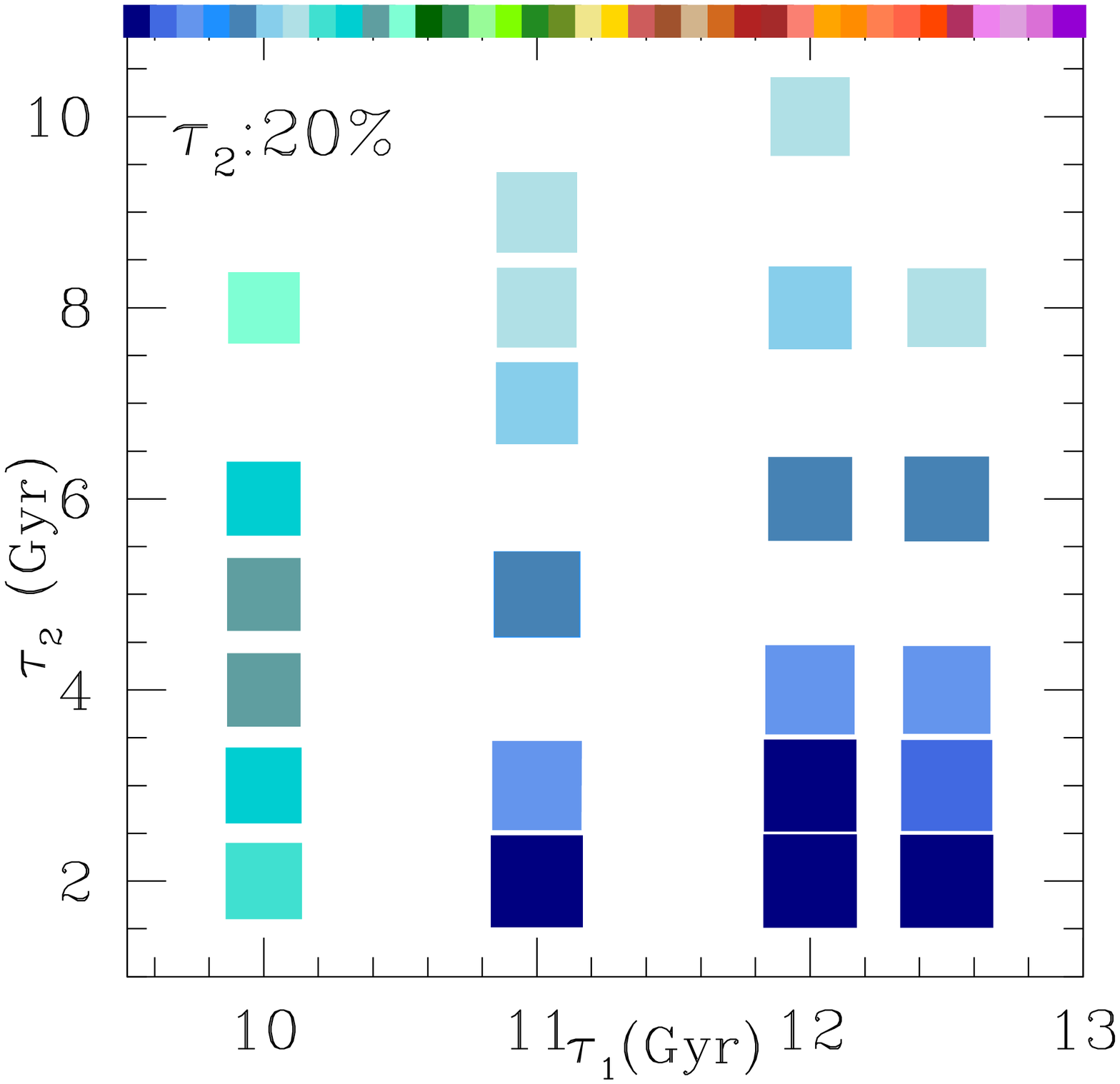}
\includegraphics[angle=0]{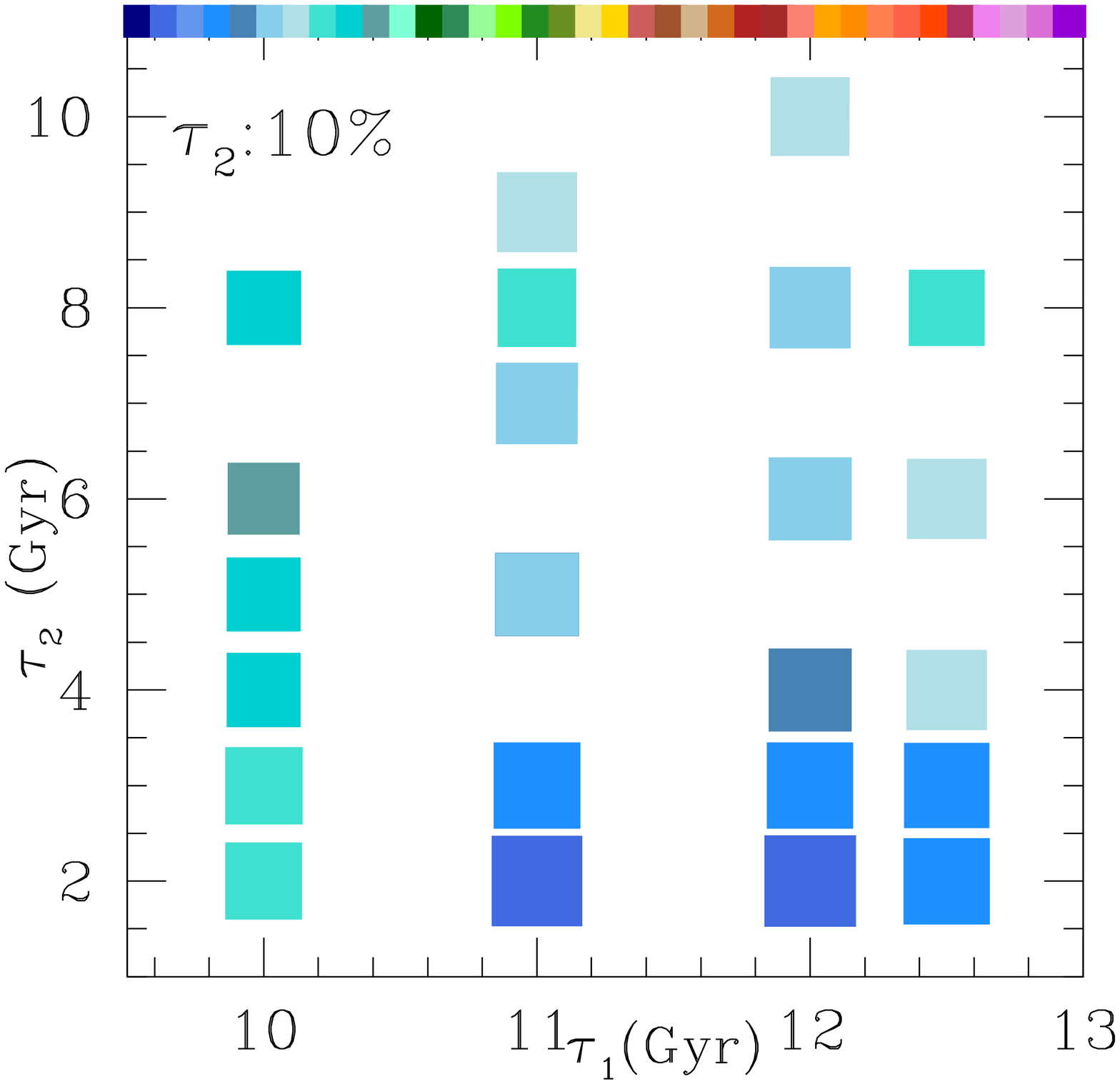}
}
\caption[]{The $\chi^2$ diagnostic for the full CMD fit for 2 burst
simulations, built using alpha-enhanced stellar evolutionary models,
as compared to observations  is plotted as a function of 
the old (x-axis) + young (y-axis) population age. 
Each panel shows different relative fractions
of old + young population in the combined 2-burst simulation.  
The size and colour of the points are normalized to the full
range of the $\chi^2$ values of the CMD fit for all 2-burst models with input observed MDF.
The larger the symbol, and darker blue its colour, the smaller the $\chi^2$. 
}
\label{fig:cmdchi2_2burst_perc}
\end{figure*}

\begin{figure*}
\centering
\resizebox{\hsize}{!}{
\includegraphics[angle=0]{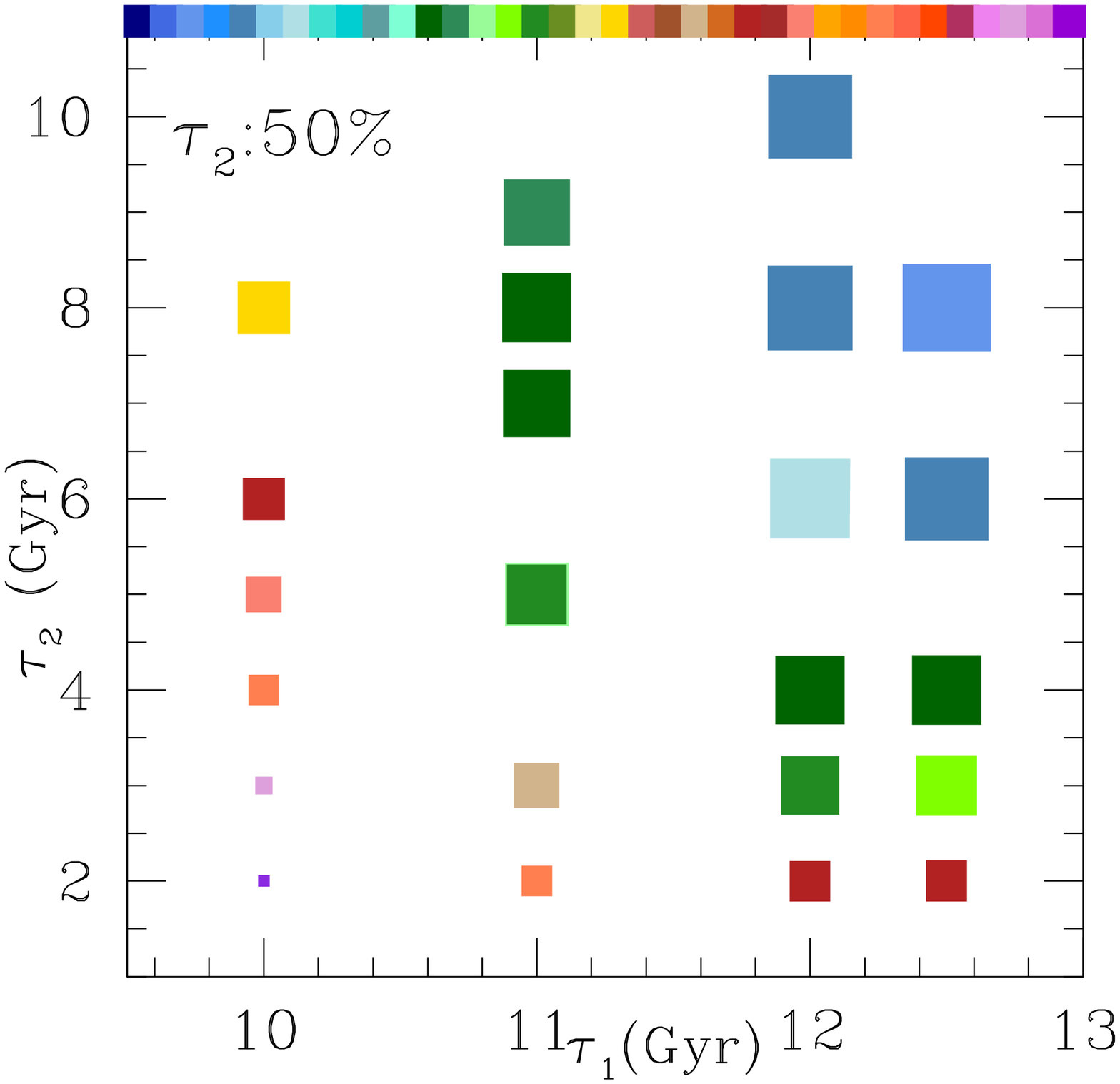}
\includegraphics[angle=0]{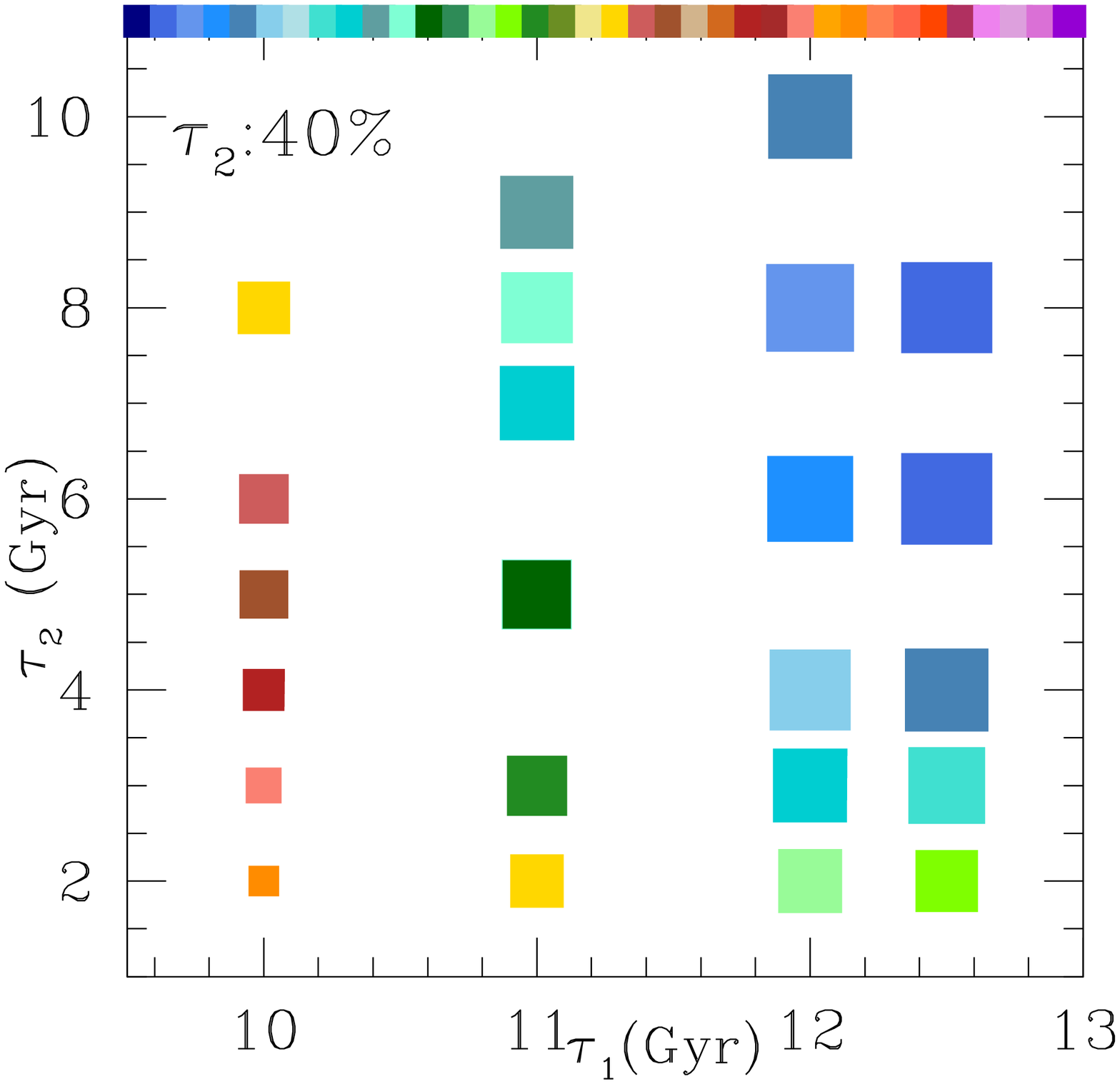}
\includegraphics[angle=0]{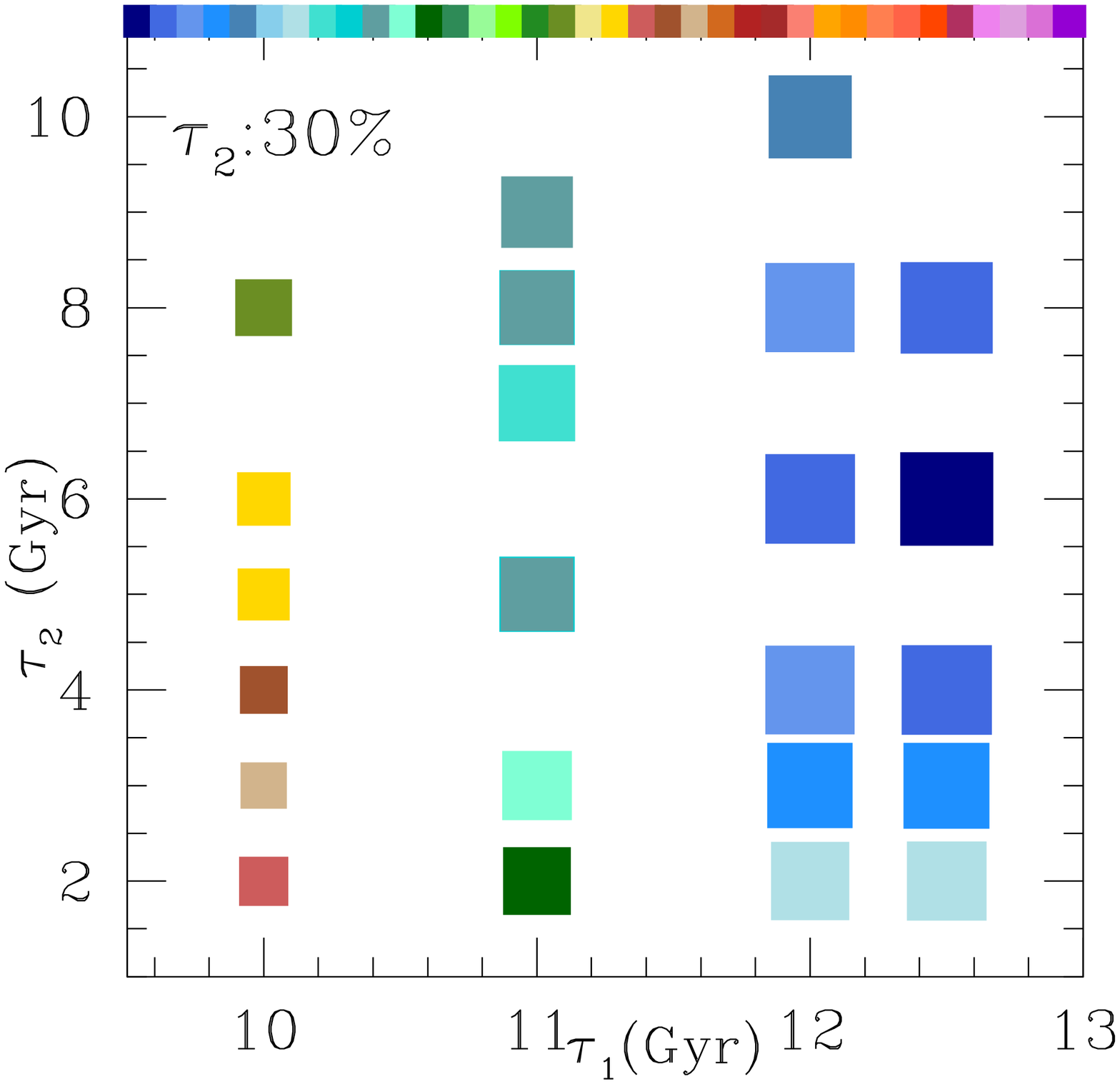}
\includegraphics[angle=0]{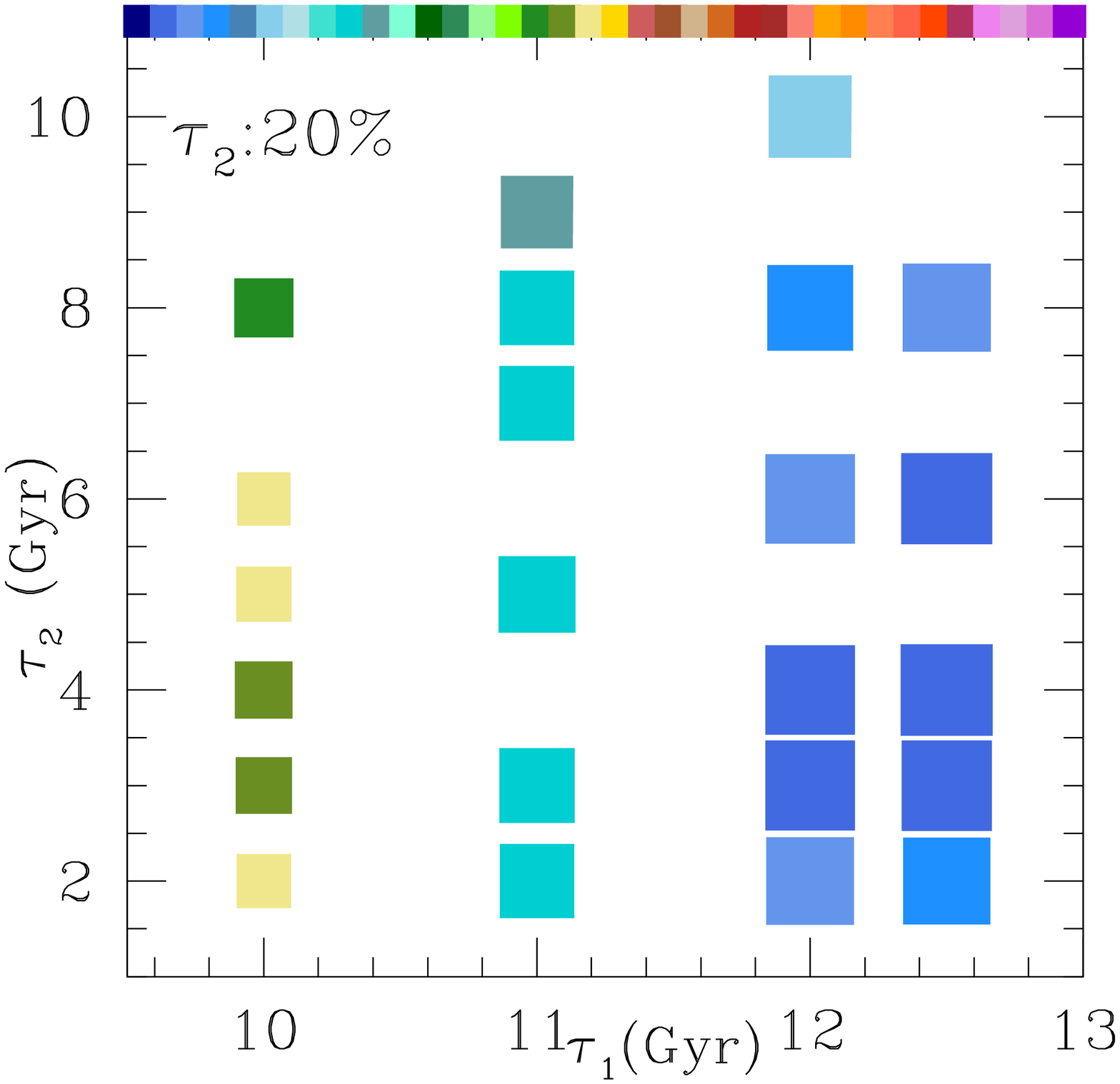}
\includegraphics[angle=0]{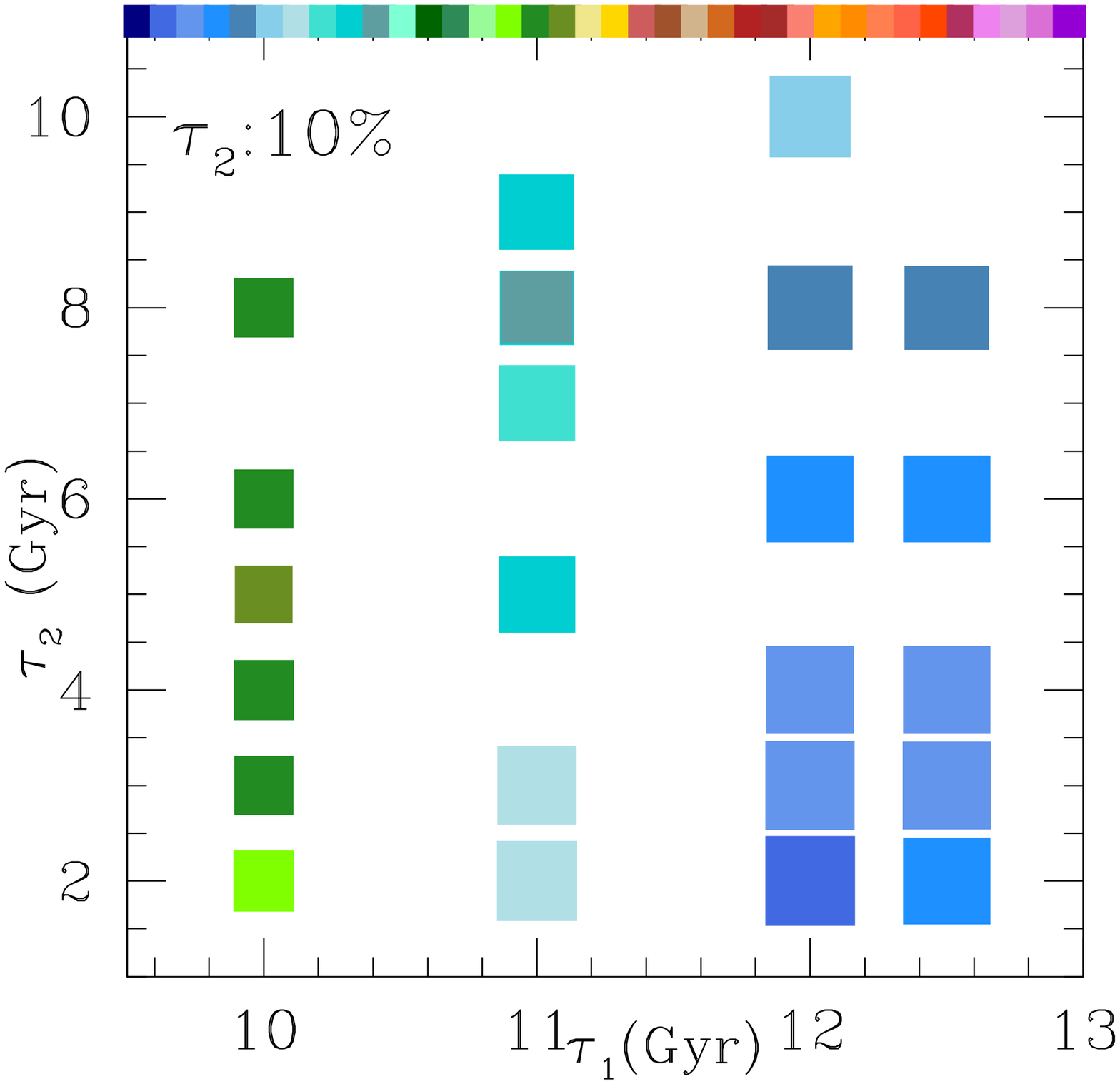}
}
\caption[]{The $\chi^2$ diagnostic for the I-band LF fit for 2 burst 
simulations,  built using alpha-enhanced stellar evolutionary models,
as compared to observations  is plotted as a function of 
the old (x-axis) + young (y-axis) population age. 
Each panel shows different relative fractions of old + young population in the 
combined 2-burst simulation.  
The size and colour of the points are normalized to the full
range of the $\chi^2$ values of the LF$_I$ fit for all 2-burst models with input observed MDF.
The larger the symbol, and darker blue its colour, the smaller the $\chi^2$. 
}
\label{fig:LFichi2_2burst_perc}
\end{figure*}

\begin{figure*}
\centering
\resizebox{\hsize}{!}{
\includegraphics[angle=0]{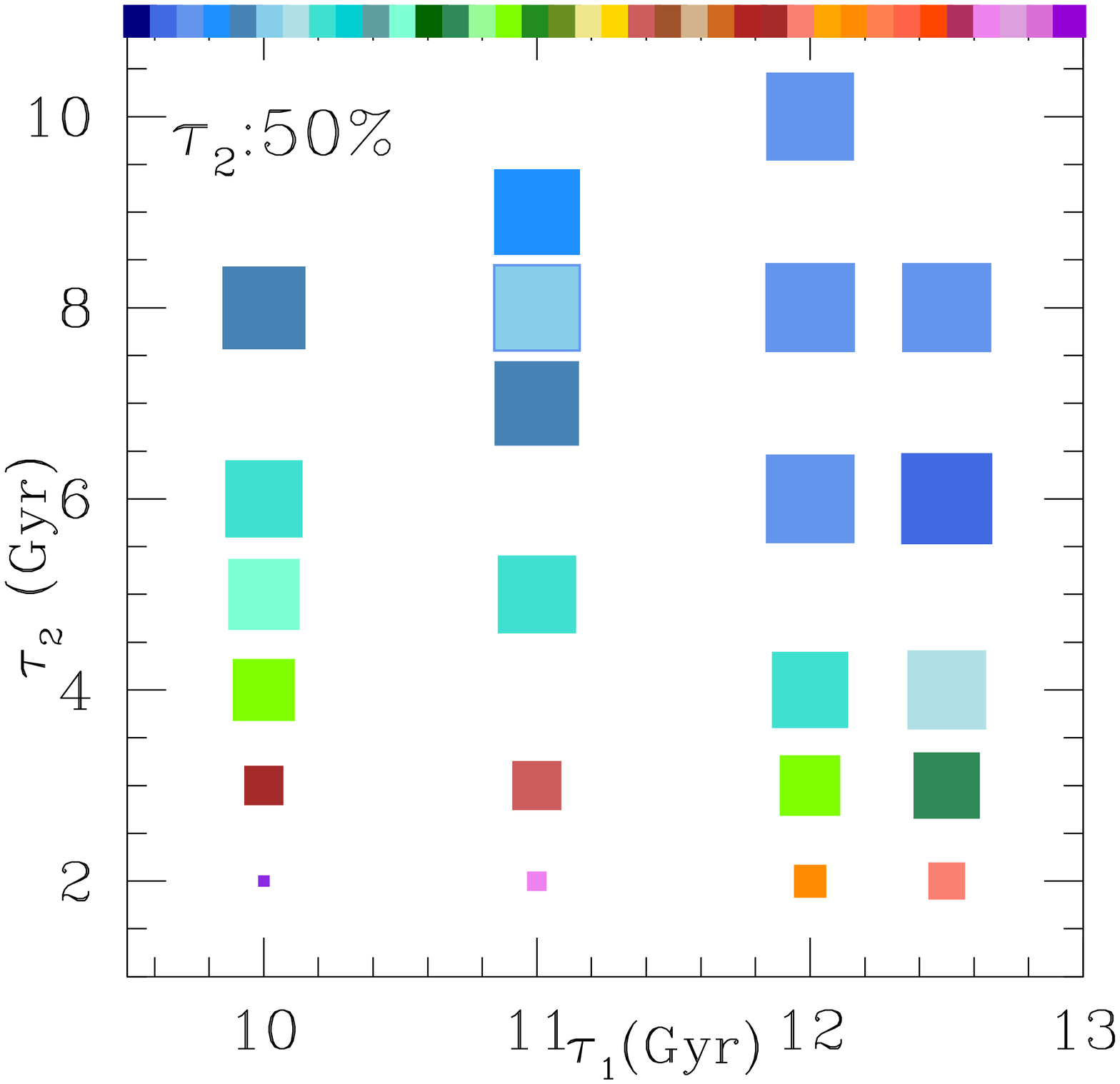}
\includegraphics[angle=0]{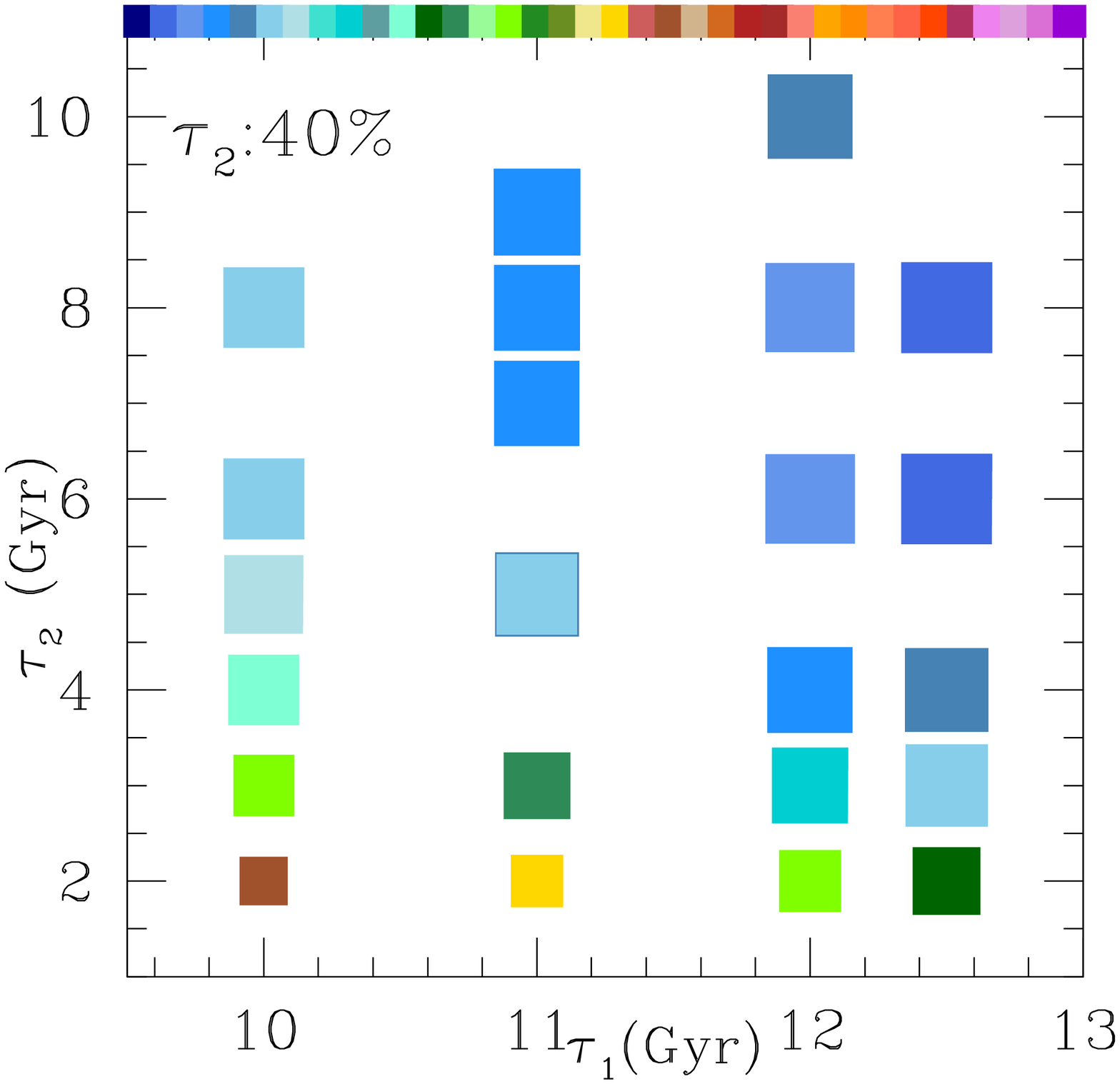}
\includegraphics[angle=0]{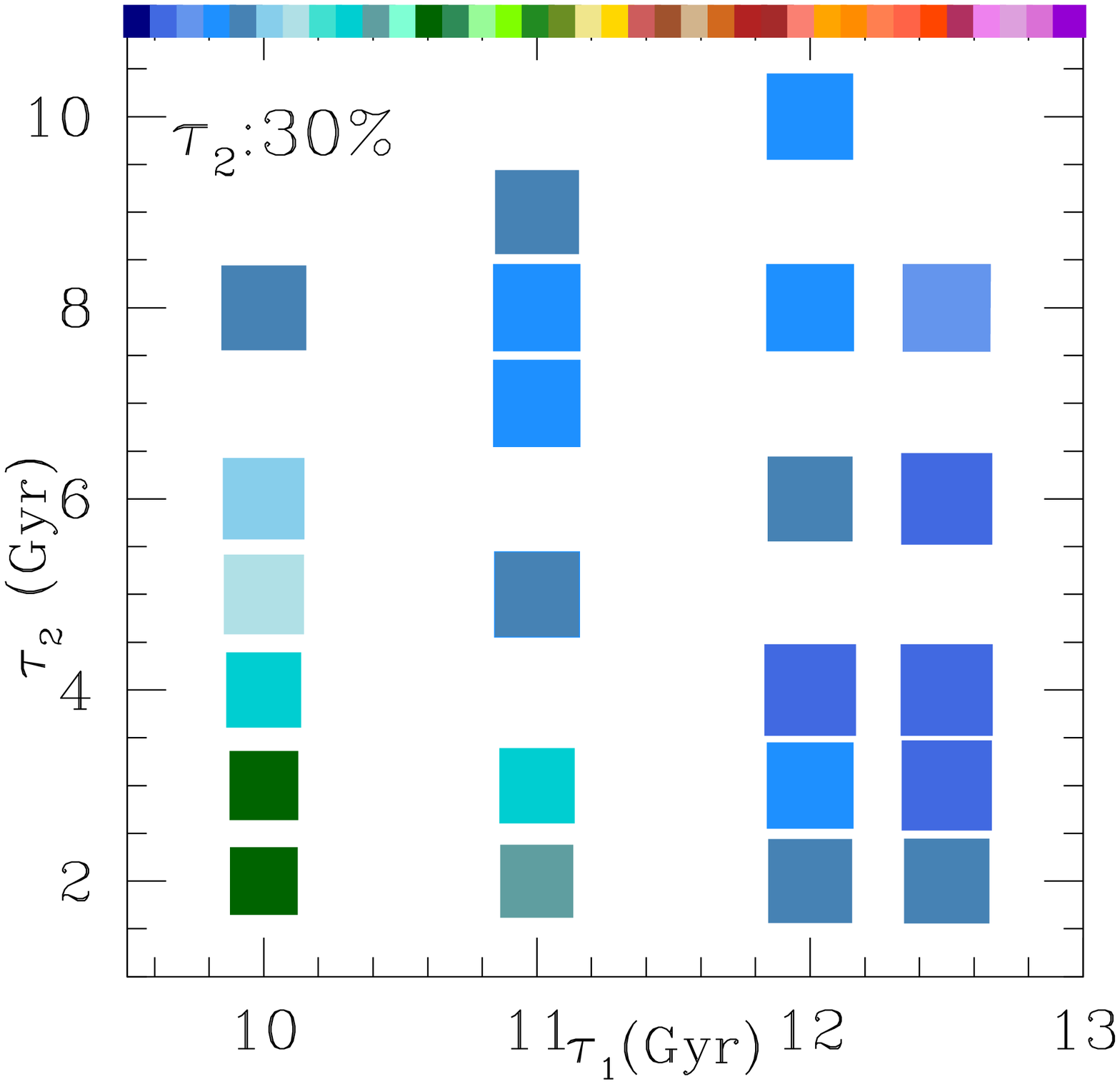}
\includegraphics[angle=0]{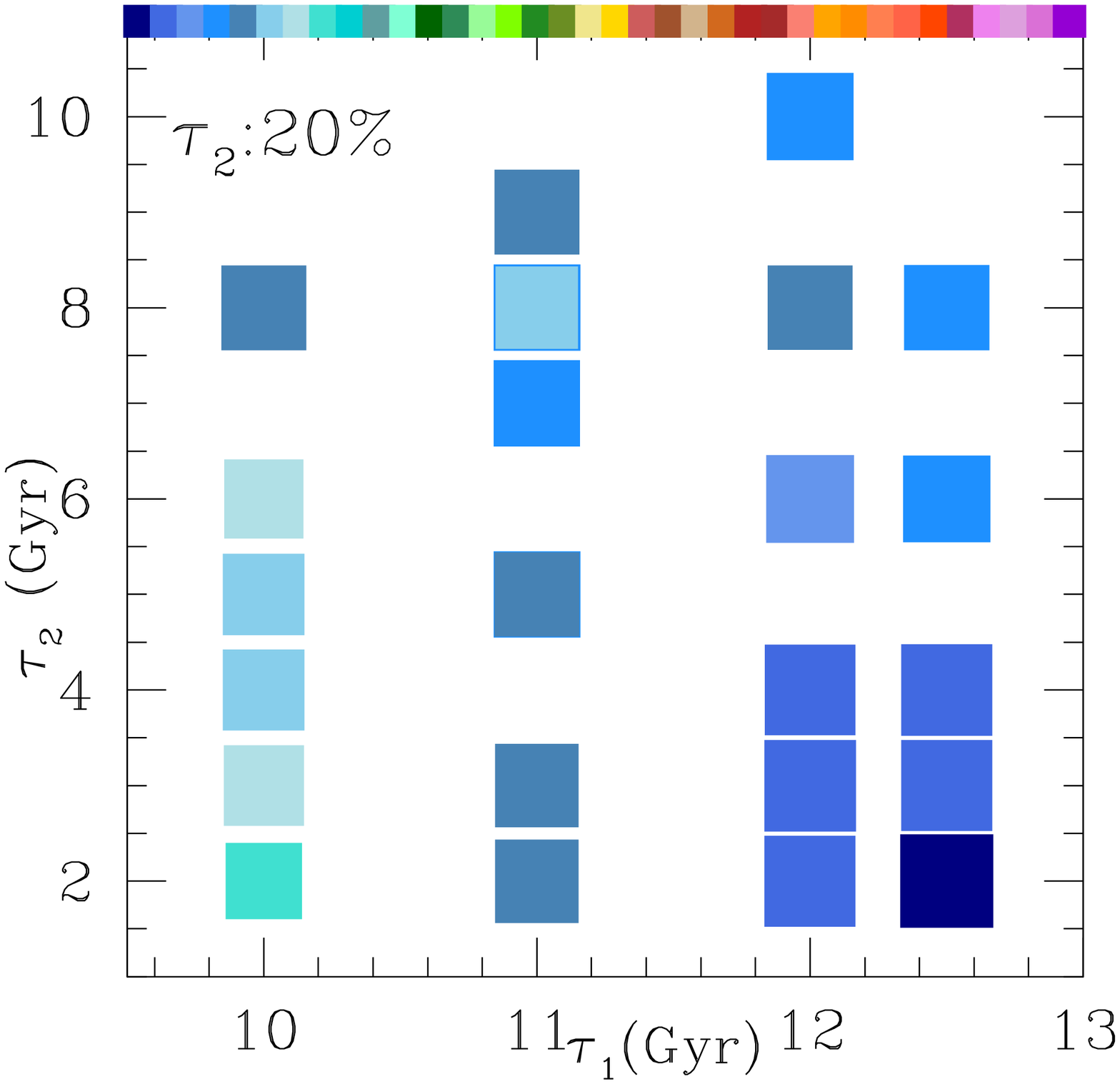}
\includegraphics[angle=0]{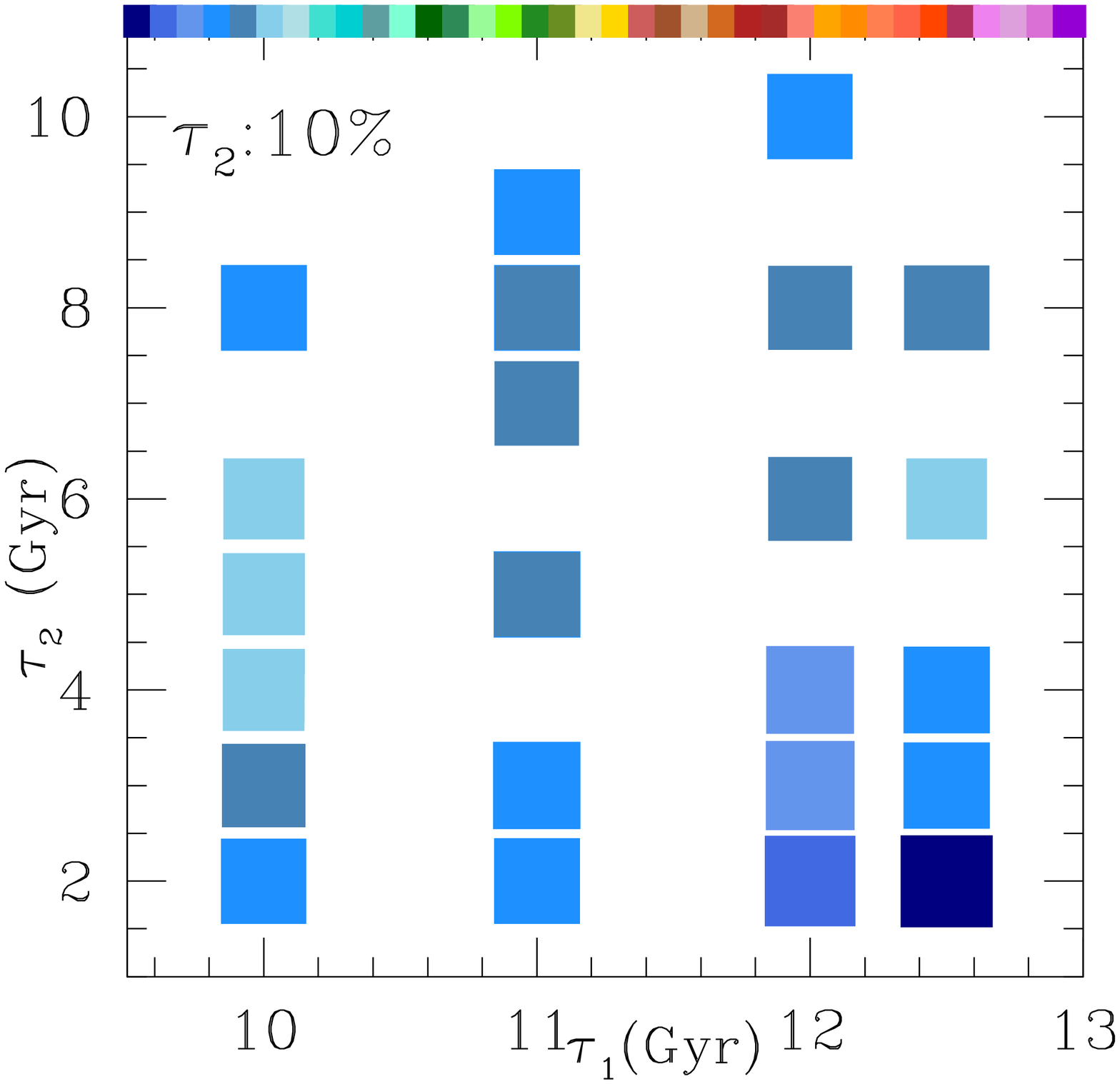}
}
\caption[]{The $\chi^2$ diagnostic for the V-band LF  fit for 2 burst simulations, 
built using alpha-enhanced stellar evolutionary models,
as compared to observations  is plotted as a function
of the old (x-axis) + young (y-axis) population age. 
Each panel shows different relative fractions
of old + young population in the combined 2-burst simulation.  
The size and colour of the points are normalized to the full
range of the $\chi^2$ values of the LF$_V$ fit for all 2-burst models with input observed MDF.
The larger the symbol, and darker blue its colour, the smaller the $\chi^2$. 
}
\label{fig:LFvchi2_2burst_perc}
\end{figure*}

As already mentioned above, the single-age
models produce widths of the RC and AGB bump features 
that are too narrow. In the next section we explore whether the fits
are improved by adding a second age component, hence simulating a 
two bursts star formation history. 
This is motivated also from other observations: 
NGC~5128 is likely to have experienced a history of satellite accretions
(minor mergers), but also some previous 
observations of resolved stellar populations and globular clusters 
have implied a smaller population of younger stars with ages 
close to $\sim 3-5$~Gyr \citep{soria+96,marleau+00,rejkuba+03,woodley+10a}.

\subsection{Two burst simulations}

A series of two burst simulations was created 
first by randomly drawing from an old single age
simulation a certain percentage P1 of the total number of stars. Then, we add
to the list a percentage P2 of a younger population, again drawing stars 
randomly from a single age simulation. By definition $P1 + P2 = 100$\%,
and both components were given the input observed MDF. When drawing the
stars randomly from the parent single age simulations, we verify that the 
final combined simulation has MDF bins populated such that it matches the observed
MDF. Therefore, since the metal-poor bins have fewer stars, and since the old age
(P1) simulation is first extracted, the metal-poor bins on average have an older age.
We note however, that the combined CMD also contains some metal-rich stars 
from the old (P1) episode.

The combinations simulated in this way have
relative percentages of  90-10, 80-20, 70-30, 60-40, and 50-50 old+young stars. 
The old component was allowed to range between 10.0 and 12.5 Gyr, and the 
young component between 2 and 10 Gyr. In addition to mixing alpha enhanced 
simulations (for both old and young age), we also considered that the younger 
population might have lower alpha enhancement, and thus
we combined old alpha enhanced models with younger solar scaled simulations. 
Table~\ref{tab:chisq2age}, given fully in the electronic format, lists 
all 2-burst simulations we considered and it shows also the $\chi^2$ values 
for our fit diagnostics. Table~\ref{tab:bestfitmodels2} lists the best fitting models 
separately for each diagnostic.
Here we summarize the main conclusions based on the inspection of these
diagnostics and careful inspection of simulated CMDs.

\begin{table*}
\caption[]{The best fitting models for each diagnostic are listed for 
the two burst combinations with the input observed MDF.
Diagnostics are $\chi^2$ for the full CMD, $\chi^2$ of the LF fit for I, and
V-bands, and in the last two columns 
$\Delta N / \sqrt{N_{obs}}= \mathrm{N}_{obs}-\mathrm{N}_{sim}/ \sqrt{N_{obs}}$ 
for the RC and AGBb boxes in our grid.  
}
\label{tab:bestfitmodels2}
\begin{tabular}{cccrrrrrrrrr}
\hline
\multicolumn{12}{c}{Two burst models with input observed MDF} \\
\hline \hline
\multicolumn{1}{c}{combined} & \multicolumn{1}{c}{old} &\multicolumn{1}{c}{young} & 
\multicolumn{1}{c}{age1} &\multicolumn{1}{c}{age2} & 
\multicolumn{1}{c}{\%} &\multicolumn{1}{c}{\%} & 
\multicolumn{1}{c}{$\chi^2$}  & \multicolumn{1}{c}{$\chi^2$} &
\multicolumn{1}{c}{$\chi^2$}  & \multicolumn{1}{c}{$\frac{\Delta N}{\sqrt{N_{obs}}}$} &
\multicolumn{1}{c}{$\frac{\Delta N}{\sqrt{N_{obs}}}$}
\\
\multicolumn{1}{c}{simulation}& \multicolumn{1}{c}{simulation} &\multicolumn{1}{c}{simulation} &
\multicolumn{1}{c}{(Gyr)}&\multicolumn{1}{c}{(Gyr)}& \multicolumn{1}{c}{(old)}&\multicolumn{1}{c}{(young)}& 
\multicolumn{1}{c}{CMD} & \multicolumn{1}{c}{LF$_I$} &
\multicolumn{1}{c}{LF$_V$} & \multicolumn{1}{c}{RC} &
\multicolumn{1}{c}{AGBb}
\\
\hline
\multicolumn{12}{l}{CMD:}\\
 cmb112 &  aen022 &  aen041 & 12.0 &  2.5 & 80& 20&   65.0 &   13.4 &	12.3 &   -3.8 &   -1.6 \\ 
 cmb137 &  aen022 &  aen042 & 12.0 &    2 & 80& 20&   65.4 &   15.2 &   12.6 &    1.7 &   -3.1 \\ 
 cmb142 &  aen021 &  aen042 & 12.5 &    2 & 80& 20&   65.6 &   16.2 &   11.2 &    7.1 &   -3.9 \\ 
 cmb372 &  aen022 &  aen040 & 12.0 &    3 & 80& 20&   67.0 &   12.4 &   12.3 &   -4.5 &   -1.2 \\ 
\multicolumn{12}{l}{LF$_I$:}\\
 cmb383 &  aen021 &  aen034 & 12.5 &    6 & 70& 30&   74.3 &    8.3 &   12.3 &   -5.1 &    1.0 \\ 
 cmb387 &  aen021 &  aen038 & 12.5 &    4 & 80& 20&   77.5 &   10.1 &   12.5 &   -0.0 &   -1.8 \\ 
 cmb378 &  aen021 &  aen030 & 12.5 &    8 & 70& 30&   86.1 &   10.2 &   14.8 &  -11.1 &    2.0 \\ 
 cmb472 &  aen022 &  sol040 & 12.0 &    3 & 80& 20&   80.7 &   10.5 &   14.1 &   -9.4 &    0.2 \\ 
\multicolumn{12}{l}{LF$_V$:}\\
 cmb142 &  aen021 &  aen042 & 12.5 &    2 & 80& 20&   65.6 &   16.2 &   11.2 &    7.1 &   -3.9 \\ 
 cmb141 &  aen021 &  aen042 & 12.5 &    2 & 90& 10&   80.2 &   16.6 &   11.9 &    0.1 &   -0.7 \\ 
 cmb402 &  aen026 &  sol030 & 10.0 &    8 & 80& 20&  118.3 &   52.1 &   11.9 &  -35.7 &    1.5 \\ 
 cmb388 &  aen021 &  aen038 & 12.5 &    4 & 70& 30&   73.9 &   12.6 &   12.1 &    1.6 &   -3.1 \\ 
\multicolumn{12}{l}{RC:}\\
 cmb387 &  aen021 &  aen038 & 12.5 &    4 & 80& 20&   77.5 &   10.1 &   12.5 &    0.0 &   -1.8 \\ 
 cmb141 &  aen021 &  aen042 & 12.5 &    2 & 90& 10&   80.2 &   16.6 &   11.9 &    0.1 &   -0.7 \\ 
 cmb129 &  aen026 &  aen042 & 10.0 &    2 & 60& 40&  165.3 &   89.7 &   41.1 &    0.4 &   -5.0 \\ 
 cmb325 &  aen026 &  aen040 & 10.0 &    3 & 50& 50&  180.9 &  107.1 &   47.2 &   -0.2 &   -3.6 \\ 
\multicolumn{12}{l}{AGBb:}\\
 cmb136 &  aen022 &  aen042 & 12.0 &    2 & 90& 10&   69.0 &   12.6 &   12.8 &   -7.6 &    0.0 \\ 
 cmb382 &  aen021 &  aen034 & 12.5 &    6 & 80& 20&   83.8 &   10.9 &   15.3 &   -5.5 &    0.0 \\ 
 cmb489 &  aen021 &  sol038 & 12.5 &    4 & 60& 40&   85.6 &   13.7 &   12.7 &   -2.7 &    0.0 \\ 
 cmb211 &  aen024 &  aen036 & 11.0 &    5 & 90& 10&   87.9 &   30.6 &   15.6 &  -25.4 &    0.0 \\ 
\hline
\end{tabular}
\end{table*}

Figures~\ref{fig:cmdchi2_2burst_perc}, \ref{fig:LFichi2_2burst_perc}, and 
\ref{fig:LFvchi2_2burst_perc} show the $\chi^2$ diagnostics for 2-burst alpha
enhanced simulations in three-dimensional form.  Here the age of the older
component is plotted along the x-axis and the younger component along the y-axis.
Each panel shows all the models with a particular old/young ratio (P1/P2).  
As an example, in the first panel, the point located at (x=11, y=5) refers to a simulation
with a 50\% 11-Gyr component and 50\% 5-Gyr component.  
The third dimension, which represents the quality of the $\chi^2$ fit  is given 
by the \emph{size} and  \emph{colour} of each small square.
The larger the symbol, and darker blue its colour, the smaller the $\chi^2$. 
In short, the best-fit solution regions of these figures are the ones where
the biggest and darkest squares are sitting.  These tend to be on the lower right,
with a dominant old component and a minor younger component.

Careful inspection of these figures and Tables~\ref{tab:chisq2age} 
and \ref{tab:bestfitmodels2}  reveals that 
the 2-burst simulations which best reproduce the observed CMD and luminosity 
functions are those with an old component of 12-12.5~Gyr that is alpha 
enhanced, along with a younger component of  2-6 Gyr which is also alpha 
enhanced. The proportion of the younger population should be between  
20\% and 30\%.  
This younger component needs to be present to give the best fits,
but it cannot dominate the system.  Said differently, 
the simulations that have 90\% or more old population have worse fits 
regardless of the age of the younger component.
On the contrary, if the younger component makes up more than 30\% of the total, 
then the age of the young component needs to be relatively old, 
$\sim 8$~Gyr or more, in order to be competitive with the best-fit cases.
All these indicators clearly show that the bulk of the population has 
to be old.

While the overall trend is valid for all diagnostics we note that different 
diagnostics indicate somewhat different values for the best fitting simulations. 
The small dependence of the $\chi^2$ values on the random extractions can be
appreciated from the comparison of results for single age simulations for 13, 11, and
10 Gyr single burst simulations (Table~\ref{tab:chisq1age}) as well as for 
double burst simulations for 11+8 and 11+5 Gyr old combinations 
(Table~\ref{tab:chisq2age}). Looking at the individual diagnostics for the 
best fitting models in Table~\ref{tab:bestfitmodels2} we notice that  
the two most sensitive diagnostics (Figure~\ref{fig:diagnostics_chi2}), 
the full CMD fit and the I-band luminosity function fit,  provide the lower and the 
upper limit for the age of the young component. The full CMD fit prefers a
20\% contribution of 2-3~Gyr
old population, while the I-band LF can accommodate up to 30\%  of 6~Gyr old 
stars for the best fitting model. The best fitting $\chi^2$ values for the somewhat
less sensitive diagnostics tend towards the lower limit for the young component.
Given the small difference of I-band LF fit $chi^2$ values for the models cmb112 (that provides
the best fit to the full CMD) and the models cmb383 and cmb387 (the two best 
fitting models for the I-band LF in Table~\ref{tab:bestfitmodels2}), as well as taking into
account the larger variation in the full CMD fit $\chi^2$ values for the same models and the
results from the other diagnostics, we conclude that on average the best fitting models
require a young population of $\sim 2-4$~Gyr. 

\begin{figure}
\centering
\resizebox{\hsize}{!}{
\includegraphics[angle=0]{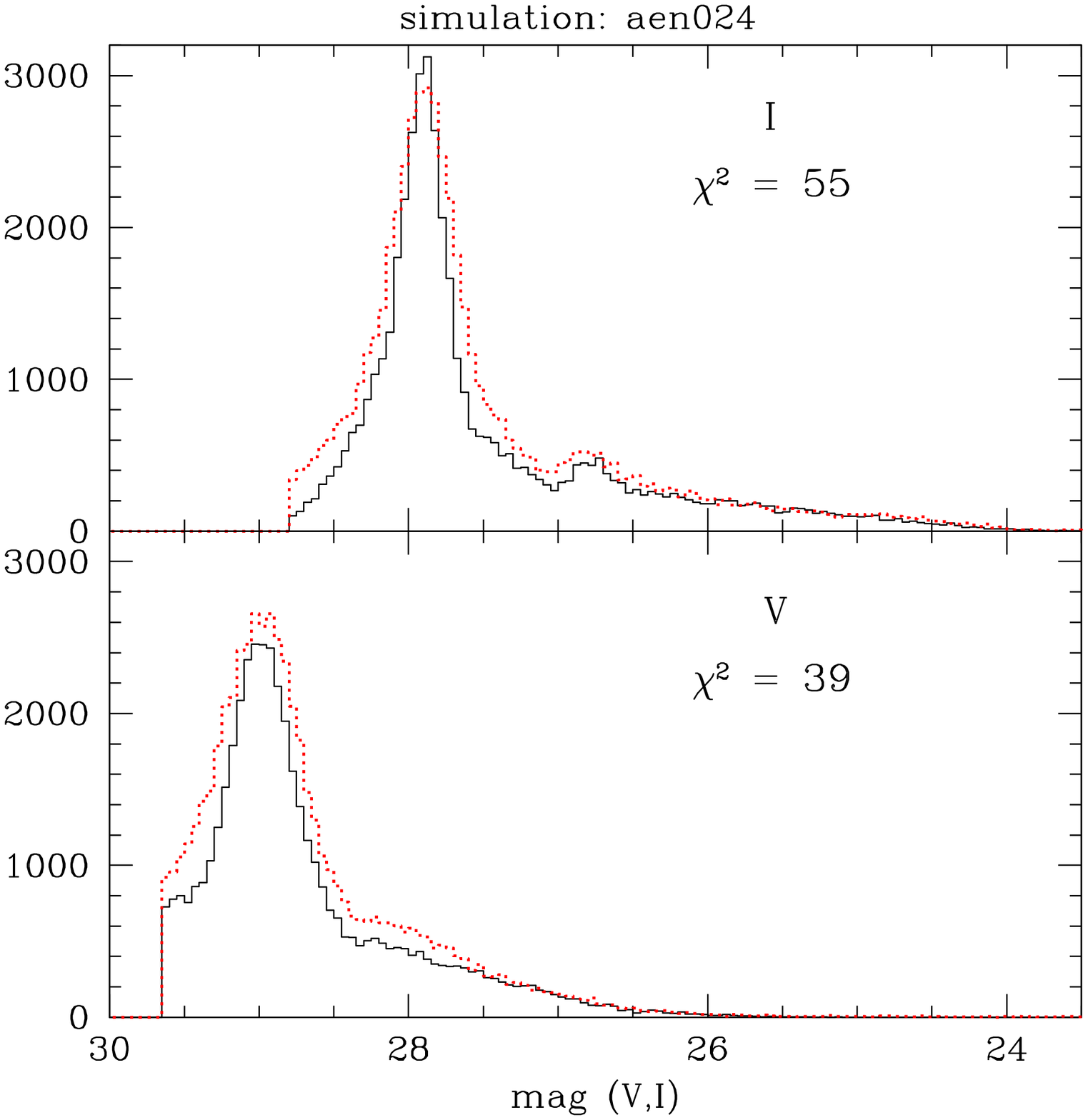}
\includegraphics[angle=0]{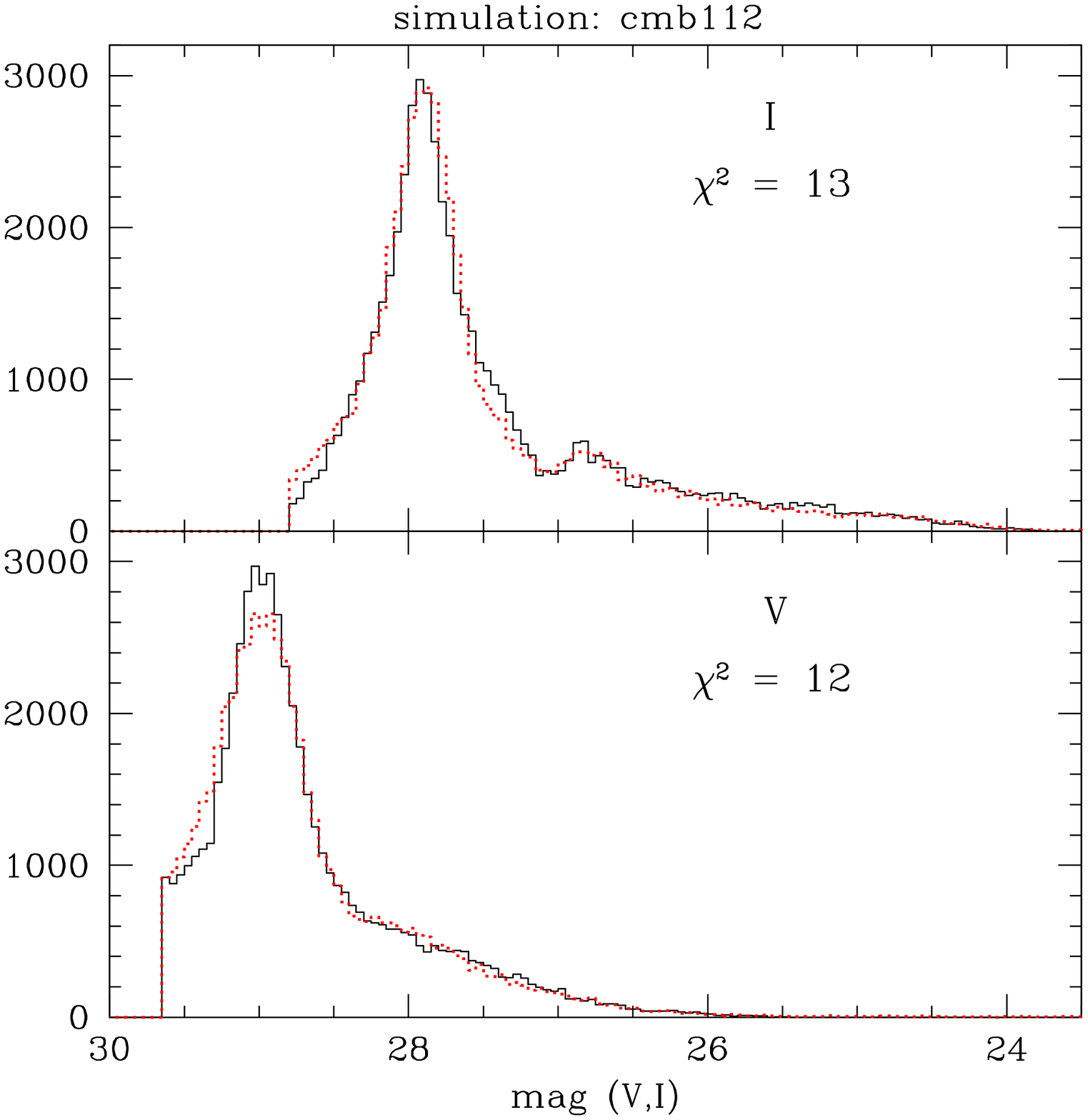}
}
\resizebox{\hsize}{!}{
\includegraphics[angle=0]{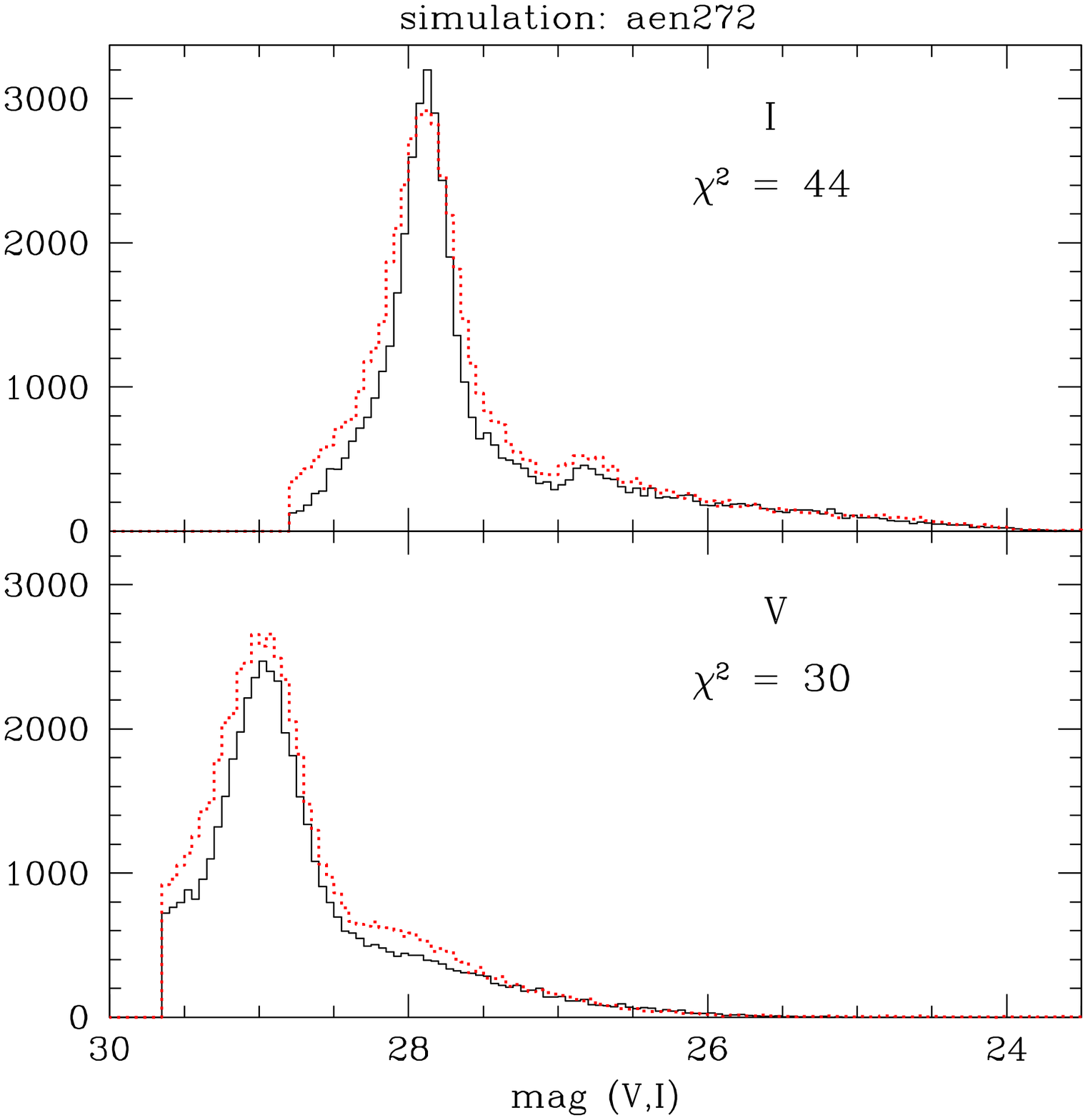}
\includegraphics[angle=0]{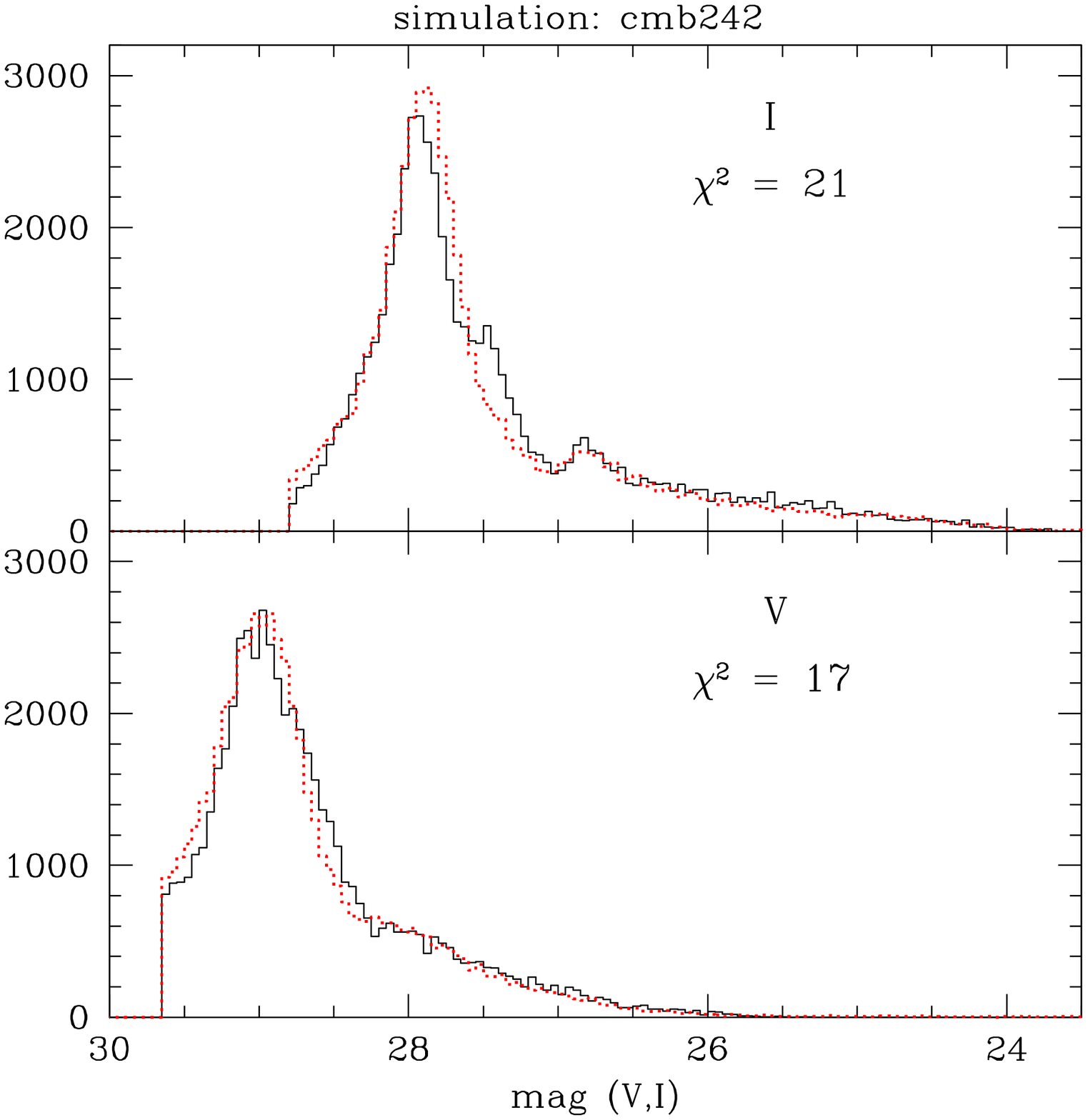}
}
\caption[]{Comparison of the LF fits between the observations 
(dotted red line) and the simulations (solid black line), where only data 
above 50\% completeness limits are considered. The left panels plot the 
best fitting LFs for the single age simulation that has 11 Gyr, and the right
panels plot the LFs for the best fitting 2-burst
simulation that has 80\% 12 Gyr population and 20\% 2.5 Gyr old population.
The top panels are for the comparison with simulations that have input observed
MDF, and the bottom panels show simulated LFs with input closed box enrichment.
The single age closed box model is shown on the left and the two-burst model 
composed of input closed box simulations is on the right.
}
\label{fig:bestfitLF}
\end{figure}

\begin{figure}
\centering
\resizebox{\hsize}{!}{
\includegraphics[angle=0]{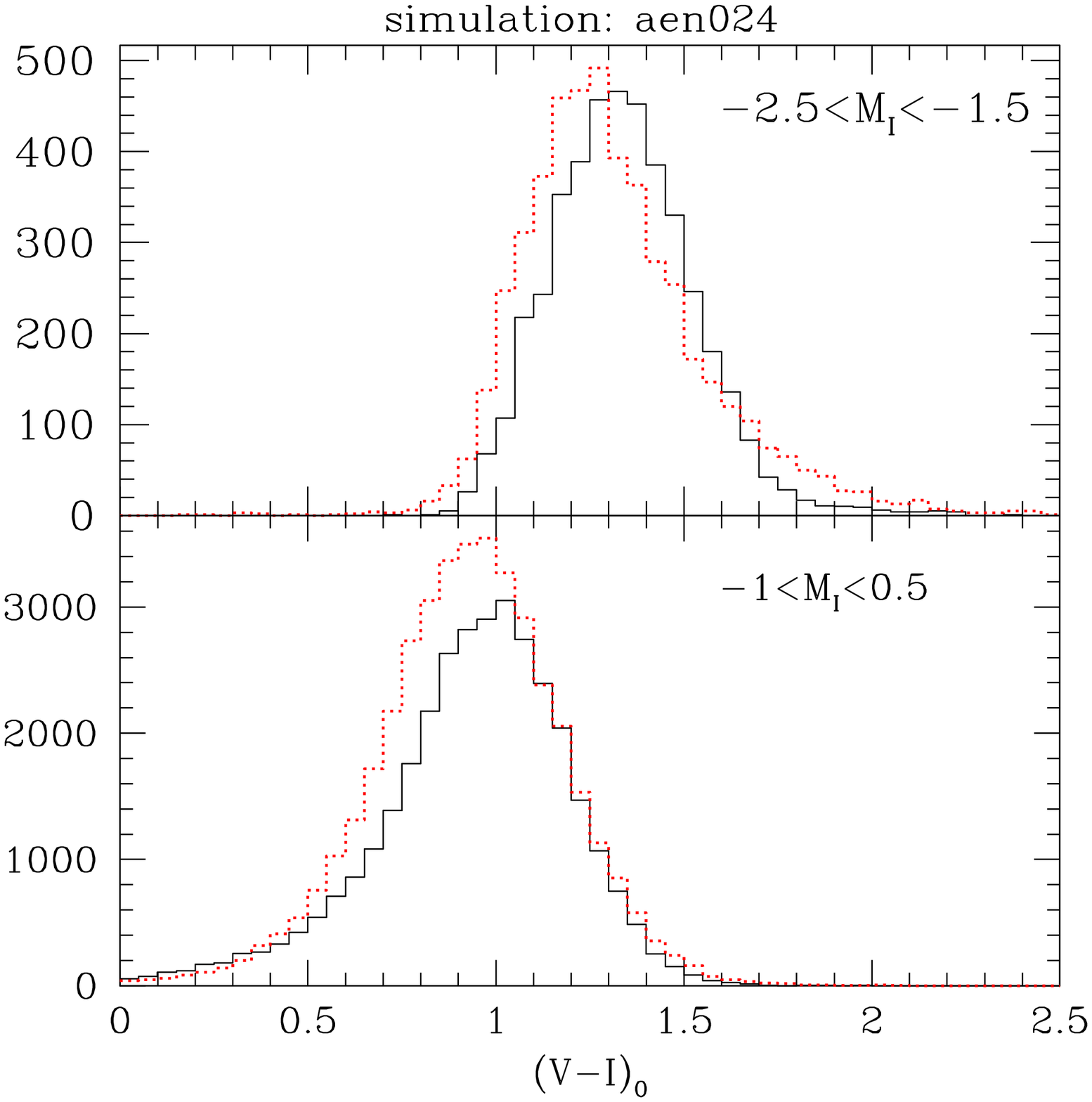}
\includegraphics[angle=0]{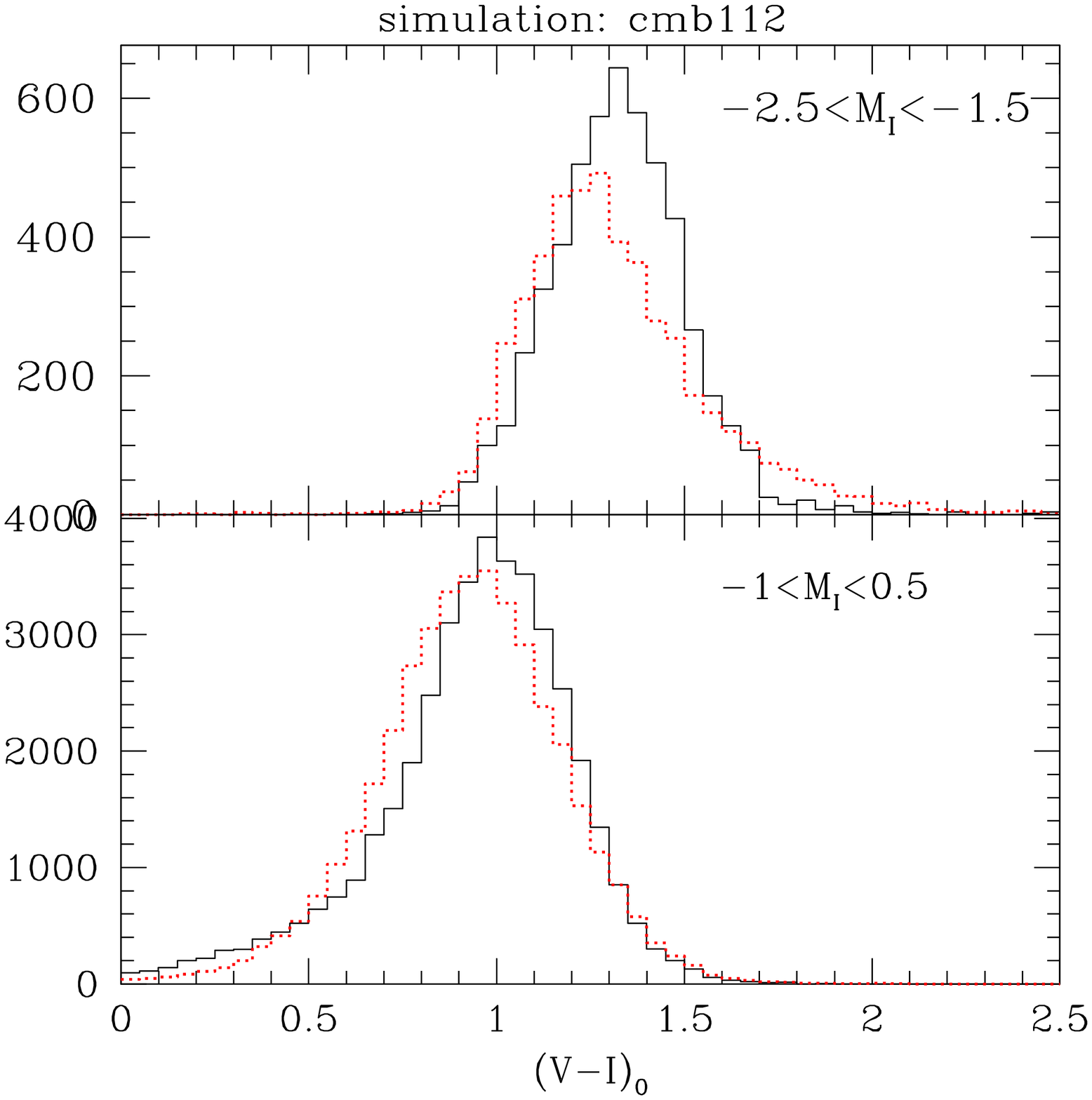}}
\resizebox{\hsize}{!}{
\includegraphics[angle=0]{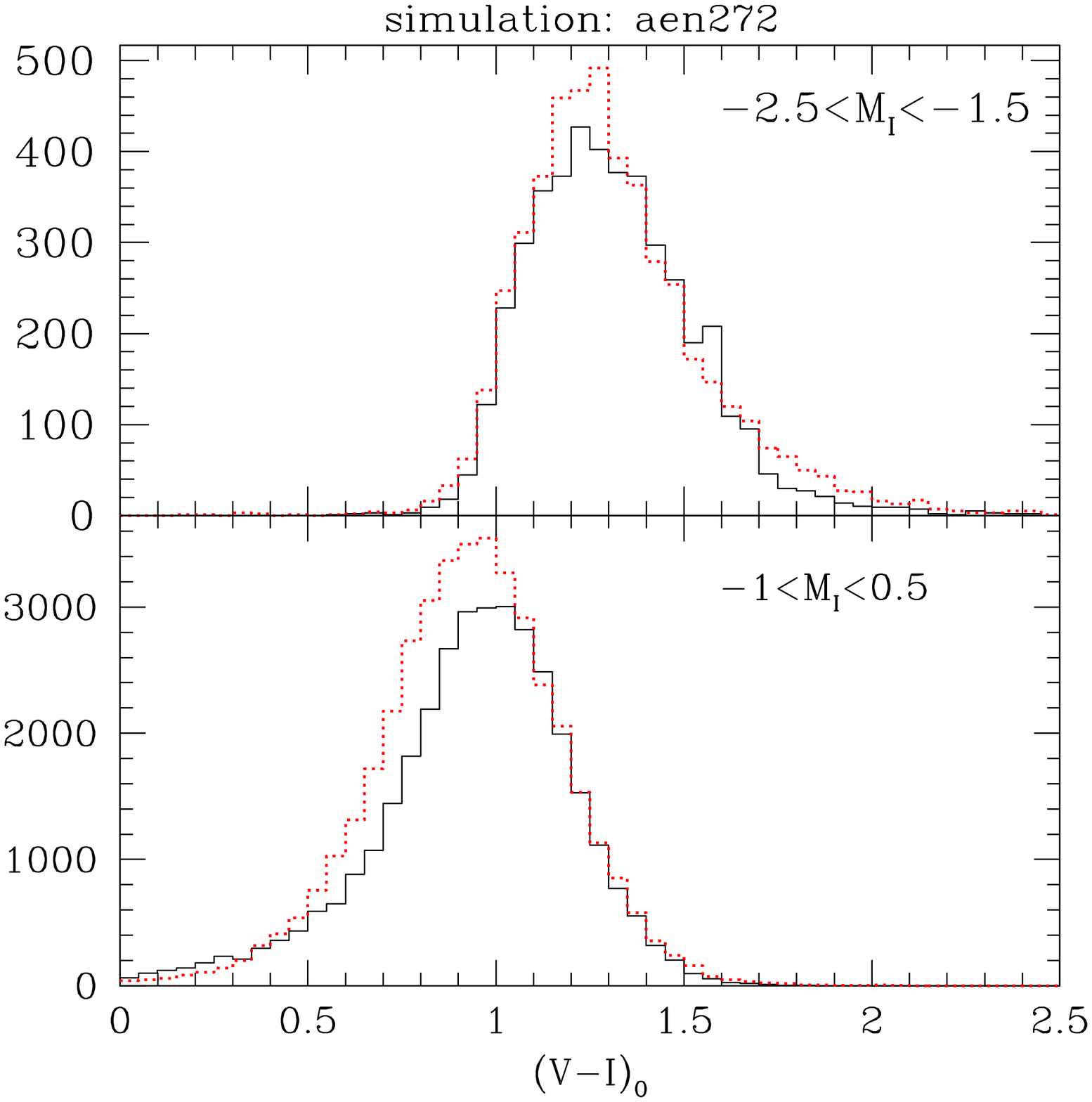}
\includegraphics[angle=0]{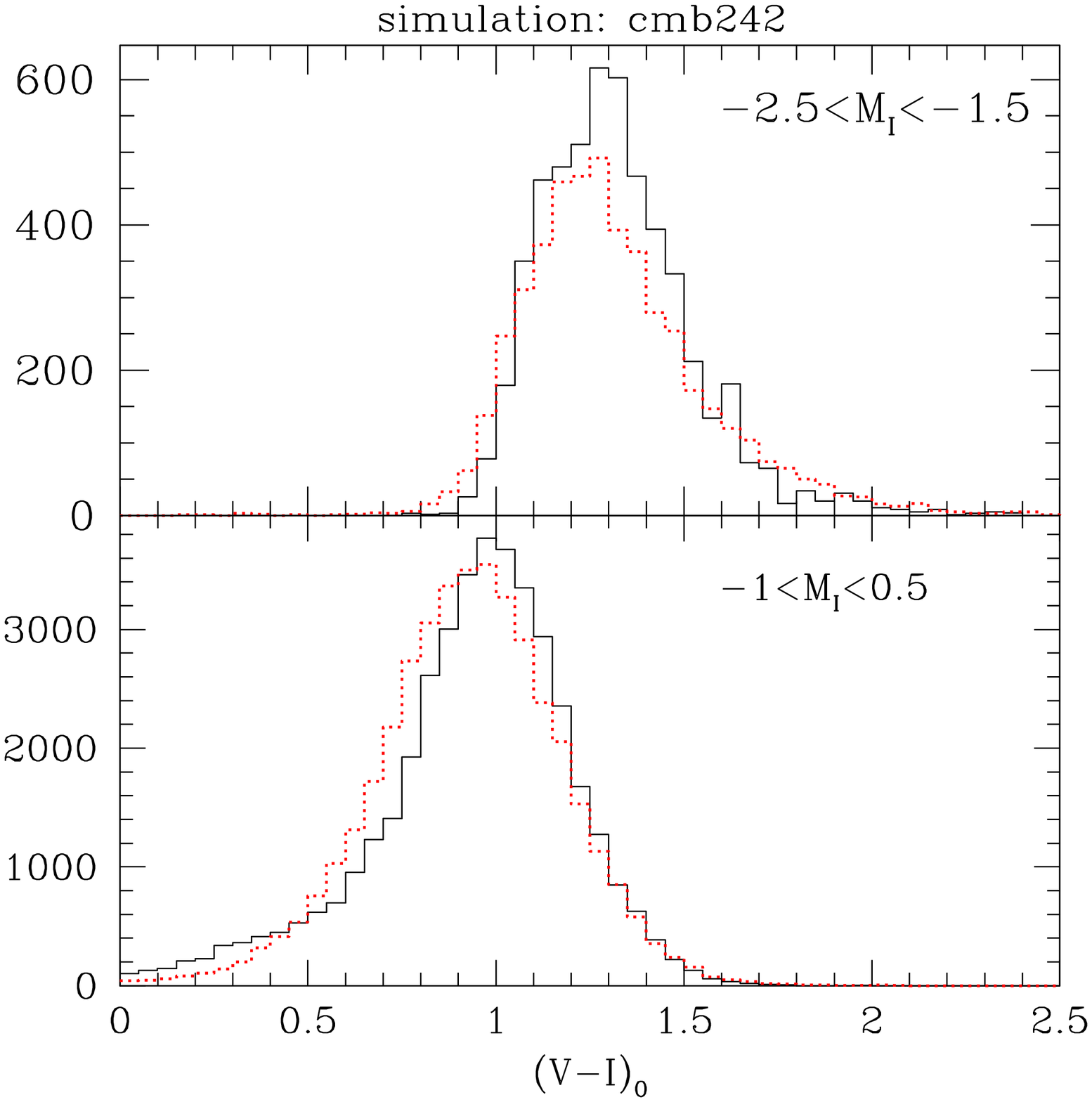}
}
\caption[]{Comparison of V-I colour distributions between the data 
(dotted red line) and the simulations (solid black line), where only data 
in the specific luminosity range are considered. 
Upper panels show the colour distributions for the red giant branch stars 
that are selected in the magnitude range $-2.5 < M_I < -1.5$ (where the 
bolometric corrections for the upper cool part of the RGB are not a problem), 
while the lower panels show the colour distributions for the part of the
luminosity range dominated by helium burning stars (red clump) selected to 
be within range $-1 < M_I < +0.5$ mag. Upper left diagrams are for the best 
fitting single age simulation (11 Gyr old model) with input observed MDF. 
Upper right diagrams are for the best fitting two burst model which is made by 
combining single age simulations with input observed MDFs
(80\% 12 + 20\% 3 Gyr). The bottom panels show the best fitting simulations with
input closed box metallicity distributions - the single age closed box model on
the left and the two-burst model composed of input closed box simulations
is on the right.
}
\label{fig:VIhistograms}
\end{figure}

\begin{figure}
\centering
\resizebox{\hsize}{!}{
\includegraphics[angle=270]{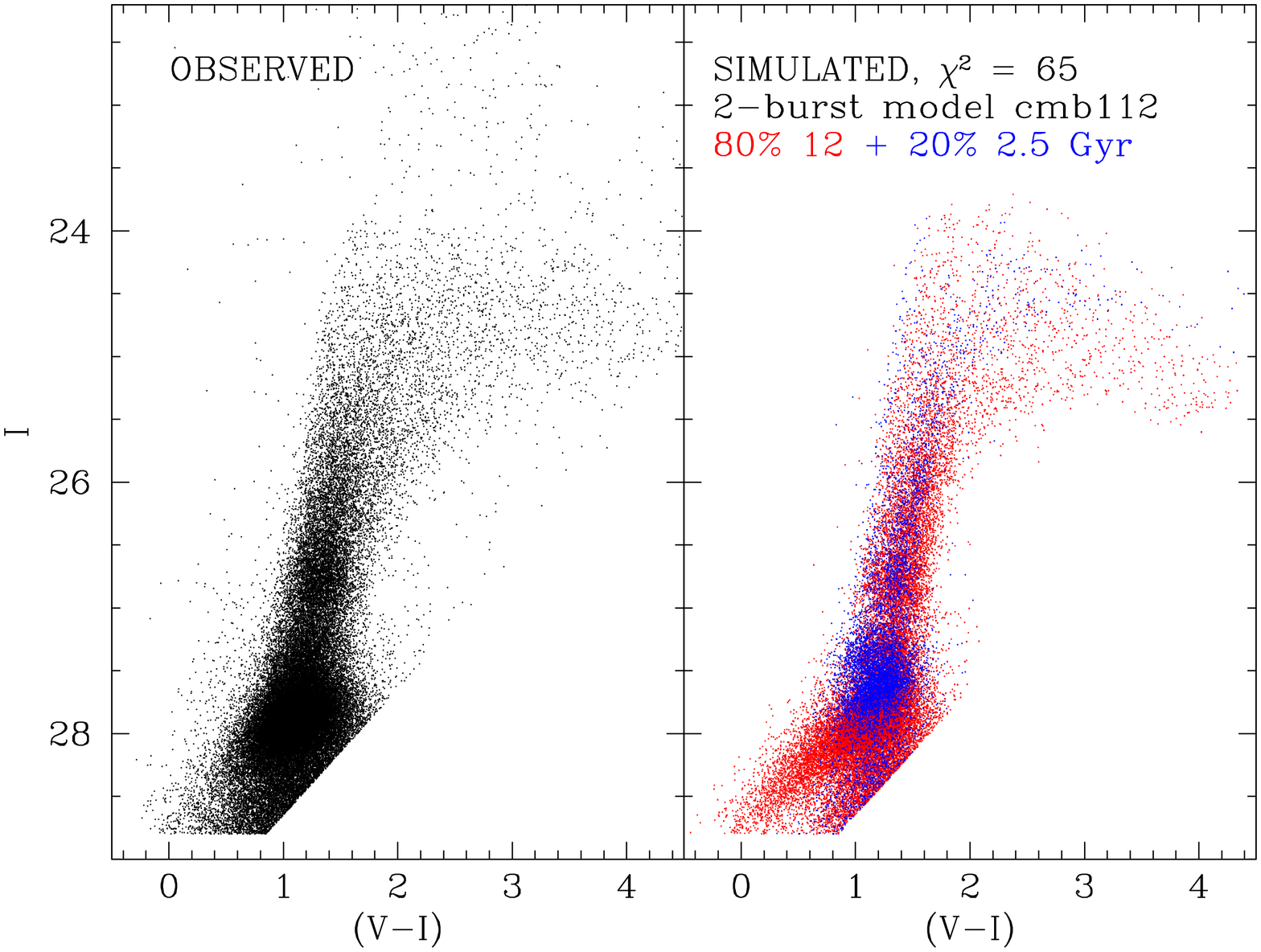}}
\caption[]{Comparison between the observed CMD (left) and the best fitting 
2-burst simulation (right). The simulated stars are colour coded according to their
age: 12 Gyr old stars are red, and 3 Gyr old stars are blue.
}
\label{fig:2burstCMD}
\end{figure}

The $\chi^2$ values for the whole CMD fit for 
the best fitting 2-burst models are similar to those of the best 
fitting single age models: $\chi^2=65$ for 80\% 12 Gyr + 20\% 2.5 Gyr 
model and $\chi^2=66$ for the 11 Gyr single age simulation. The fact
that there is almost no improvement in the full CMD fit
between the single age and two-burst best fitting models, again confirms
that the bulk of the stellar population in the observed CMD is old.

In contrast, the luminosity function fits are 
significantly improved over the single-age simulations (Figure~\ref{fig:bestfitLF}).
The $\chi^2$ values for  the single age 12 Gyr old alpha enhanced simulation 
were 36 and 26 for the  I and V-band luminosity functions.
By contrast, for the 2-burst  simulation with 
70\% 12.5 Gyr + 30\% 6 Gyr the LF fits
give $\chi^2 = 8$ and 12 for the I and V-band respectively. 
For the 12+2.5 Gyr model that has an 80\% old population 
(that best fits the whole CMD), the LFs give $\chi^2=13$ and $12$ for I and 
V-band, respectively. However, the V-band LF has too many 
stars with respect to the data at the magnitude corresponding 
to the RC maximum (Figure~\ref{fig:bestfitLF}).

The improvement with respect to single age simulations 
is visible also in the colour distribution of red clump and RGB regions 
(Figure~\ref{fig:VIhistograms}), as well as in the number of AGB bump 
and RC stars in their respective boxes in the CMD (Table~\ref{tab:bestfitmodels2}).
Therefore the two burst star formation history is clearly favored over a single star 
formation event.

To understand the basic effect of adding a second component, we may
ask just where in the CMD the younger component is contributing differently
from the ``baseline'' old component.
Figure~\ref{fig:2burstCMD} shows the best fitting 2-burst simulated 
CMD compared to the observations. The simulated CMD is colour coded 
according to the ages of simulated stars:  here, we see that the 
``young'' component (blue) contributes most strongly to the brighter, 
redder end of the red clump population. Without those stars,
the luminosity function of the red clump is too narrow in magnitude to match
the data and the model solution is not as successful.

\subsection{Simulations with closed box chemical evolution}
\label{sect:closedbox}

\begin{table*}
\caption[]{The best fitting models for each diagnostic are listed for 
the single age and two burst simulations with input closed box MDF.
Diagnostics are $\chi^2$ for the full CMD, $\chi^2$ of the LF fit for I, and
V-bands, and in the last two columns 
$\Delta N / \sqrt{N_{obs}}= \mathrm{N}_{obs}-\mathrm{N}_{sim}/ \sqrt{N_{obs}}$ 
for the RC and AGBb boxes in our grid.  
}
\label{tab:bestfitmodels1cb}
\begin{tabular}{crrrrrrrrr}
\hline
\multicolumn{10}{c}{Single age models with closed box MDF} \\
\hline
\hline
\multicolumn{1}{c}{simulation} & \multicolumn{1}{c}{age} &
\multicolumn{1}{c}{$Z_{min}$} & \multicolumn{1}{c}{$Z_{max}$} & \multicolumn{1}{c}{yield} &
\multicolumn{1}{c}{$\chi^2$}  & \multicolumn{1}{c}{$\chi^2$} &
\multicolumn{1}{c}{$\chi^2$}  & 
\multicolumn{1}{c}{$\frac{\Delta N}{\sqrt{N_{obs}}}$} &
\multicolumn{1}{c}{$\frac{\Delta N}{\sqrt{N_{obs}}}$}
\\
\multicolumn{1}{c}{ID}& \multicolumn{1}{c}{Gyr}&
                      &                          &                                         &
\multicolumn{1}{c}{CMD} & \multicolumn{1}{c}{LF$_I$} &
\multicolumn{1}{c}{LF$_V$} & \multicolumn{1}{c}{RC} &
\multicolumn{1}{c}{AGBb}
\\
\hline
\multicolumn{10}{l}{CMD:}\\
 aen272 & 10.5 & 0.0001 & 0.0400 & 0.0130 &   51.5 &   44.0 &   29.7 &   -0.7 &   12.5 \\ 
 aen273 & 11.0 & 0.0001 & 0.0400 & 0.0140 &   54.7 &   35.3 &   21.3 &    2.8 &    9.5 \\ 
 aen279 & 10.0 & 0.0001 & 0.0400 & 0.0130 &   54.8 &   66.4 &   38.5 &    0.2 &   11.9 \\ 
\multicolumn{10}{l}{LF$_I$:}\\
 aen276 & 11.0 & 0.0001 & 0.0400 & 0.0170 &   62.4 &   27.4 &   12.0 &   -5.3 &    9.6 \\ 
 aen277 & 11.0 & 0.0001 & 0.0400 & 0.0180 &   69.2 &   33.0 &   15.3 &   -3.4 &   10.6 \\ 
 aen274 & 11.0 & 0.0001 & 0.0400 & 0.0150 &   60.7 &   34.5 &   21.3 &    2.5 &   11.2 \\ 
\multicolumn{10}{l}{LF$_V$:}\\
 aen276 & 11.0 & 0.0001 & 0.0400 & 0.0170 &   62.4 &   27.4 &   12.0 &   -5.3 &    9.6 \\ 
 aen277 & 11.0 & 0.0001 & 0.0400 & 0.0180 &   69.2 &   33.0 &   15.3 &   -3.4 &   10.6 \\ 
 aen278 & 11.0 & 0.0001 & 0.0400 & 0.0190 &   74.6 &   34.7 &   16.8 &   -1.8 &   11.3 \\ 
\multicolumn{10}{l}{RC:}\\
 aen279 & 10.0 & 0.0001 & 0.0400 & 0.0130 &   54.8 &   66.4 &   38.5 &    0.2 &   11.9 \\ 
 aen254 &  8.0 & 0.0020 & 0.0200 & 0.0130 &  129.0 &  168.7 &  102.9 &    0.6 &    5.6 \\ 
 aen272 & 10.5 & 0.0001 & 0.0400 & 0.0130 &   51.5 &   44.0 &   29.7 &   -0.7 &   12.5 \\ 
\multicolumn{10}{l}{AGBb:}\\
 aen320 &  2.5 & 0.0080 & 0.0400 & 0.0130 &  498.5 &  501.8 &  239.0 &   48.4 &   -0.1 \\ 
 aen303 &  3.0 & 0.0050 & 0.0300 & 0.0130 &  444.9 &  496.4 &  256.0 &   52.0 &   -0.8 \\ 
 aen306 &  4.0 & 0.0050 & 0.0200 & 0.0130 &  383.7 &  430.2 &  252.0 &   41.9 &    0.5 \\ 
\hline
\end{tabular}
\newline
\begin{tabular}{cccrrrrrrrrrll}
\multicolumn{12}{c}{Two burst models with closed box MDF} \\
\hline \hline
\multicolumn{1}{c}{combined} & \multicolumn{1}{c}{old} &\multicolumn{1}{c}{young} & 
\multicolumn{1}{c}{age1} &\multicolumn{1}{c}{age2} & 
\multicolumn{1}{c}{\%} &\multicolumn{1}{c}{\%} & 
\multicolumn{1}{c}{$\chi^2$}  & \multicolumn{1}{c}{$\chi^2$} &
\multicolumn{1}{c}{$\chi^2$}  & \multicolumn{1}{c}{$\frac{\Delta N}{\sqrt{N_{obs}}}$} &
\multicolumn{1}{c}{$\frac{\Delta N}{\sqrt{N_{obs}}}$} & $Z_{min}, Z_{max}, y$ & $Z_{min}, Z_{max}, y$ 
\\
\multicolumn{1}{c}{simulation}& \multicolumn{1}{c}{P1} &\multicolumn{1}{c}{P2} &
\multicolumn{1}{c}{Gyr}&\multicolumn{1}{c}{Gyr}& \multicolumn{1}{c}{P1}&\multicolumn{1}{c}{P2}& 
\multicolumn{1}{c}{CMD} & \multicolumn{1}{c}{LF$_I$} &
\multicolumn{1}{c}{LF$_V$} & \multicolumn{1}{c}{RC} &
\multicolumn{1}{c}{AGBb} & P1 & P2 
\\
\hline
\multicolumn{12}{l}{CMD:}\\
cmb242&aen270&aen316& 12.0 &  2 & 80& 20&   56.7 &   21.3 & 17.4 & 8.8 & -1.6 & 0.0001, 0.04, 0.013 & 0.002, 0.02, 0.013\\ 
cmb792&aen270&aen259& 12.0 &  3 & 80& 20&   57.9 &   14.5 & 11.0 & 4.0 & -2.1 & 0.0001, 0.04, 0.013 & 0.002, 0.02, 0.013\\ 
cmb257&aen270&aen321& 12.0 &2.5 & 80& 20&   60.5 &   18.4 & 16.4 & 5.8 & -3.9 & 0.0001, 0.04, 0.013 & 0.005, 0.02, 0.013\\ 
cmb222&aen270&aen307& 12.0 &  2 & 80& 20&   61.5 &   15.3 & 10.6 & 4.1 & -0.5 & 0.0001, 0.04, 0.013 & 0.0001,  0.04, 0.013\\ 
\multicolumn{12}{l}{LF$_I$:}\\
cmb971&aen270&aen304& 12.0 & 4 & 80& 20&   85.2 & 8.9 &	11.5 &-7.0 & 0.1 & 0.0001, 0.04, 0.013 &0.005, 0.03, 0.013\\ 
cmb822&aen282&aen292& 12.0 & 5 & 80& 20&   81.9 & 9.2 &	13.0 &-5.9 & 0.4 & 0.0001, 0.04, 0.013 &0.0001, 0.04, 0.013\\ 
cmb787&aen270&aen268& 12.0 & 4 & 80& 20&   87.9 & 9.3 &	13.4 &-7.5 &-0.2 & 0.0001, 0.04, 0.013 &0.005, 0.04, 0.02\\ 
cmb943&aen270&aen298& 12.0 & 4 & 80& 20&   82.9 & 9.3 &	 9.9 &-4.8 & 0.2 & 0.0001, 0.04, 0.013 &0.002, 0.04, 0.013\\ 
\multicolumn{12}{l}{LF$_V$:}\\
cmb221&aen270&aen307& 12.0 & 2 & 90& 10&   73.5 &   13.5 &  9.2 &-3.8 & 2.1 & 0.0001, 0.04, 0.013 & 0.0001, 0.04, 0.013\\ 
cmb241&aen270&aen316& 12.0 & 2 & 90& 10&   64.5 &   15.4 &  9.3 &-2.0 & 1.3 & 0.0001, 0.04, 0.013 & 0.002, 0.02, 0.013\\ 
cmb236&aen270&aen315& 12.0 & 2 & 90& 10&   64.0 &   15.8 &  9.6 &-1.9 &-1.3 & 0.0001, 0.04, 0.013 & 0.005, 0.02, 0.013\\ 
cmb943&aen270&aen298& 12.0 & 4 & 80& 20&   82.9 &    9.3 &  9.9 &-4.8 & 0.2 & 0.0001, 0.04, 0.013 & 0.002, 0.04, 0.013\\ 
\multicolumn{12}{l}{RC:}\\
cmb986&aen270&aen293& 12.0 & 3 & 80& 20&   73.5 &   11.6 &   10.0 &-0.3 & 1.1 & 0.0001, 0.04, 0.013 &0.0001, 0.04, 0.013\\ 
cmb768&aen205&aen259& 10.0 & 3 & 70& 30&  197.0 &   62.4 &  117.5 & 0.0 &-8.1 & 0.0001, 0.01, 0.008 &0.002, 0.02, 0.013\\ 
cmb766&aen205&aen259& 10.0 & 3 & 90& 10&  245.4 &   36.1 &  131.7 & 0.2 &-2.6 & 0.0001, 0.01, 0.008 &0.002, 0.02, 0.013\\ 
cmb744&aen279&aen259& 10.0 & 3 & 60& 40&  160.6 &   87.6 &   49.8 & 0.4 &-4.0 & 0.0001, 0.04, 0.013 &0.002, 0.02, 0.013\\ 
\multicolumn{12}{l}{AGBb:}\\
cmb940&aen270&aen297& 12.0 & 3 & 80& 20&   81.7 &   10.7 & 10.6 &   -1.1 & 0.0 & 0.0001, 0.04, 0.013 & 0.002, 0.04, 0.015\\ 
cmb838&aen272&aen292& 10.5 & 5 & 70& 30&   90.2 &   44.3 & 17.0 &  -22.9 & 0.0 & 0.0001, 0.04, 0.013 & 0.0001, 0.04, 0.013\\ 
cmb713&aen273&aen267& 11.0 & 5 & 70& 30&   93.8 &   29.2 & 15.1 &  -21.3 & 0.0 & 0.0001, 0.04, 0.014 & 0.005, 0.04, 0.02\\ 
cmb785&aen270&aen266& 12.0 & 6 & 50& 50&  128.1 &   22.0 & 14.4 &  -11.9 & 0.0 & 0.0001, 0.04, 0.013 & 0.005, 0.04, 0.02\\ 
\hline
\end{tabular}
\end{table*}

To explore single age and two-burst models with an alternative, physically
motivated input MDF
we compare the observations with models that follow the classic closed box
chemical enrichment. While strictly speaking, a closed-box model cannot 
be an instantaneous burst (as are the models in the previous sections), 
we assume here that the duration of the closed-box enrichment 
sequence is ``fast'' relative to the time resolution of the model grid, 
which is near 1 Gyr. This is consistent with the adoption of alpha enhanced stellar evolution models. 

We explored a wide range of closed-box yields, minimum and maximum 
metallicities, and found the best fit to the observed CMD is provided
by models with ages 10.5-11~Gyr,  yield $y \sim 0.65-0.7 Z_\odot$, 
and metallicity spanning the full range of the adopted set of models, 
with the minimum $Z_{min}=0.0001$ and maximum metallicity 
$Z_{max}=+0.04$. The $\chi^2$ values of the best fitting single age model 
with the input closed box enrichment (model aen272, age 10.5~Gyr,) are:
$\chi^2(\mathrm{CMD})=52$, $\chi^2(\mathrm{LF}_I)=44$, and 
$\chi^2(\mathrm{LF}_V)=30$. The 11~Gyr old 
population model with slightly higher effective yield provide 
even smaller $\chi^2$ values of 27 and 12 for the I and V-band LFs, 
respectively (see Table~\ref{tab:chisq1ageCB} for details). 
As found in the case of single burst models with input observed MDF, the LF fits favor slightly
older age than the full CMD fit. However, overall the result is essentially the
same with the best fitting single age model of 10.5-11~Gyr.

We note that for the full CMD fit, LF fit (Fig.~\ref{fig:bestfitLF}) 
and colour distribution comparison with
data (Fig.~\ref{fig:VIhistograms}) the closed box single 
age models are a slightly better match than 
the single age simulations with the input observed MDF.
Also, as found for the models with the input observed MDF, the 
simulations with alpha enhanced isochrones provide better fit to the observations.

As done for models with the input MDF  we combine the single age closed box
models in order to explore the two-burst scenario. However, now 
in addition to the parameter of age, we have three more parameters: the
effective yield, minimum and maximum metallicity. Therefore the number of 
possible two burst combinations is significantly increased. 

The full set of models that have been constructed by combining two closed box
single age simulations, by randomly extracting a fraction P1 of old stars and a
fraction  P2 of 
younger stars, in the same way as described above for two
burst simulations with input observed MDF, is provided in the electronic format in
Table~\ref{tab:chi2cb2burst}.

Summarizing the results, we confirm the finding from the two-burst
input MDF simulations above: the best fitting models have $\sim 80$\% of 12~Gyr
old population mixed with $\sim 20$\% 2-4~Gyr old stars. The $\chi^2$ of the 
full CMD fit for the two burst model does not improve over the single age models
with input closed box enrichment, but the luminosity functions fit the data much
better. Also the numbers of RC and AGB bump stars in the respective 
boxes on the CMD are in better agreement with the observations for two burst
models. In particular this is a significant difference for the number of AGB bump 
stars that in single age models is systematically lower than in the observations.

Exploring the different minimum and maximum metallicity for the old and for the
young component we gain in addition some insight in the possible 
age-metallicity relation. 
The best fitting two-burst closed box model is cmb242 that is a combination of
80\% 12 Gyr old model aen270 that has effective yield 0.013, and that spans the
full scale of metallicity, from $Z_{min}=0.0001$ to $Z_{max}=0.04$. The young
component contributing 20\% of the stars is best represented by model aen316,
which has the same effective yield, but a higher minimum 
metallicity $Z_{min}=0.002$. The maximum metallicity extends to the solar
value $Z_{max}=0.02$. 
The simulations that had 
the young component with wider metallicity distribution, and in particular with
a metal-poor component had worse CMD fits. Similarly, the simulations constructed
with the old component that does not span the full metallicity range provide worse
fits. Therefore we can conclude that it is not only necessary to have
the bulk of population with age older than $\sim 10$~Gyr, but also that the 
old stars must cover the full range of metallicity.

\section{Discussion}

\begin{figure}
\centering
\resizebox{\hsize}{!}{
\includegraphics[angle=270]{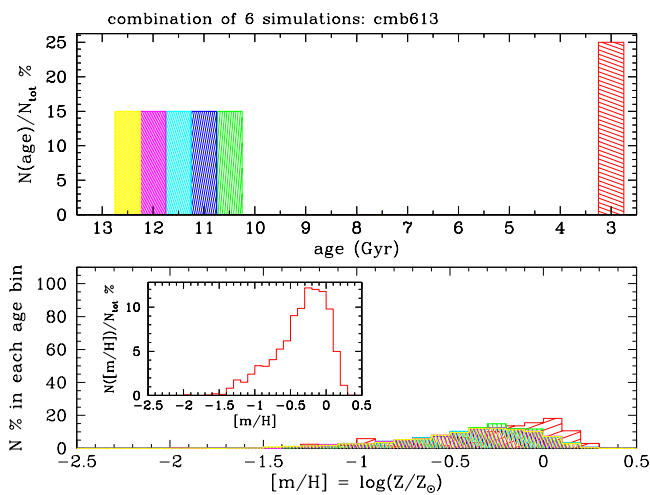}}
\caption[]{Age distribution (top panel) and metallicity distribution (bottom
panel) for an example of complex star formation history that has stars formed
first between 10.5-12.5~Gyr, and then a second very short burst at 3~Gyr. The
top panel shows the percentage of the stars in each age bin as a function of age,
while in the bottom panel the percentage of stars is given for each metallicity
bin, normalized to the total number of stars in a given age. In the insert on the
bottom panel the simulated total metallicity distribution is shown.
}
\label{fig:agemet_cmb613}
\end{figure}

The best fitting mean age of the halo stars in NGC~5128 is $11 \pm 1$~Gyr. 
This is older than the mean luminosity-weighted age of $\sim 8$ Gyr we 
derived in Paper~I from the
comparison of the observed luminosity function with the luminosity functions 
computed from BASTI models. In part the difference may be explained 
by the fact that in Paper~I we did not take into account the effects of 
photometric scatter in the models, and in part by the fact that we 
used a different model grid for the MDF derivation (the alpha-enhanced Victoria 
isochrones from \citet{vandenberg+00}) with respect to the LF modeling
\citep[BASTI][]{pietrinferni+04}.
In the present paper we self-consistently used the same stellar evolutionary 
models to derive the empirical MDF and in the simulations. 
One additional (though small) difference is the fact that in our new MDF, given in 
Table~\ref{tab:MDF_AGBcorr}, we make a correction for the average AGB 
contribution. 

The two-burst models with $\sim$70-80\% of 12 Gyr old population combined 
with 30-20\% 2-3 Gyr old second (younger) population give us the
best match to the observed CMD. The 2-burst model LFs as well as number counts
around RC and AGB bump features significantly improve 
the fit to the data over single-age models, and provide similar constraints 
to the full CMD fits: 70-80\% of the population has 12-12.5 old stars, 
and the younger component of 30-20\% has ages between 2-6 Gyr. 

The simulations allow us to estimate the total mass transformed into stars in the target field which accounts for the observed number of stars. For a flattened Salpeter IMF 
($\phi(M) \propto M^{-1.3}$ between 0.1 and 0.5 $M_\odot$ and $\phi(M) \propto M^{-2.35}$ between 0.5 and 120 $M_\odot$) the best fitting single burst simulations indicate that such mass amounts to $\simeq 4.5 \times 10^{7}$ $M_\odot$.  The best fitting double burst models yield a slightly smaller value of the total star formation in our field, i.e. $\simeq 4 \times 10^{7}$ $M_\odot$, since young populations are more efficient in producing post main sequence stars per unit mass. The mass fraction involved in the young component is 
sensitive to its precise age, and amounts to $\sim 0.1$ if the young burst occurred 4 to 5 Gyr ago, or to $\sim 0.06$ if it occurred 2 to 3 Gyr ago. 

We turn now to consider more complex star formation histories with the specific aim 
of testing some interesting scenarios. 

\subsection{Comparison of the stellar and globular cluster age and metallicity distributions}

\citet{woodley+10a} presented 
the most recent age and metallicity distributions of a large sample of globular clusters in 
NGC~5128 based on high quality Lick index measurements. 
The majority of their clusters
are located in the inner halo and bulge ($R_{gc} \lesssim 15$ kpc), 
whereas our sample of stars is a ``pencil beam''
at one particular location in the halo. Nevertheless it is instructive to 
compare the age and metallicity distributions of two populations.

Fitting the age distribution histogram of their observed clusters with Gaussians
they derive  the best fitting mean age of the clusters for a single 
Gaussian fit of $8.58$~Gyr with  $\sigma = 2.78$ Gyr, 
younger than the mean age of the stars we find in this paper, 
but consistent with our results in Paper~I. The best fitting bimodal 
distribution of clusters has 71\% of the clusters with 
$\tau_1 = 10.12$~Gyr ($\sigma =1.44$) and  
29\% with age $\tau_2 = 4.85$~Gyr ($\sigma = 1.32$). This is remarkably 
close to the proportions we find here for the halo stars in our
two-burst models for the stellar CMD. The minor residual differences in
the exact age values may well be due simply to the fact that the cluster
ages and stellar ages were derived through different methodology and 
with reference to different stellar model grids.  
On the other hand the 
majority of the globular clusters observed by \citet{woodley+10a} are 
more metal-poor than the bulk of the stellar halo.

\begin{table*}
\caption[]{The three best fitting models for each diagnostic are listed for 
the three burst combinations with the input observed MDF.
Diagnostics are $\chi^2$ for the full CMD, $\chi^2$ of the LF fit for I, and
V-bands, and in the last two columns 
$\Delta N / \sqrt{N_{obs}}= \mathrm{N}_{obs}-\mathrm{N}_{sim}/ \sqrt{N_{obs}}$ 
for the RC and AGBb boxes in our grid.  
}
\label{tab:bestfitmodels3}
\begin{tabular}{ccccrrrrrrrrrrr}
\multicolumn{15}{c}{Three burst models with input observed MDF} \\
\hline \hline
\multicolumn{1}{c}{combined} & \multicolumn{1}{c}{P1} &\multicolumn{1}{c}{P2} &
\multicolumn{1}{c}{P3} & \multicolumn{1}{c}{age1} &
\multicolumn{1}{c}{age2} &\multicolumn{1}{c}{age3} & 
\multicolumn{1}{c}{\%} &\multicolumn{1}{c}{\%} & \multicolumn{1}{c}{\%}&
\multicolumn{1}{c}{$\chi^2$}  & \multicolumn{1}{c}{$\chi^2$} &
\multicolumn{1}{c}{$\chi^2$}  & \multicolumn{1}{c}{$\frac{\Delta N}{\sqrt{N_{obs}}}$} &
\multicolumn{1}{c}{$\frac{\Delta N}{\sqrt{N_{obs}}}$}
\\
\multicolumn{1}{c}{simulation}& \multicolumn{1}{c}{simulation} &\multicolumn{1}{c}{simulation} &\multicolumn{1}{c}{simulation} &
\multicolumn{1}{c}{(Gyr)}&\multicolumn{1}{c}{(Gyr)}&\multicolumn{1}{c}{(Gyr)}&
\multicolumn{1}{c}{(P1)}&\multicolumn{1}{c}{(P2)}& \multicolumn{1}{c}{(P3)}& 
\multicolumn{1}{c}{CMD} & \multicolumn{1}{c}{LF$_I$} &
\multicolumn{1}{c}{LF$_V$} & \multicolumn{1}{c}{RC} &
\multicolumn{1}{c}{AGBb}
\\
\hline
\multicolumn{15}{l}{CMD:}\\
 new526 &  aen022 &  aen032 &  aen040 & 12.0 &  7.0 &  3.0 & 70& 10& 20&   64.7 &   12.2 &   14.4 &   -5.3 &   -2.5 \\ 
 new538 &  aen022 &  aen034 &  aen040 & 12.0 &  6.0 &  3.0 & 70& 10& 20&   67.2 &   14.0 &   13.4 &   -3.4 &   -1.9 \\ 
 cmb503 &  aen021 &  aen028 &  aen040 & 12.5 &  9.0 &  3.0 & 68& 14& 18&   67.9 &   10.6 &   10.8 &   -1.2 &   -2.6 \\ 
\multicolumn{15}{l}{LF$_I$:}\\
 cmb536 &  aen020 &  aen032 &  aen036 & 13.0 &  7.0 &  5.0 & 68& 14& 18&   78.8 &    7.6 &   10.7 &    0.1 &   -2.7 \\ 
 cmb507 &  aen021 &  aen028 &  aen038 & 12.5 &  9.0 &  4.0 & 68& 14& 18&   75.7 &    8.2 &   14.4 &   -6.2 &   -0.7 \\ 
 cmb524 &  aen020 &  aen030 &  aen036 & 13.0 &  8.0 &  5.0 & 68& 14& 18&   83.8 &    8.2 &   10.2 &    1.3 &   -2.4 \\ 
\multicolumn{15}{l}{LF$_V$:}\\
 cmb544 &  aen020 &  aen034 &  aen038 & 13.0 &  6.0 &  4.0 & 68& 14& 18&   77.1 &   11.7 &    9.0 &    5.2 &   -4.0 \\ 
 new540 &  aen020 &  aen034 &  aen040 & 13.0 &  6.0 &  3.0 & 70& 10& 20&   78.3 &   15.1 &    9.0 &    9.4 &   -5.7 \\ 
 cmb548 &  aen020 &  aen034 &  aen036 & 13.0 &  6.0 &  5.0 & 68& 14& 18&   80.6 &   10.1 &    9.3 &    3.8 &   -3.8 \\ 
\multicolumn{15}{l}{RC:}\\
 new575 &  aen021 &  aen032 &  aen040 & 12.5 &  7.0 &  3.0 & 80&  5& 15&   76.1 &   11.1 &   14.1 &   -0.1 &   -1.4 \\ 
 cmb536 &  aen020 &  aen032 &  aen036 & 13.0 &  7.0 &  5.0 & 68& 14& 18&   78.8 &    7.6 &   10.7 &    0.1 &   -2.7 \\ 
 cmb512 &  aen020 &  aen028 &  aen036 & 13.0 &  9.0 &  5.0 & 68& 14& 18&   82.2 &    8.8 &   11.2 &   -0.3 &   -2.4 \\ 
\multicolumn{15}{l}{AGBb:}\\
 new506 &  aen022 &  aen028 &  aen038 & 12.0 &  9.0 &  4.0 & 70& 10& 20&   71.1 &   12.1 &   12.9 &  -10.5 &    0.0 \\ 
 new534 &  aen022 &  aen032 &  aen036 & 12.0 &  7.0 &  5.0 & 70& 10& 20&   73.4 &   12.4 &   14.7 &  -12.4 &    0.1 \\ 
 cmb511 &  aen021 &  aen028 &  aen036 & 12.5 &  9.0 &  5.0 & 68& 14& 18&   76.5 &    8.8 &   12.8 &   -6.7 &   -0.1 \\ 
\hline
\end{tabular}
\end{table*}

In addition to the single age and bimodal age distribution \citet{woodley+10a}
also fit their globular cluster age distribution with three Gaussians. They find
that 68\% of the NGC~5128 globular clusters are old ($\tau_1>8$~Gyr), 14\%
have intermediate age ($\tau_2=5-8$~Gyr), and 18\% have young ages
($\tau_3<5$~Gyr). We explored the combination of 
old + intermediage-age + young single age models using both input MDF and
closed box MDF.  For these three-bursts star formation histories the fraction 
of the old component ($11-13$~Gyr) ranges between 60-80\%, the fraction of the
intermediate-age component ($6-9$~Gyr) between 5-15\%, and the young component 
($3-5$~Gyr) contributes 10-20\% of the stars. None of the many combinations
provided significantly better fits to the overall CMD and LFs with respect 
to the two burst simulations, but we note that the combination having similar
percentages of the old, intermediate-age and young stars as was found for the
clusters, is also consistent with the observed distribution of stars in the halo
field CMD (see Table~\ref{tab:bestfitmodels3}). 

The stellar 
MDF measured in four fields in NGC~5128, from 8 to 40 kpc \citep{harris+99, harris+harris00, 
harris+harris02, rejkuba+05}, does not show large differences. In all fields it is 
deficient in metal-poor stars, and has a peak at $\mathrm{[M/H]} \simeq -0.6$~dex. 
This metal-rich peak is very close to the metal-rich peak of the bimodal 
globular cluster metallicity distribution \citep{rejkuba01,woodley+10a}, while the metal-poor peak 
of the globular cluster MDF ($\mathrm{[Fe/H]} \sim -1.2$~dex) does not have the 
corresponding peak in the stellar MDF. This has already been noted
in previous studies of NGC~5128 \citep{harris+harris01,harris+harris02}, as well as
in other galaxies \citep[e.g.]{harris+harris01, forte+07}. It implies that the efficiency
of cluster formation relative to stars, measured by the globular cluster specific frequency
$S_N$, is a function of metallicity and that metal-poor clusters have greater formation efficiency
with respect to stars \citep{harris+harris02}. 
\citet{forte+07} and \citet{peng+08} discuss the possible implications of this proposed difference in efficiencies based on Monte Carlo based models and observations 
of many different galaxies. 

One interesting implication predicted  by  \citet{forte+07} is the
coexistence of two distinct stellar populations characterized by widely different metallicities
and spatial distributions. The metal-poor stellar halo is expected to be much more extended and
start dominating over the metal-rich component only at large galactocentric distances. The inner, more metal-rich halo component is expected to be extremely heterogeneous and to dominate the inner region of galaxies \citep{forte+07}. This is
remarkably reminiscent of the emerging picture mentioned in the introduction, where stellar halos appear to have two components with the metal-poor population dominating at large distances in Milky Way, M31, NGC~3379 \citep{carollo+07,chapman+06,kalirai+06,harris+07b}. 
Moreover, small-scale sub-structures appear to be quite frequent in large galaxy 
halos \citep{bell+08,ibata+09,mouhcine+10b}. 
However current data do not confirm whether this double nature of stellar halos is universal and the metal-poor component indeed dominating also in NGC~5128 at galactocentric distances larger than $\sim 12 \mathrm{R}_{eff}$. For that new observations are necessary.

\subsection{Duration of the star formation episodes}

Besides exploring the age of the old/young components, it is also interesting
to investigate whether it is possible to put constraints on the duration of the
star formation burst(s). To test this we made a number of combinations of single
age simulations with input observed MDF to test:
\begin{itemize}
\item Flat age distribution with constant star formation ranging over an extended
period. Here we tested the constant star formation between 3-13, 5-13, and 9-13
Gyr.
\item Flat age distribution for the old component with extended star formation
between 9-13 Gyr, with a 20-40\% contribution of young and intermediate-age stars
with ages between 3-5 Gyr.
\item Declining star formation lasting over more than 3~Gyr and starting $12$~Gyr
ago.
\item Bell shaped (first increasing, then decreasing), flat or declining 
star formation for the old population lasting 2-4~Gyr, 
with a very short younger burst contributing 20\% of the stars.
\end{itemize}

This, admittedly limited in parameter space, set of more complex
star formation histories shows much larger $\chi^2$ values ($>95$  for 
CMD fit) for all combinations when the old population was formed over an 
extended period longer than 3~Gyr, for both flat and declining age distributions. 
Better agreement with the observations is 
found for simulations where the old population has a mean age
between 10-12.5~Gyr and a bell-shaped or flat distribution of ages, 
with the majority of the first generation of stars born within a short 
$\la 2$~Gyr period. In
addition to the old population the younger component with star formation lasting
only $\sim 1$~Gyr is necessary  in order to
bring the full CMD $\chi^2$ fit down to values similar to those obtained with the
best fitting single and two burst simulations. An example of such a complex star
formation history that has $\chi^2=63$ for the full CMD fit, and  $\chi^2=18$ for
the luminosity function fits is shown in Figure~\ref{fig:agemet_cmb613}.

\subsection{Age-metallicity relation}

We explored the possibility of a variation of metallicity in lock steps 
with the age in the sense of increasing metallicity with decreasing age. 
The complex simulations were constructed by using
in input several single age simulations with the flat metallicity distribution, from which we 
extracted stars, filling the metal-poor bins of the MDF 
with old stars and more metal-rich bins with increasingly younger stars.
All of these cases gave significantly worse $\chi^2$ with respect to simulations that 
had old populations spanning the whole metallicity range. This is not too
surprising, since the age-metallicity anti-correlation keeps the RGB narrow in 
colour. This confirms again the result found from 2-burst closed box simulations, that 
required the old component to span the full metallicity range from $Z=0.0001$ to $Z=0.04$.

\section{Summary and conclusions}

In this paper we have used a series of stellar population models to study the
age distribution (ADF) and metallicity distribution (MDF) of the outer-halo stars in
NGC 5128.  Because it is only 3.8 Mpc distant, 
this target provides the best available opportunity to probe directly
into the stellar population of a giant E galaxy.
The reference data consist of our previously published deep HST/ACS
photometry in $(V,I)$ which cover the RGB, HB (RC), and AGB stages \citep{rejkuba+05}.  
We generate a large range of synthetic colour-magnitude diagrams from the
Teramo stellar model grid \citep{pietrinferni+04, pietrinferni+06, cordier+07} 
and use a variety of diagnostics to constrain
the best-fitting ADF and MDF.  The simulations are compared with the observations
and the following diagnostics are used to judge how well a simulation performs
with respect to the others: $\chi^2$ of the full CMD fit based on a custom made
grid, $\chi^2$ fits of the V and I-band luminosity functions, and relative number
of RC and AGB bump stars based on star counts in the appropriate boxes on the CMD. The
most sensitive diagnostics are the full CMD fit as well as the I-band LF.

Tables~\ref{tab:bestfitmodels1mdf},
\ref{tab:bestfitmodels2}, \ref{tab:bestfitmodels1cb},  and \ref{tab:bestfitmodels3} list 
the values of our diagnostics for the three best fitting model CMDs for 
single age, two-burst, and three-burst simulations. Based on these and discussion
above for additional more complex simulations the summary of our findings is as
follows. The most convincing conclusions are:

\begin{enumerate}
\item Almost irrespective of the metallicity distribution adopted to simulate our field, 
the observational CMD requires an old age. This is mostly driven by the position of the RC. 
The best-fitting mean  age for the halo stars is $11 \pm 1$ Gyr.

\item Again, almost irrespective  of the adopted metallicity distribution, the data are better 
fit with two episodes of star formation in which the old component dominates. This is 
mostly driven by the luminosity function of the RC stars. 
The best matches to the data are models with $\simeq 80$\% of the stars
at roughly 12 Gyr age, and only $\simeq 20$\% in the range  $2-4$ Gyr.  

\item The old component must span the full metallicity range 
($Z_{min}=0.0001$ to $Z_{max}=0.04$). This is driven by the 
width of the RGB. 

\end{enumerate}

In addition, we find formally better fits to both single age and two-burst models for
input closed box MDF, with respect to the input observed MDF. This is driven
by the colour distribution of the RC and the bright RGB.  We however do 
not emphasize this point due to the uncertainties in the bolometric corrections 
and the additional parameter of 
element overabundance of the models.
We also note that our best fitting multiple burst models have the young component which
does not extend all the way to the low metallicity end ($Z \ga 0.002$), suggestive that the young component is, on the average, more metal rich than the old one.

The  alpha enhanced isochrones 
provide superior fit to the data with respect to scaled solar ones. This is 
evident from the $\chi^2$ values of the overall CMD fit. Our tests of model to 
model comparisons (Sect. 3.3 and Fig.~\ref{fig:diagnostics_chi2}) 
clearly show that (i) when adequate input alpha enhanced tracks are used, the CMD $\chi^2$ diagnostic correctly reaches a value of 1 at the right age; (ii) when the input tracks have an inadequate $\alpha$ element distribution, the CMD $\chi^2$ diagnostic saturates at a relatively high value. Therefore, we suspect that more precise results could be obtained from simulations based on isochrones with a different chemical pattern, possibly with higher [$\alpha$/Fe]
ratios. This however would hardly accommodate younger age for the old component, since, as
[$\alpha$/Fe] increases, the isochrones become bluer. 

We conclude that the age of the bulk of the stars in our NGC 5128 halo field are
$\sim$ 11-12 Gyr old; 
a modest component with younger age ($\sim 2-4$ Gyr) is also present.
\emph{If} this region of NGC 5128 can be taken as representative
of the halo, we would conclude that most of its stars and clusters formed
at a very early time, in agreement with observational
discoveries of old early-type galaxies at high redshift
\citep{cimatti+04, daddi+05, renzini06, kriek+08} as well as with lower redshift
studies \citep[e.g.\ ][]{silva+bothun98, kuntschner+02, thomas+05}. 
This is also in agreement with fast monolithic collapse models of \citet{ikuta07}, 
who considered the  CMD morphology differences between  fast early monolithic collapse and
slow hierarchical merging early-type galaxy formation scenarios.
Our solutions leave room for a significant merger or accretion event a few Gyr in the past,
but they do not support the idea that the bulk of NGC 5128  stars formed in
a ``major merger'' at $z<2$.  

The ``anomalous'' features in NGC~5128, such as the central dust lane with its star forming
regions and the ring of young massive 
stars \citep{graham79, moellenhoff81, quillen+93, kainulainen+09} are significantly younger
than the 2-4 Gyr younger halo component. Therefore it is possible that NGC~5128 has suffered several merger episodes. Possible evidence of an older accretion event in the halo is provided by the  very diffuse star cluster candidate identified in our field \citep{mouhcine+10}. However,  accurate deep photometry over the wide area of the halo would be necessary to find the stellar streams or subgroups of star clusters associated with the merger event that contributed the 20-30\% younger component that is implied by the best fitting age distribution in our data.
The turn off magnitude of the young component occurs at
M$_I \simeq 2$, M$_{J,K} \simeq 1.5$. Therefore the presence of this young burst, its age and its stellar mass will be testable with JWST.

\begin{acknowledgements}
GLHH and WEH are pleased to thank ESO for support during visiting fellowships in 2009 and 2010. WEH acknowledges support from the Natural Sciences and Engineering Research Council of Canada. 
LG thanks ESO support during visiting felloship in 2010.
We thank the anonymous referee for a thoughtful and thorough report that helped to improve
the paper.
\end{acknowledgements}

\bibliographystyle{aa}


\bibliography{/Users/mrejkuba/Work/publications/Article/mybiblio}

\appendix

\section{Single age simulations diagnostics}

This section lists the comparison between the single age simulations and
observations for input observed MDF (Table~\ref{tab:chisq1age}) and for
closed box model with a range of yields, and initial/final
metallicities (Table~\ref{tab:chisq1ageCB}). All the simulations that are made using
solar scaled isochrones have names starting with "sol*" and those that have
alpha enhanced models are named "aen*".
The $\chi^2$ goodness of the fit of the full CMD, the V- and I- band
LFs as well as the comparison of the number of stars within the RC and AGBb boxes
are used as diagnostics of the fit. Those diagnostics that indicate the best
fit between the observations and models are indicated in bold faced letters. 
The bold text refers to the best three models listed in Table~\ref{tab:bestfitmodels1mdf} 
and the first  part of Table~\ref{tab:bestfitmodels1cb}.
This appendix and tables are only provided in the electronic edition.

\begin{table*}
\caption[]{Diagnostics for all single age simulations with input observed MDF
compared to observations. 
$\Delta N / \sqrt{N_{obs}}= \mathrm{N}_{obs}-\mathrm{N}_{sim}/ \sqrt{N_{obs}}$ 
in columns 6, 7, 13, and 14 is the 
number difference between the observed and simulated stars in the parts of the CMD
that are dominated by the RC (columns 6, 13) or AGBb stars (columns 7, 14), 
weighted by the Poissonian fluctuation in the number of stars expected from the 
observations.
The first set (left) is for the models that included alpha enhancement (aen*), 
the second (right) for
solar scaled models (sol*). The preferred models, those having smallest $\chi^2$
(or $\Delta N$ close to 0), 
for each diagnostic are indicated with bold-faced fonts. This table is given fully
in the electronic version.
}
\label{tab:chisq1age}
\begin{tabular}{crrrrrr|crrrrrr}
\hline \hline
\multicolumn{1}{c}{(1)} & \multicolumn{1}{c}{(2)} & 
\multicolumn{1}{c}{(3)} & \multicolumn{1}{c}{(4)} & 
\multicolumn{1}{c}{(5)} & \multicolumn{1}{c}{(6)} & 
\multicolumn{1}{c|}{(7)} &
\multicolumn{1}{c}{(8)} & \multicolumn{1}{c}{(9)} & 
\multicolumn{1}{c}{(10)} & \multicolumn{1}{c}{(11)} & 
\multicolumn{1}{c}{(12)} & \multicolumn{1}{c}{(13)} & 
\multicolumn{1}{c}{(14)} 
\\
\multicolumn{1}{c}{simulation} & \multicolumn{1}{c}{age} &
\multicolumn{1}{c}{$\chi^2$}  & \multicolumn{1}{c}{$\chi^2$} &
\multicolumn{1}{c}{$\chi^2$}  & 
\multicolumn{1}{c}{$\frac{\Delta N}{\sqrt{N_{obs}}}$} &
\multicolumn{1}{c|}{$\frac{\Delta N}{\sqrt{N_{obs}}}$} &
\multicolumn{1}{c}{simulation} & \multicolumn{1}{c}{age} &
\multicolumn{1}{c}{$\chi^2$}  & \multicolumn{1}{c}{$\chi^2$} &
\multicolumn{1}{c}{$\chi^2$}  & 
\multicolumn{1}{c}{$\frac{\Delta N}{\sqrt{N_{obs}}}$} &
\multicolumn{1}{c}{$\frac{\Delta N}{\sqrt{N_{obs}}}$}
\\
\multicolumn{1}{c}{ID}& \multicolumn{1}{c}{(Gyr)}&
\multicolumn{1}{c}{CMD} & \multicolumn{1}{c}{LF$_I$} &
\multicolumn{1}{c}{LF$_V$} & \multicolumn{1}{c}{RC} &
\multicolumn{1}{c|}{AGBb} &
\multicolumn{1}{c}{ID}& \multicolumn{1}{c}{(Gyr)}&
\multicolumn{1}{c}{CMD} & \multicolumn{1}{c}{LF$_I$} &
\multicolumn{1}{c}{LF$_V$} & \multicolumn{1}{c}{RC} &
\multicolumn{1}{c}{AGBb}
\\
\hline
 aen015 & 13.0 &  108.7   &   73.7   &   52.2   &   29.0   &	8.9  &	sol015 & 13.0 &  151.0 &   58.5 &   57.5   &   22.5   &  8.7 \\
 aen020 & 13.0 &  104.4   &   65.5   &   42.8   &   28.6   &	6.1  &	sol020 & 13.0 &  150.9 &   72.4 &   69.2   &   28.3   &  9.6 \\
 aen021 & 12.5 &   94.7   &   58.3   &   44.2   &   21.5   &	8.2  &	sol021 & 12.5 &  142.3 &   52.6 &   51.0   &   17.5   &  9.3 \\
 aen022 & 12.0 &   76.4   &{\bf 36.2}&{\bf 25.8}&    8.4   &{\bf 5.3}&  sol022 & 12.0 &  128.0 &   53.7 &   51.4   &   15.2   &  9.8 \\
 aen023 & 11.5 &   75.3   &   42.7   &   34.4   &    5.1   &	7.5  &	sol023 & 11.5 &  115.6 &   53.9 &   47.6   &	6.2   & 11.0 \\
 aen018 & 11.0 &{\bf 65.6}&   53.6   &   40.5   &{\bf -0.9}&    8.8  &  sol018 & 11.0 &  107.2 &   48.4 &   36.7   &   -3.7   &  9.1 \\
 aen024 & 11.0 &{\bf 66.2}&   55.3   &   39.1   &{\bf  1.1}&	7.4  &  sol024 & 11.0 &  110.4 &   49.8 &   41.7   &   -1.4   & 10.5 \\
 aen025 & 10.5 &{\bf 65.2}&   55.7   &   34.5   &  -10.6   &    7.0  &	sol025 & 10.5 &  108.0 &   52.4 &   32.9   &  -10.9   &  8.0 \\
 aen016 & 10.0 &   67.1   &   72.4   &   41.0   &  -10.1   &    7.4  &	sol016 & 10.0 &  105.0 &   60.3 &   32.7   &  -13.4   & 11.0 \\
 aen017 & 10.0 &   67.7   &   74.1   &   39.3   &   -9.4   &    8.6  &	sol017 & 10.0 &  107.7 &   56.2 &   32.0   &  -15.0   &  9.2 \\
 aen026 & 10.0 &   68.9   &   69.9   &   38.4   &  -11.5   &    8.4  &	sol026 & 10.0 &  108.7 &   62.4 &   37.1   &   -7.7   &  8.4 \\
 aen027 &  9.5 &   76.0   &   82.7   &   39.0   &  -15.0   &    9.1  &	sol027 &  9.5 &  109.8 &   65.2 &   30.4   &  -17.2   &  8.2 \\
 aen028 &  9.0 &   81.4   &  114.4   &   50.9   &   -7.9   &    7.5  &	sol028 &  9.0 &  111.7 &   73.5 &   29.4   &  -16.9   &  8.9 \\
 aen029 &  8.5 &   90.3   &  131.3   &   54.9   &   -6.8   &    8.1  &	sol029 &  8.5 &  115.9 &   85.4 &   29.0   &  -16.2   &  8.0 \\
 aen030 &  8.0 &   95.7   &  140.4   &   56.5   &{\bf -0.5}&    6.9  &  sol030 &  8.0 &  115.1 &   78.9 &{\bf 22.9}&  -15.8   &	 7.4 \\
 aen031 &  7.5 &  107.1   &  158.6   &   68.0   &    6.5   &	8.4  &	sol031 &  7.5 &  115.4 &   96.7 &   26.6   &  -10.9   &	 8.9 \\
 aen019 &  7.0 &  130.7   &  199.3   &   83.8   &   13.2   &	7.6  &	sol019 &  7.0 &  116.6 &  113.3 &   34.3   &   -3.3   &	 8.7 \\
 aen032 &  7.0 &  128.6   &  210.7   &   94.6   &   17.0   &   10.0  &	sol032 &  7.0 &  116.1 &  110.9 &   32.4   &   -6.2   &	 8.8 \\
 aen033 &  6.5 &  149.6   &  225.3   &   94.9   &   19.1   &	8.2  &	sol033 &  6.5 &  122.0 &  131.9 &   36.8   &{\bf -0.1}&  7.8 \\
 aen034 &  6.0 &  173.0   &  241.4   &  103.5   &   24.2   &	6.8  &	sol034 &  6.0 &  133.1 &  146.5 &   41.3   &	3.2   &	 8.7 \\
 aen035 &  5.5 &  210.8   &  292.7   &  125.1   &   29.4   &	7.3  &	sol035 &  5.5 &  140.5 &  166.8 &   47.7   &	9.6   &	 7.3 \\
 aen036 &  5.0 &  237.3   &  331.6   &  157.7   &   39.1   &	8.0  &	sol036 &  5.0 &  153.4 &  207.5 &   68.6   &   19.0   &	 6.6 \\
 aen037 &  4.5 &  285.6   &  370.8   &  180.1   &   45.2   &	7.7  &	sol037 &  4.5 &  177.6 &  254.1 &   90.7   &   26.3   &	 9.7 \\
 aen038 &  4.0 &  320.0   &  419.4   &  226.8   &   51.9   &	7.1  &	sol038 &  4.0 &  199.7 &  282.2 &  105.4   &   30.6   &	 7.1 \\
 aen039 &  3.5 &  375.1   &  513.2   &  282.4   &   58.7   &	7.3  &	sol039 &  3.5 &  231.7 &  349.8 &  147.8   &   38.6   &	 6.7 \\
 aen040 &  3.0 &  423.9   &  608.7   &  364.8   &   66.8   &	8.1  &	sol040 &  3.0 &  274.4 &  426.2 &  210.5   &   48.1   &	 6.1 \\
 aen041 &  2.5 &  511.2   &  684.6   &  428.0   &   73.6   &    7.5  &  sol041 &  2.5 &  347.8 &  526.8 &  275.0   &   59.7   &  7.5 \\
 aen042 &  2.0 &  643.7   &  801.0   &  568.3   &   80.9   &   10.2  &  sol042 &  2.0 &  427.7 &  640.7 &  379.7   &   68.7   &  6.7 \\
\hline
\end{tabular}
\end{table*}

\begin{table*}
\caption[]{Diagnostics for all single age simulations with closed box input MDF
compared to observations. Diagnostics are the same as in Table~\ref{tab:chisq1age}.
The preferred models for each diagnostic are indicated with bold-faced fonts. 
}
\label{tab:chisq1ageCB}
\begin{tabular}{crrrrrrrrr}
\hline \hline
\multicolumn{1}{c}{(1)} & \multicolumn{1}{c}{(2)} & 
\multicolumn{1}{c}{(3)} & \multicolumn{1}{c}{(4)} & 
\multicolumn{1}{c}{(5)} & \multicolumn{1}{c}{(6)} & \multicolumn{1}{c}{(7)} &
\multicolumn{1}{c}{(8)} & \multicolumn{1}{c}{(9)} & \multicolumn{1}{c}{(10)} 
\\
\multicolumn{1}{c}{simulation} & \multicolumn{1}{c}{age} &
\multicolumn{1}{c}{$Z_{min}$} & \multicolumn{1}{c}{$Z_{max}$} & \multicolumn{1}{c}{yield} &
\multicolumn{1}{c}{$\chi^2$}  & \multicolumn{1}{c}{$\chi^2$} &
\multicolumn{1}{c}{$\chi^2$}  & 
\multicolumn{1}{c}{$\frac{\Delta N}{\sqrt{N_{obs}}}$} &
\multicolumn{1}{c}{$\frac{\Delta N}{\sqrt{N_{obs}}}$}
\\
\multicolumn{1}{c}{ID}& \multicolumn{1}{c}{(Gyr)}&
                      &                          &                                         &
\multicolumn{1}{c}{CMD} & \multicolumn{1}{c}{LF$_I$} &
\multicolumn{1}{c}{LF$_V$} & \multicolumn{1}{c}{RC} &
\multicolumn{1}{c}{AGBb}
\\
\hline
 sol200 & 12.0 & 0.0001 & 0.0100 & 0.0060 &  183.2 &  129.7 &  115.6 &   46.7 &    7.7 \\ 
 sol201 & 10.0 & 0.0001 & 0.0100 & 0.0060 &  134.6 &   86.5 &  147.2 &   20.3 &   10.1 \\ 
 sol202 &  8.0 & 0.0080 & 0.0400 & 0.0130 &  861.0 &  224.9 &  644.9 &  -49.5 &   12.9 \\ 
 sol203 &  5.0 & 0.0080 & 0.0400 & 0.0130 &  475.0 &  182.0 &  272.8 &  -27.2 &   10.7 \\ 
 sol204 & 12.0 & 0.0001 & 0.0100 & 0.0080 &  173.5 &  133.4 &  120.9 &   45.1 &    8.4 \\ 
 sol205 & 10.0 & 0.0001 & 0.0100 & 0.0080 &  121.3 &   86.3 &  134.4 &   17.2 &    9.1 \\ 
 sol206 &  8.0 & 0.0080 & 0.0400 & 0.0100 &  714.7 &  218.9 &  491.1 &  -47.0 &   11.7 \\ 
 sol207 &  5.0 & 0.0080 & 0.0400 & 0.0100 &  421.2 &  176.2 &  286.4 &  -24.5 &    8.1 \\ 
 sol208 &  3.0 & 0.0080 & 0.0400 & 0.0100 &  334.1 &  292.6 &  188.2 &   10.6 &    0.5 \\ 
 aen200 & 12.0 & 0.0001 & 0.0100 & 0.0060 &  325.5 &  183.5 &  174.7 &   62.9 &   12.2 \\ 
 aen201 & 10.0 & 0.0001 & 0.0100 & 0.0060 &  207.3 &   96.5 &  240.7 &   32.8 &   13.9 \\ 
 aen202 &  8.0 & 0.0080 & 0.0400 & 0.0130 &  459.2 &  197.6 &  282.5 &  -43.4 &    8.3 \\ 
 aen203 &  5.0 & 0.0080 & 0.0400 & 0.0130 &  387.9 &  212.2 &  168.5 &   -9.2 &    3.7 \\ 
 aen204 & 12.0 & 0.0001 & 0.0100 & 0.0080 &  290.9 &  173.6 &  171.1 &   60.5 &   10.8 \\ 
 aen205 & 10.0 & 0.0001 & 0.0100 & 0.0080 &  186.8 &  111.7 &  240.5 &   30.1 &   14.9 \\ 
 aen206 &  8.0 & 0.0080 & 0.0400 & 0.0100 &  380.4 &  181.0 &  298.0 &  -38.4 &    5.8 \\ 
 aen207 &  5.0 & 0.0080 & 0.0400 & 0.0100 &  347.2 &  218.4 &  139.7 &   -4.5 &    4.4 \\ 
 aen208 &  3.0 & 0.0080 & 0.0400 & 0.0100 &  436.9 &  440.2 &  218.9 &   41.1 &    0.9 \\ 
 aen230 & 12.0 & 0.0010 & 0.0100 & 0.0020 &  471.5 &  291.9 &  267.7 &   77.6 &   15.2 \\ 
 aen231 & 11.0 & 0.0010 & 0.0100 & 0.0020 &  400.8 &  132.4 &  285.6 &   61.7 &   15.1 \\ 
 aen232 & 10.0 & 0.0010 & 0.0100 & 0.0020 &  302.8 &  121.9 &  356.7 &   44.0 &   16.1 \\ 
 aen233 &  9.0 & 0.0010 & 0.0100 & 0.0020 &  300.0 &  217.0 &  385.8 &   29.2 &   14.7 \\ 
 aen234 &  8.0 & 0.0010 & 0.0100 & 0.0020 &  341.9 &  317.8 &  453.1 &   40.6 &   16.3 \\ 
 aen241 & 11.0 & 0.0010 & 0.0100 & 0.0100 &  162.9 &   77.0 &  153.0 &   27.7 &    7.1 \\ 
 aen242 & 10.0 & 0.0010 & 0.0100 & 0.0100 &  146.2 &  116.2 &  187.8 &   13.2 &    6.9 \\ 
 aen243 &  9.0 & 0.0010 & 0.0100 & 0.0100 &  168.7 &  198.0 &  229.7 &   11.7 &    8.5 \\ 
 aen244 &  8.0 & 0.0010 & 0.0100 & 0.0100 &  206.5 &  274.7 &  275.9 &   23.7 &    8.3 \\ 
 aen245 &  7.0 & 0.0010 & 0.0100 & 0.0100 &  272.5 &  367.3 &  337.4 &   39.4 &   10.4 \\ 
 aen246 &  6.0 & 0.0010 & 0.0100 & 0.0100 &  369.2 &  465.7 &  411.4 &   54.2 &   11.2 \\ 
 aen247 &  5.0 & 0.0010 & 0.0100 & 0.0100 &  471.2 &  601.9 &  531.0 &   68.0 &   11.7 \\ 
 aen251 & 11.0 & 0.0020 & 0.0200 & 0.0130 &   77.8 &   67.4 &   66.2 &   -6.0 &    4.5 \\ 
 aen252 & 10.0 & 0.0020 & 0.0200 & 0.0130 &   95.8 &  102.8 &   77.4 &  -11.8 &    4.5 \\ 
 aen253 &  9.0 & 0.0020 & 0.0200 & 0.0130 &  107.8 &  125.4 &   86.9 &   -9.3 &    5.5 \\ 
 aen254 &  8.0 & 0.0020 & 0.0200 & 0.0130 &  129.0 &  168.7 &  102.9 & {\bf 0.6} &    5.6 \\ 
 aen255 &  7.0 & 0.0020 & 0.0200 & 0.0130 &  171.8 &  230.4 &  132.5 &   13.0 &    5.7 \\ 
 aen256 &  6.0 & 0.0020 & 0.0200 & 0.0130 &  222.2 &  297.4 &  171.0 &   29.6 &    5.6 \\ 
 aen257 &  5.0 & 0.0020 & 0.0200 & 0.0130 &  282.4 &  360.5 &  217.7 &   41.6 &    5.1 \\ 
 aen258 &  4.0 & 0.0020 & 0.0200 & 0.0130 &  372.2 &  510.8 &  314.0 &   56.2 &    6.3 \\ 
 aen259 &  3.0 & 0.0020 & 0.0200 & 0.0130 &  473.4 &  707.6 &  487.1 &   74.5 &    7.1 \\ 
 aen261 & 11.0 & 0.0050 & 0.0400 & 0.0200 &  401.5 &   85.1 &  182.9 &  -51.4 &    7.8 \\ 
 aen262 & 10.0 & 0.0050 & 0.0400 & 0.0200 &  372.2 &   91.0 &  146.8 &  -47.5 &    7.3 \\ 
 aen263 &  9.0 & 0.0050 & 0.0400 & 0.0200 &  367.5 &   96.8 &  153.0 &  -45.7 &    8.7 \\ 
 aen264 &  8.0 & 0.0050 & 0.0400 & 0.0200 &  337.5 &  105.0 &  115.7 &  -33.6 &    9.6 \\ 
 aen265 &  7.0 & 0.0050 & 0.0400 & 0.0200 &  301.8 &  125.0 &   90.5 &  -22.5 &    8.4 \\ 
 aen266 &  6.0 & 0.0050 & 0.0400 & 0.0200 &  295.2 &  157.7 &   86.2 &  -12.4 &    5.9 \\ 
 aen267 &  5.0 & 0.0050 & 0.0400 & 0.0200 &  301.6 &  193.0 &   78.7 &    5.1 &    7.1 \\ 
 aen268 &  4.0 & 0.0050 & 0.0400 & 0.0200 &  316.2 &  302.6 &  106.3 &   23.6 &    6.6 \\ 
 aen269 &  3.0 & 0.0050 & 0.0400 & 0.0200 &  372.8 &  418.7 &  167.5 &   44.0 &    5.0 \\ 
 aen271 & 11.0 & 0.0001 & 0.0400 & 0.0130 &   66.3 &   42.2 &   34.4 &   11.2 &   11.9 \\ 
 aen281 & 10.0 & 0.0001 & 0.0400 & 0.0160 &   59.6 &   55.8 &   22.9 &   -7.5 &   13.4 \\ 
 aen291 &  8.0 & 0.0001 & 0.0400 & 0.0130 &   82.1 &  147.4 &   62.0 &    7.2 &   12.8 \\ 
 aen272 & 10.5 & 0.0001 & 0.0400 & 0.0130 & {\bf 51.5} &   44.0 &   29.7 & {\bf -0.7} &   12.5 \\ 
\hline
\end{tabular}
\end{table*}

\addtocounter{table}{-1}

\begin{table*}
\caption[]{cont.}
\begin{tabular}{crrrrrrrrr}
\hline \hline
\multicolumn{1}{c}{(1)} & \multicolumn{1}{c}{(2)} & 
\multicolumn{1}{c}{(3)} & \multicolumn{1}{c}{(4)} & 
\multicolumn{1}{c}{(5)} & \multicolumn{1}{c}{(6)} & \multicolumn{1}{c}{(7)} &
\multicolumn{1}{c}{(8)} & \multicolumn{1}{c}{(9)} & \multicolumn{1}{c}{(10)} 
\\
\multicolumn{1}{c}{simulation} & \multicolumn{1}{c}{age} &
\multicolumn{1}{c}{$Z_{min}$} & \multicolumn{1}{c}{$Z_{max}$} & \multicolumn{1}{c}{yield} &
\multicolumn{1}{c}{$\chi^2$}  & \multicolumn{1}{c}{$\chi^2$} &
\multicolumn{1}{c}{$\chi^2$}  & 
\multicolumn{1}{c}{$\frac{\Delta N}{\sqrt{N_{obs}}}$} &
\multicolumn{1}{c}{$\frac{\Delta N}{\sqrt{N_{obs}}}$}
\\
\multicolumn{1}{c}{ID}& \multicolumn{1}{c}{(Gyr)}&
                      &                          &                                         &
\multicolumn{1}{c}{CMD} & \multicolumn{1}{c}{LF$_I$} &
\multicolumn{1}{c}{LF$_V$} & \multicolumn{1}{c}{RC} &
\multicolumn{1}{c}{AGBb}
\\
\hline
 aen270 & 12.0 & 0.0001 & 0.0400 & 0.0130 &   94.7 &   50.2 &   33.1 &   23.2 &   11.6 \\ 
 aen273 & 11.0 & 0.0001 & 0.0400 & 0.0140 & {\bf 54.7} &   35.3 &   21.3 &    2.8 &    9.5 \\ 
 aen274 & 11.0 & 0.0001 & 0.0400 & 0.0150 &   60.7 & {\bf 34.5} &   21.3 &    2.5 &   11.2 \\ 
 aen275 & 11.0 & 0.0001 & 0.0400 & 0.0160 &   65.9 &   36.2 &   19.6 &    1.4 &    9.7 \\ 
 aen276 & 11.0 & 0.0001 & 0.0400 & 0.0170 &   62.4 & {\bf 27.4} & {\bf 12.0} &   -5.3 &    9.6 \\ 
 aen277 & 11.0 & 0.0001 & 0.0400 & 0.0180 &   69.2 & {\bf 33.0} & {\bf 15.3} &   -3.4 &   10.6 \\ 
 aen278 & 11.0 & 0.0001 & 0.0400 & 0.0190 &   74.6 &   34.7 & {\bf 16.8} &   -1.8 &   11.3 \\ 
 aen279 & 10.0 & 0.0001 & 0.0400 & 0.0130 & {\bf 54.8} &   66.4 &   38.5 & {\bf 0.2} &   11.9 \\ 
 aen280 & 10.0 & 0.0001 & 0.0400 & 0.0100 &   62.7 &   69.0 &   63.9 &    7.6 &   11.9 \\ 
 aen281 & 10.0 & 0.0001 & 0.0400 & 0.0160 &   59.6 &   55.8 &   22.9 &   -7.5 &   13.4 \\ 
 aen282 & 12.0 & 0.0001 & 0.0400 & 0.0130 &   92.3 &   48.4 &   30.9 &   23.6 &   10.1 \\ 
 aen283 & 12.0 & 0.0001 & 0.0400 & 0.0100 &  116.2 &   66.5 &   46.7 &   30.3 &   10.2 \\ 
 aen284 & 12.0 & 0.0001 & 0.0400 & 0.0160 &   90.2 &   42.5 &   26.1 &   18.2 &   10.5 \\ 
 aen285 & 12.0 & 0.0001 & 0.0100 & 0.0130 &  248.8 &  154.1 &  153.2 &   55.1 &   11.7 \\ 
 aen286 & 12.0 & 0.0001 & 0.0100 & 0.0100 &  267.4 &  166.7 &  173.7 &   57.0 &   13.1 \\ 
 aen287 & 12.0 & 0.0001 & 0.0100 & 0.0160 &  243.7 &  151.6 &  153.6 &   54.2 &   12.4 \\ 
 aen288 & 10.0 & 0.0001 & 0.0100 & 0.0130 &  159.4 &   98.4 &  196.2 &   23.9 &   11.0 \\ 
 aen289 & 10.0 & 0.0001 & 0.0100 & 0.0100 &  171.9 &  102.4 &  215.7 &   27.2 &   10.8 \\ 
 aen290 & 10.0 & 0.0001 & 0.0100 & 0.0160 &  158.1 &  104.2 &  206.1 &   23.9 &   11.2 \\ 
 aen291 &  8.0 & 0.0001 & 0.0400 & 0.0130 &   82.1 &  147.4 &   62.0 &    7.2 &   12.8 \\ 
 aen292 &  5.0 & 0.0001 & 0.0400 & 0.0130 &  227.2 &  326.5 &  153.9 &   43.7 &   10.5 \\ 
 aen293 &  3.0 & 0.0001 & 0.0400 & 0.0130 &  444.7 &  565.7 &  318.6 &   66.6 &   11.7 \\ 
 aen294 &  8.0 & 0.0020 & 0.0400 & 0.0130 &  116.9 &  132.7 &   45.5 &   -9.5 &    7.3 \\ 
 aen295 &  5.0 & 0.0020 & 0.0400 & 0.0130 &  233.3 &  274.1 &  105.1 &   29.7 &    7.5 \\ 
 aen296 &  3.0 & 0.0020 & 0.0400 & 0.0130 &  418.8 &  483.3 &  254.8 &   57.9 &    5.7 \\ 
 aen297 &  3.0 & 0.0020 & 0.0400 & 0.0150 &  366.2 &  494.5 &  239.2 &   57.9 &    7.5 \\ 
 aen298 &  4.0 & 0.0020 & 0.0400 & 0.0130 &  291.4 &  392.1 &  180.6 &   45.9 &    6.0 \\ 
 aen299 &  3.0 & 0.0020 & 0.0300 & 0.0130 &  416.4 &  587.6 &  348.0 &   66.5 &    7.0 \\ 
 aen300 &  4.0 & 0.0020 & 0.0300 & 0.0130 &  313.3 &  407.1 &  206.9 &   48.2 &    8.0 \\ 
 aen301 &  3.0 & 0.0050 & 0.0400 & 0.0130 &  376.0 &  467.4 &  226.1 &   51.9 &    5.5 \\ 
 aen302 &  4.0 & 0.0050 & 0.0400 & 0.0130 &  313.4 &  323.5 &  144.4 &   31.1 &    5.1 \\ 
 aen303 &  3.0 & 0.0050 & 0.0300 & 0.0130 &  444.9 &  496.4 &  256.0 &   52.0 & {\bf -0.8} \\ 
 aen304 &  4.0 & 0.0050 & 0.0300 & 0.0130 &  354.1 &  336.1 &  157.0 &   32.0 &    2.5 \\ 
 aen305 &  3.0 & 0.0050 & 0.0200 & 0.0130 &  502.8 &  571.9 &  356.6 &   61.2 &   -2.1 \\ 
 aen306 &  4.0 & 0.0050 & 0.0200 & 0.0130 &  383.7 &  430.2 &  252.0 &   41.9 & {\bf 0.5} \\ 
 aen307 &  2.0 & 0.0001 & 0.0400 & 0.0130 &  702.8 &  780.9 &  514.8 &   81.2 &   12.1 \\ 
 aen308 &  2.5 & 0.0001 & 0.0400 & 0.0130 &  537.6 &  680.7 &  417.9 &   74.2 &   12.3 \\ 
 aen309 &  2.0 & 0.0001 & 0.0400 & 0.0170 &  567.9 &  801.0 &  472.2 &   79.1 &   11.1 \\ 
 aen310 &  2.0 & 0.0001 & 0.0400 & 0.0100 &  845.3 &  878.3 &  620.2 &   84.8 &   13.5 \\ 
 aen311 &  2.0 & 0.0010 & 0.0400 & 0.0130 &  645.2 &  737.4 &  464.0 &   78.5 &    7.8 \\ 
 aen312 &  2.0 & 0.0020 & 0.0400 & 0.0130 &  552.8 &  769.4 &  479.5 &   77.8 &    7.0 \\ 
 aen313 &  2.0 & 0.0050 & 0.0400 & 0.0130 &  481.3 &  702.2 &  396.6 &   71.8 &    4.1 \\ 
 aen314 &  2.0 & 0.0080 & 0.0400 & 0.0130 &  503.9 &  614.2 &  318.3 &   63.8 &    0.9 \\ 
 aen315 &  2.0 & 0.0050 & 0.0200 & 0.0130 &  660.7 &  819.9 &  605.8 &   80.7 &    1.0 \\ 
 aen316 &  2.0 & 0.0020 & 0.0200 & 0.0130 &  727.2 &  930.1 &  734.2 &   86.9 &    8.0 \\ 
 aen317 &  2.5 & 0.0010 & 0.0400 & 0.0130 &  449.8 &  705.2 &  423.3 &   74.1 &   12.9 \\ 
 aen318 &  2.5 & 0.0020 & 0.0400 & 0.0130 &  454.5 &  624.6 &  354.6 &   69.8 &    6.5 \\ 
 aen319 &  2.5 & 0.0050 & 0.0400 & 0.0130 &  453.3 &  544.8 &  272.1 &   58.7 &    2.9 \\ 
 aen320 &  2.5 & 0.0080 & 0.0400 & 0.0130 &  498.5 &  501.8 &  239.0 &   48.4 &{\bf -0.1} \\ 
 aen321 &  2.5 & 0.0050 & 0.0200 & 0.0130 &  538.9 &  704.1 &  477.4 &   73.2 &    1.2 \\ 
 aen322 &  2.5 & 0.0020 & 0.0200 & 0.0130 &  599.3 &  776.6 &  554.0 &   79.4 &    3.9 \\ 
\hline
\end{tabular}
\end{table*}

\section{Double burst simulations diagnostics}

This section lists the comparison between the combined simulations that simulate 
two burst star formation history. 
All the single age simulations that are made using
solar scaled isochrones have names starting with "sol*" and those that have
alpha enhanced models are named "aen*". The combined simulations have names starting with 
"cmb*".
The combinations are made by extracting randomly
P1 \% of stars from simulation 1 (old age) and P2 \% of stars from simulation 2
(young age) and verifying that the observed MDF is populated correctly. The input
simulations are single age simulations with either input observed 
MDF (Table~\ref{tab:chisq1age}) or with 
closed box model with a range of yields, and initial/final
metallicities (Table~\ref{tab:chisq1ageCB}). 
All the single age simulations used in input that are made using
solar scaled isochrones have names starting with "sol*" and those that have
alpha enhanced models are named "aen*".
The $\chi^2$ goodness of the fit of the full CMD, the V- and I- band
LFs as well as the comparison of the number of stars within the RC and AGBb boxes
are used as diagnostics of the fit. Those diagnostics that indicate the best
fit between the observations and models are indicated in bold faced letters 
and correspond to the best three models reported in the second part of 
Table~\ref{tab:bestfitmodels1cb} and in Table~\ref{tab:bestfitmodels2}.
This appendix and tables are only provided in the electronic edition.

\begin{table*}
\caption[]{Diagnostics for all 2-burst simulations composed by combining two
single age simulations with input observed MDF compared to observations.
The preferred models, those having smallest $\chi^2$ (or $\Delta N$ close to 0), 
for each diagnostic are indicated with bold-faced fonts. This table is given fully in 
the electronic version.
}
\label{tab:chisq2age}
\begin{tabular}{cccrrrrrrrrr}
\hline \hline
\multicolumn{1}{c}{(1)} & \multicolumn{1}{c}{(2)} & 
\multicolumn{1}{c}{(3)} & \multicolumn{1}{c}{(4)} & 
\multicolumn{1}{c}{(5)} & \multicolumn{1}{c}{(6)} & 
\multicolumn{1}{c}{(7)} 
&\multicolumn{1}{c}{(8)} & \multicolumn{1}{c}{(9)} & \multicolumn{1}{c}{(10)} & \multicolumn{1}{c}{(11)}
&\multicolumn{1}{c}{(12)} 
\\
\multicolumn{1}{c}{combined} & \multicolumn{1}{c}{old} &\multicolumn{1}{c}{young} & 
\multicolumn{1}{c}{age1} &\multicolumn{1}{c}{age2} & 
\multicolumn{1}{c}{\%} &\multicolumn{1}{c}{\%} & 
\multicolumn{1}{c}{$\chi^2$}  & \multicolumn{1}{c}{$\chi^2$} &
\multicolumn{1}{c}{$\chi^2$}  & \multicolumn{1}{c}{$\frac{\mathrm{N}_{obs}-\mathrm{N}_{sim}}{\sqrt{N}}$} &
\multicolumn{1}{c}{$\frac{\mathrm{N}_{obs}-\mathrm{N}_{sim}}{\sqrt{N}}$}
\\
\multicolumn{1}{c}{simulation}& \multicolumn{1}{c}{simulation} &\multicolumn{1}{c}{simulation} &
\multicolumn{1}{c}{(Gyr)}&\multicolumn{1}{c}{(Gyr)}& \multicolumn{1}{c}{(old)}&\multicolumn{1}{c}{(young)}& 
\multicolumn{1}{c}{CMD} & \multicolumn{1}{c}{LF$_I$} &
\multicolumn{1}{c}{LF$_V$} & \multicolumn{1}{c}{RC} &
\multicolumn{1}{c}{AGBb}
\\
\hline
 cmb376 &  aen021 &  aen030 & 12.5 &    8 & 90& 10&  102.6 &   20.3 &   16.7 &   -5.8 &    1.4 \\ 
 cmb377 &  aen021 &  aen030 & 12.5 &    8 & 80& 20&   96.6 &   14.5 &   16.4 &   -9.2 &    1.2 \\ 
 cmb378 &  aen021 &  aen030 & 12.5 &    8 & 70& 30&   86.1 & {\bf 10.2} &   14.8 &  -11.1 &    2.0 \\ 
 cmb379 &  aen021 &  aen030 & 12.5 &    8 & 60& 40&   78.9 &   10.7 &   12.8 &  -13.5 &    1.6 \\ 
 cmb380 &  aen021 &  aen030 & 12.5 &    8 & 50& 50&   79.8 &   15.0 &   14.0 &  -15.6 &    1.7 \\
\hline
\end{tabular}
\end{table*}

\addtocounter{table}{-1}

\begin{table*}
\caption[]{continue.
}
\begin{tabular}{cccrrrrrrrrr}
\hline \hline
\multicolumn{1}{c}{(1)} & \multicolumn{1}{c}{(2)} & 
\multicolumn{1}{c}{(3)} & \multicolumn{1}{c}{(4)} & 
\multicolumn{1}{c}{(5)} & \multicolumn{1}{c}{(6)} & 
\multicolumn{1}{c}{(7)} 
&\multicolumn{1}{c}{(8)} & \multicolumn{1}{c}{(9)} & \multicolumn{1}{c}{(10)} & \multicolumn{1}{c}{(11)}
&\multicolumn{1}{c}{(12)} 
\\
\multicolumn{1}{c}{combined} & \multicolumn{1}{c}{old} &\multicolumn{1}{c}{young} & 
\multicolumn{1}{c}{age1} &\multicolumn{1}{c}{age2} & 
\multicolumn{1}{c}{\%} &\multicolumn{1}{c}{\%} & 
\multicolumn{1}{c}{$\chi^2$}  & \multicolumn{1}{c}{$\chi^2$} &
\multicolumn{1}{c}{$\chi^2$}  & \multicolumn{1}{c}{$\frac{\mathrm{N}_{obs}-\mathrm{N}_{sim}}{\sqrt{N}}$} &
\multicolumn{1}{c}{$\frac{\mathrm{N}_{obs}-\mathrm{N}_{sim}}{\sqrt{N}}$}
\\
\multicolumn{1}{c}{simulation}& \multicolumn{1}{c}{simulation} &\multicolumn{1}{c}{simulation} &
\multicolumn{1}{c}{(Gyr)}&\multicolumn{1}{c}{(Gyr)}& \multicolumn{1}{c}{(old)}&\multicolumn{1}{c}{(young)}& 
\multicolumn{1}{c}{CMD} & \multicolumn{1}{c}{LF$_I$} &
\multicolumn{1}{c}{LF$_V$} & \multicolumn{1}{c}{RC} &
\multicolumn{1}{c}{AGBb}
\\
\hline
 cmb351 &  aen022 &  aen026 & 12.0 &   10 & 90& 10&   96.8 &   23.7 &   16.4 &  -13.9 &   -0.2 \\ 
 cmb352 &  aen022 &  aen026 & 12.0 &   10 & 80& 20&   96.1 &   22.1 &   15.3 &  -15.0 &    0.5 \\ 
 cmb353 &  aen022 &  aen026 & 12.0 &   10 & 70& 30&   93.2 &   21.6 &   15.8 &  -16.7 &    2.4 \\ 
 cmb354 &  aen022 &  aen026 & 12.0 &   10 & 60& 40&   86.8 &   19.9 &   16.8 &  -21.4 &    1.9 \\ 
 cmb355 &  aen022 &  aen026 & 12.0 &   10 & 50& 50&   81.3 &   20.7 &   14.7 &  -24.4 &    1.0 \\
\hline
\end{tabular}
\end{table*}

\addtocounter{table}{-1}

\begin{table*}
\caption[]{continue.
}
\begin{tabular}{cccrrrrrrrrr}
\hline \hline
\multicolumn{1}{c}{(1)} & \multicolumn{1}{c}{(2)} & 
\multicolumn{1}{c}{(3)} & \multicolumn{1}{c}{(4)} & 
\multicolumn{1}{c}{(5)} & \multicolumn{1}{c}{(6)} & 
\multicolumn{1}{c}{(7)} 
&\multicolumn{1}{c}{(8)} & \multicolumn{1}{c}{(9)} & \multicolumn{1}{c}{(10)} & \multicolumn{1}{c}{(11)}
&\multicolumn{1}{c}{(12)} 
\\
\multicolumn{1}{c}{combined} & \multicolumn{1}{c}{old} &\multicolumn{1}{c}{young} & 
\multicolumn{1}{c}{age1} &\multicolumn{1}{c}{age2} & 
\multicolumn{1}{c}{\%} &\multicolumn{1}{c}{\%} & 
\multicolumn{1}{c}{$\chi^2$}  & \multicolumn{1}{c}{$\chi^2$} &
\multicolumn{1}{c}{$\chi^2$}  & \multicolumn{1}{c}{$\frac{\mathrm{N}_{obs}-\mathrm{N}_{sim}}{\sqrt{N}}$} &
\multicolumn{1}{c}{$\frac{\mathrm{N}_{obs}-\mathrm{N}_{sim}}{\sqrt{N}}$}
\\
\multicolumn{1}{c}{simulation}& \multicolumn{1}{c}{simulation} &\multicolumn{1}{c}{simulation} &
\multicolumn{1}{c}{(Gyr)}&\multicolumn{1}{c}{(Gyr)}& \multicolumn{1}{c}{(old)}&\multicolumn{1}{c}{(young)}& 
\multicolumn{1}{c}{CMD} & \multicolumn{1}{c}{LF$_I$} &
\multicolumn{1}{c}{LF$_V$} & \multicolumn{1}{c}{RC} &
\multicolumn{1}{c}{AGBb}
\\
\hline
 cmb471 &  aen022 &  sol040 & 12.0 &    3 & 90& 10&   83.6 &   17.1 &   15.4 &  -12.7 &   -0.3 \\ 
 cmb472 &  aen022 &  sol040 & 12.0 &    3 & 80& 20&   80.7 & {\bf 10.5} &   14.1 &   -9.4 &    0.2 \\ 
 cmb473 &  aen022 &  sol040 & 12.0 &    3 & 70& 30&   71.8 &   15.5 &   15.0 &   -5.9 &   -3.9 \\ 
 cmb474 &  aen022 &  sol040 & 12.0 &    3 & 60& 40&   83.9 &   22.4 &   16.0 &   -5.8 &   -3.1 \\ 
 cmb475 &  aen022 &  sol040 & 12.0 &    3 & 50& 50&  103.8 &   37.1 &   18.7 &   -1.6 &   -5.3 \\
\hline
\end{tabular}
\end{table*}

\addtocounter{table}{-1}
\begin{table*}
\caption[]{continue.
}
\begin{tabular}{cccrrrrrrrrr}
\hline \hline
\multicolumn{1}{c}{(1)} & \multicolumn{1}{c}{(2)} & 
\multicolumn{1}{c}{(3)} & \multicolumn{1}{c}{(4)} & 
\multicolumn{1}{c}{(5)} & \multicolumn{1}{c}{(6)} & 
\multicolumn{1}{c}{(7)} 
&\multicolumn{1}{c}{(8)} & \multicolumn{1}{c}{(9)} & \multicolumn{1}{c}{(10)} & \multicolumn{1}{c}{(11)}
&\multicolumn{1}{c}{(12)} 
\\
\multicolumn{1}{c}{combined} & \multicolumn{1}{c}{old} &\multicolumn{1}{c}{young} & 
\multicolumn{1}{c}{age1} &\multicolumn{1}{c}{age2} & 
\multicolumn{1}{c}{\%} &\multicolumn{1}{c}{\%} & 
\multicolumn{1}{c}{$\chi^2$}  & \multicolumn{1}{c}{$\chi^2$} &
\multicolumn{1}{c}{$\chi^2$}  & \multicolumn{1}{c}{$\frac{\mathrm{N}_{obs}-\mathrm{N}_{sim}}{\sqrt{N}}$} &
\multicolumn{1}{c}{$\frac{\mathrm{N}_{obs}-\mathrm{N}_{sim}}{\sqrt{N}}$}
\\
\multicolumn{1}{c}{simulation}& \multicolumn{1}{c}{simulation} &\multicolumn{1}{c}{simulation} &
\multicolumn{1}{c}{(Gyr)}&\multicolumn{1}{c}{(Gyr)}& \multicolumn{1}{c}{(old)}&\multicolumn{1}{c}{(young)}& 
\multicolumn{1}{c}{CMD} & \multicolumn{1}{c}{LF$_I$} &
\multicolumn{1}{c}{LF$_V$} & \multicolumn{1}{c}{RC} &
\multicolumn{1}{c}{AGBb}
\\
\hline
 cmb431 &  aen018 &  sol030 & 11.0 &    8 & 90& 10&   98.5 &   33.9 &   14.8 &  -27.4 &    2.0 \\ 
 cmb432 &  aen018 &  sol030 & 11.0 &    8 & 80& 20&   98.5 &   33.4 &   16.7 &  -27.4 &    4.1 \\ 
 cmb433 &  aen018 &  sol030 & 11.0 &    8 & 70& 30&  106.2 &   33.2 &   15.0 &  -28.9 &    4.1 \\ 
 cmb434 &  aen018 &  sol030 & 11.0 &    8 & 60& 40&  117.7 &   37.9 &   15.8 &  -28.4 &    4.7 \\ 
 cmb435 &  aen018 &  sol030 & 11.0 &    8 & 50& 50&  127.1 &   36.9 &   15.1 &  -28.0 &    2.8 \\
\hline
\end{tabular}
\end{table*}

\addtocounter{table}{-1}
\begin{table*}
\caption[]{continue.
}
\begin{tabular}{cccrrrrrrrrr}
\hline \hline
\multicolumn{1}{c}{(1)} & \multicolumn{1}{c}{(2)} & 
\multicolumn{1}{c}{(3)} & \multicolumn{1}{c}{(4)} & 
\multicolumn{1}{c}{(5)} & \multicolumn{1}{c}{(6)} & 
\multicolumn{1}{c}{(7)} 
&\multicolumn{1}{c}{(8)} & \multicolumn{1}{c}{(9)} & \multicolumn{1}{c}{(10)} & \multicolumn{1}{c}{(11)}
&\multicolumn{1}{c}{(12)} 
\\
\multicolumn{1}{c}{combined} & \multicolumn{1}{c}{old} &\multicolumn{1}{c}{young} & 
\multicolumn{1}{c}{age1} &\multicolumn{1}{c}{age2} & 
\multicolumn{1}{c}{\%} &\multicolumn{1}{c}{\%} & 
\multicolumn{1}{c}{$\chi^2$}  & \multicolumn{1}{c}{$\chi^2$} &
\multicolumn{1}{c}{$\chi^2$}  & \multicolumn{1}{c}{$\frac{\mathrm{N}_{obs}-\mathrm{N}_{sim}}{\sqrt{N}}$} &
\multicolumn{1}{c}{$\frac{\mathrm{N}_{obs}-\mathrm{N}_{sim}}{\sqrt{N}}$}
\\
\multicolumn{1}{c}{simulation}& \multicolumn{1}{c}{simulation} &\multicolumn{1}{c}{simulation} &
\multicolumn{1}{c}{(Gyr)}&\multicolumn{1}{c}{(Gyr)}& \multicolumn{1}{c}{(old)}&\multicolumn{1}{c}{(young)}& 
\multicolumn{1}{c}{CMD} & \multicolumn{1}{c}{LF$_I$} &
\multicolumn{1}{c}{LF$_V$} & \multicolumn{1}{c}{RC} &
\multicolumn{1}{c}{AGBb}
\\
\hline
 cmb101 &  aen026 &  aen041 & 10.0 &  2.5 & 90& 10&   97.3 &   50.7 &	14.9 &  -32.0 &   -0.3 \\ 
 cmb102 &  aen026 &  aen041 & 10.0 &  2.5 & 80& 20&  103.2 &   60.2 &	21.5 &  -23.9 &   -3.3 \\ 
 cmb103 &  aen026 &  aen041 & 10.0 &  2.5 & 70& 30&  116.4 &   69.5 &	27.0 &  -14.1 &   -5.2 \\ 
 cmb104 &  aen026 &  aen041 & 10.0 &  2.5 & 60& 40&  147.4 &   87.0 &	37.7 &   -5.8 &   -3.0 \\ 
 cmb105 &  aen026 &  aen041 & 10.0 &  2.5 & 50& 50&  203.8 &  111.1 &	53.0 &    4.7 &   -6.7 \\
\hline
\end{tabular}
\end{table*}

\begin{table*}
\caption[]{Diagnostics for all 2-burst simulations with input closed box MDF
compared to observations.
The preferred models, those having smallest $\chi^2$ (or $\Delta N$ close to 0), 
for each diagnostic are indicated with bold-faced fonts. This table is given fully in 
the electronic version.
}
\label{tab:chi2cb2burst}
\begin{tabular}{cccrrrrrrrrr}
\hline \hline
\multicolumn{1}{c}{(1)} & \multicolumn{1}{c}{(2)} & 
\multicolumn{1}{c}{(3)} & \multicolumn{1}{c}{(4)} & 
\multicolumn{1}{c}{(5)} & \multicolumn{1}{c}{(6)} & 
\multicolumn{1}{c}{(7)} 
&\multicolumn{1}{c}{(8)} & \multicolumn{1}{c}{(9)} & \multicolumn{1}{c}{(10)} & \multicolumn{1}{c}{(11)}
&\multicolumn{1}{c}{(12)} 
\\
\multicolumn{1}{c}{combined} & \multicolumn{1}{c}{old} &\multicolumn{1}{c}{young} & 
\multicolumn{1}{c}{age1} &\multicolumn{1}{c}{age2} & 
\multicolumn{1}{c}{\%} &\multicolumn{1}{c}{\%} & 
\multicolumn{1}{c}{$\chi^2$}  & \multicolumn{1}{c}{$\chi^2$} &
\multicolumn{1}{c}{$\chi^2$}  & \multicolumn{1}{c}{$\frac{\mathrm{N}_{obs}-\mathrm{N}_{sim}}{\sqrt{N}}$} &
\multicolumn{1}{c}{$\frac{\mathrm{N}_{obs}-\mathrm{N}_{sim}}{\sqrt{N}}$}
\\
\multicolumn{1}{c}{simulation}& \multicolumn{1}{c}{simulation} &\multicolumn{1}{c}{simulation} &
\multicolumn{1}{c}{(Gyr)}&\multicolumn{1}{c}{(Gyr)}& \multicolumn{1}{c}{(old)}&\multicolumn{1}{c}{(young)}& 
\multicolumn{1}{c}{CMD} & \multicolumn{1}{c}{LF$_I$} &
\multicolumn{1}{c}{LF$_V$} & \multicolumn{1}{c}{RC} &
\multicolumn{1}{c}{AGBb}
\\
\hline
 cmb701 &  aen273 &  aen203 & 11.0 &    5 & 90& 10&   88.4 &   27.4 &   13.2 &  -27.0 &   -0.8 \\ 
 cmb702 &  aen273 &  aen203 & 11.0 &    5 & 80& 20&   99.3 &   28.6 &   14.0 &  -25.3 &    1.6 \\ 
 cmb703 &  aen273 &  aen203 & 11.0 &    5 & 70& 30&  105.2 &   33.1 &   16.3 &  -23.5 &    2.3 \\ 
 cmb704 &  aen273 &  aen203 & 11.0 &    5 & 60& 40&  130.3 &   40.8 &   18.8 &  -23.4 &    3.8 \\ 
 cmb705 &  aen273 &  aen203 & 11.0 &    5 & 50& 50&  171.0 &   50.7 &   18.2 &  -21.4 &    3.1 \\
\hline
\end{tabular}
\end{table*}

\addtocounter{table}{-1}

\begin{table*}
\caption[]{continue.
}
\begin{tabular}{cccrrrrrrrrr}
\hline \hline
\multicolumn{1}{c}{(1)} & \multicolumn{1}{c}{(2)} & 
\multicolumn{1}{c}{(3)} & \multicolumn{1}{c}{(4)} & 
\multicolumn{1}{c}{(5)} & \multicolumn{1}{c}{(6)} & 
\multicolumn{1}{c}{(7)} 
&\multicolumn{1}{c}{(8)} & \multicolumn{1}{c}{(9)} & \multicolumn{1}{c}{(10)} & \multicolumn{1}{c}{(11)}
&\multicolumn{1}{c}{(12)} 
\\
\multicolumn{1}{c}{combined} & \multicolumn{1}{c}{old} &\multicolumn{1}{c}{young} & 
\multicolumn{1}{c}{age1} &\multicolumn{1}{c}{age2} & 
\multicolumn{1}{c}{\%} &\multicolumn{1}{c}{\%} & 
\multicolumn{1}{c}{$\chi^2$}  & \multicolumn{1}{c}{$\chi^2$} &
\multicolumn{1}{c}{$\chi^2$}  & \multicolumn{1}{c}{$\frac{\mathrm{N}_{obs}-\mathrm{N}_{sim}}{\sqrt{N}}$} &
\multicolumn{1}{c}{$\frac{\mathrm{N}_{obs}-\mathrm{N}_{sim}}{\sqrt{N}}$}
\\
\multicolumn{1}{c}{simulation}& \multicolumn{1}{c}{simulation} &\multicolumn{1}{c}{simulation} &
\multicolumn{1}{c}{(Gyr)}&\multicolumn{1}{c}{(Gyr)}& \multicolumn{1}{c}{(old)}&\multicolumn{1}{c}{(young)}& 
\multicolumn{1}{c}{CMD} & \multicolumn{1}{c}{LF$_I$} &
\multicolumn{1}{c}{LF$_V$} & \multicolumn{1}{c}{RC} &
\multicolumn{1}{c}{AGBb}
\\
\hline
 cmb756 &  aen205 &  aen032 & 10.0 &    7 & 90& 10&  216.9 &   35.2 &   99.0 &   -6.5 &   -0.9 \\ 
 cmb757 &  aen205 &  aen032 & 10.0 &    7 & 80& 20&  138.1 &   38.2 &   74.7 &  -12.2 &   -1.7 \\ 
 cmb758 &  aen205 &  aen032 & 10.0 &    7 & 70& 30&   92.0 &   46.3 &   51.5 &  -17.8 &   -0.6 \\ 
 cmb759 &  aen205 &  aen032 & 10.0 &    7 & 60& 40&   84.5 &   50.3 &   36.3 &  -23.1 &   -1.6 \\ 
 cmb760 &  aen205 &  aen032 & 10.0 &    7 & 50& 50&  114.0 &   66.6 &   30.5 &  -29.1 &   -0.3 \\
\hline
\end{tabular}
\end{table*}

\addtocounter{table}{-1}

\begin{table*}
\caption[]{continue.
}
\begin{tabular}{cccrrrrrrrrr}
\hline \hline
\multicolumn{1}{c}{(1)} & \multicolumn{1}{c}{(2)} & 
\multicolumn{1}{c}{(3)} & \multicolumn{1}{c}{(4)} & 
\multicolumn{1}{c}{(5)} & \multicolumn{1}{c}{(6)} & 
\multicolumn{1}{c}{(7)} 
&\multicolumn{1}{c}{(8)} & \multicolumn{1}{c}{(9)} & \multicolumn{1}{c}{(10)} & \multicolumn{1}{c}{(11)}
&\multicolumn{1}{c}{(12)} 
\\
\multicolumn{1}{c}{combined} & \multicolumn{1}{c}{old} &\multicolumn{1}{c}{young} & 
\multicolumn{1}{c}{age1} &\multicolumn{1}{c}{age2} & 
\multicolumn{1}{c}{\%} &\multicolumn{1}{c}{\%} & 
\multicolumn{1}{c}{$\chi^2$}  & \multicolumn{1}{c}{$\chi^2$} &
\multicolumn{1}{c}{$\chi^2$}  & \multicolumn{1}{c}{$\frac{\mathrm{N}_{obs}-\mathrm{N}_{sim}}{\sqrt{N}}$} &
\multicolumn{1}{c}{$\frac{\mathrm{N}_{obs}-\mathrm{N}_{sim}}{\sqrt{N}}$}
\\
\multicolumn{1}{c}{simulation}& \multicolumn{1}{c}{simulation} &\multicolumn{1}{c}{simulation} &
\multicolumn{1}{c}{(Gyr)}&\multicolumn{1}{c}{(Gyr)}& \multicolumn{1}{c}{(old)}&\multicolumn{1}{c}{(young)}& 
\multicolumn{1}{c}{CMD} & \multicolumn{1}{c}{LF$_I$} &
\multicolumn{1}{c}{LF$_V$} & \multicolumn{1}{c}{RC} &
\multicolumn{1}{c}{AGBb}
\\
\hline
 cmb811 &  aen204 &  aen259 & 12.0 &    3 & 90& 10&  351.7 &   52.2 &   64.1 &   35.0 &   -4.9 \\ 
 cmb812 &  aen204 &  aen259 & 12.0 &    3 & 80& 20&  275.4 &   40.4 &   69.0 &   31.2 &   -8.4 \\ 
 cmb813 &  aen204 &  aen259 & 12.0 &    3 & 70& 30&  225.6 &   38.8 &   66.0 &   28.2 &  -11.1 \\ 
 cmb814 &  aen204 &  aen259 & 12.0 &    3 & 60& 40&  219.5 &   46.1 &   75.1 &   26.7 &  -11.7 \\ 
 cmb815 &  aen204 &  aen259 & 12.0 &    3 & 50& 50&  246.9 &   67.5 &   86.8 &   29.9 &  -14.6 \\
\hline
\end{tabular}
\end{table*}

\addtocounter{table}{-1}
\begin{table*}
\caption[]{continue.
}
\begin{tabular}{cccrrrrrrrrr}
\hline \hline
\multicolumn{1}{c}{(1)} & \multicolumn{1}{c}{(2)} & 
\multicolumn{1}{c}{(3)} & \multicolumn{1}{c}{(4)} & 
\multicolumn{1}{c}{(5)} & \multicolumn{1}{c}{(6)} & 
\multicolumn{1}{c}{(7)} 
&\multicolumn{1}{c}{(8)} & \multicolumn{1}{c}{(9)} & \multicolumn{1}{c}{(10)} & \multicolumn{1}{c}{(11)}
&\multicolumn{1}{c}{(12)} 
\\
\multicolumn{1}{c}{combined} & \multicolumn{1}{c}{old} &\multicolumn{1}{c}{young} & 
\multicolumn{1}{c}{age1} &\multicolumn{1}{c}{age2} & 
\multicolumn{1}{c}{\%} &\multicolumn{1}{c}{\%} & 
\multicolumn{1}{c}{$\chi^2$}  & \multicolumn{1}{c}{$\chi^2$} &
\multicolumn{1}{c}{$\chi^2$}  & \multicolumn{1}{c}{$\frac{\mathrm{N}_{obs}-\mathrm{N}_{sim}}{\sqrt{N}}$} &
\multicolumn{1}{c}{$\frac{\mathrm{N}_{obs}-\mathrm{N}_{sim}}{\sqrt{N}}$}
\\
\multicolumn{1}{c}{simulation}& \multicolumn{1}{c}{simulation} &\multicolumn{1}{c}{simulation} &
\multicolumn{1}{c}{(Gyr)}&\multicolumn{1}{c}{(Gyr)}& \multicolumn{1}{c}{(old)}&\multicolumn{1}{c}{(young)}& 
\multicolumn{1}{c}{CMD} & \multicolumn{1}{c}{LF$_I$} &
\multicolumn{1}{c}{LF$_V$} & \multicolumn{1}{c}{RC} &
\multicolumn{1}{c}{AGBb}
\\
\hline
 cmb866 &  aen288 &  aen292 & 10.0 &    5 & 90& 10&  199.7 &   31.4 &   88.0 &   -5.1 &   -6.4 \\ 
 cmb867 &  aen288 &  aen292 & 10.0 &    5 & 80& 20&  126.5 &   39.9 &   63.5 &  -10.6 &   -3.9 \\ 
 cmb868 &  aen288 &  aen292 & 10.0 &    5 & 70& 30&  105.7 &   49.4 &   53.4 &  -11.1 &   -4.6 \\ 
 cmb869 &  aen288 &  aen292 & 10.0 &    5 & 60& 40&  101.5 &   61.4 &   40.9 &  -18.5 &   -4.4 \\ 
 cmb870 &  aen288 &  aen292 & 10.0 &    5 & 50& 50&  138.7 &   77.9 &   38.2 &  -20.5 &   -5.1 \\
\hline
\end{tabular}
\end{table*}

\addtocounter{table}{-1}
\begin{table*}
\caption[]{continue.
}
\begin{tabular}{cccrrrrrrrrr}
\hline \hline
\multicolumn{1}{c}{(1)} & \multicolumn{1}{c}{(2)} & 
\multicolumn{1}{c}{(3)} & \multicolumn{1}{c}{(4)} & 
\multicolumn{1}{c}{(5)} & \multicolumn{1}{c}{(6)} & 
\multicolumn{1}{c}{(7)} 
&\multicolumn{1}{c}{(8)} & \multicolumn{1}{c}{(9)} & \multicolumn{1}{c}{(10)} & \multicolumn{1}{c}{(11)}
&\multicolumn{1}{c}{(12)} 
\\
\multicolumn{1}{c}{combined} & \multicolumn{1}{c}{old} &\multicolumn{1}{c}{young} & 
\multicolumn{1}{c}{age1} &\multicolumn{1}{c}{age2} & 
\multicolumn{1}{c}{\%} &\multicolumn{1}{c}{\%} & 
\multicolumn{1}{c}{$\chi^2$}  & \multicolumn{1}{c}{$\chi^2$} &
\multicolumn{1}{c}{$\chi^2$}  & \multicolumn{1}{c}{$\frac{\mathrm{N}_{obs}-\mathrm{N}_{sim}}{\sqrt{N}}$} &
\multicolumn{1}{c}{$\frac{\mathrm{N}_{obs}-\mathrm{N}_{sim}}{\sqrt{N}}$}
\\
\multicolumn{1}{c}{simulation}& \multicolumn{1}{c}{simulation} &\multicolumn{1}{c}{simulation} &
\multicolumn{1}{c}{(Gyr)}&\multicolumn{1}{c}{(Gyr)}& \multicolumn{1}{c}{(old)}&\multicolumn{1}{c}{(young)}& 
\multicolumn{1}{c}{CMD} & \multicolumn{1}{c}{LF$_I$} &
\multicolumn{1}{c}{LF$_V$} & \multicolumn{1}{c}{RC} &
\multicolumn{1}{c}{AGBb}
\\
\hline
 cmb921 &  aen288 &  aen294 & 10.0 &    8 & 90& 10&  180.7 &   33.1 &   78.0 &   -7.4 &   -2.8 \\ 
 cmb922 &  aen288 &  aen294 & 10.0 &    8 & 80& 20&  103.7 &   37.5 &   50.5 &  -13.2 &   -2.3 \\ 
 cmb923 &  aen288 &  aen294 & 10.0 &    8 & 70& 30&   76.5 &   39.6 &   35.0 &  -20.9 &   -0.2 \\ 
 cmb924 &  aen288 &  aen294 & 10.0 &    8 & 60& 40&   83.0 &   50.1 &   22.6 &  -28.0 &   -1.0 \\ 
 cmb925 &  aen288 &  aen294 & 10.0 &    8 & 50& 50&  119.3 &   57.6 &   19.6 &  -36.3 &    0.8 \\
\hline
\end{tabular}
\end{table*}

\addtocounter{table}{-1}
\begin{table*}
\caption[]{continue.
}
\begin{tabular}{cccrrrrrrrrr}
\hline \hline
\multicolumn{1}{c}{(1)} & \multicolumn{1}{c}{(2)} & 
\multicolumn{1}{c}{(3)} & \multicolumn{1}{c}{(4)} & 
\multicolumn{1}{c}{(5)} & \multicolumn{1}{c}{(6)} & 
\multicolumn{1}{c}{(7)} 
&\multicolumn{1}{c}{(8)} & \multicolumn{1}{c}{(9)} & \multicolumn{1}{c}{(10)} & \multicolumn{1}{c}{(11)}
&\multicolumn{1}{c}{(12)} 
\\
\multicolumn{1}{c}{combined} & \multicolumn{1}{c}{old} &\multicolumn{1}{c}{young} & 
\multicolumn{1}{c}{age1} &\multicolumn{1}{c}{age2} & 
\multicolumn{1}{c}{\%} &\multicolumn{1}{c}{\%} & 
\multicolumn{1}{c}{$\chi^2$}  & \multicolumn{1}{c}{$\chi^2$} &
\multicolumn{1}{c}{$\chi^2$}  & \multicolumn{1}{c}{$\frac{\mathrm{N}_{obs}-\mathrm{N}_{sim}}{\sqrt{N}}$} &
\multicolumn{1}{c}{$\frac{\mathrm{N}_{obs}-\mathrm{N}_{sim}}{\sqrt{N}}$}
\\
\multicolumn{1}{c}{simulation}& \multicolumn{1}{c}{simulation} &\multicolumn{1}{c}{simulation} &
\multicolumn{1}{c}{(Gyr)}&\multicolumn{1}{c}{(Gyr)}& \multicolumn{1}{c}{(old)}&\multicolumn{1}{c}{(young)}& 
\multicolumn{1}{c}{CMD} & \multicolumn{1}{c}{LF$_I$} &
\multicolumn{1}{c}{LF$_V$} & \multicolumn{1}{c}{RC} &
\multicolumn{1}{c}{AGBb}
\\
\hline
 cmb975 &  aen270 &  aen305 & 12.0 &    3 & 90& 10&   76.4 &   13.0 &   11.1 &   -5.1 &    1.5 \\ 
 cmb976 &  aen270 &  aen305 & 12.0 &    3 & 80& 20&   62.1 &   12.6 &   11.2 &    2.6 &   -2.2 \\ 
 cmb977 &  aen270 &  aen305 & 12.0 &    3 & 70& 30&   85.9 &   23.3 &   20.7 &   10.0 &   -4.3 \\ 
 cmb978 &  aen270 &  aen305 & 12.0 &    3 & 60& 40&  129.9 &   40.4 &   33.2 &   17.0 &   -4.4 \\ 
 cmb979 &  aen270 &  aen305 & 12.0 &    3 & 50& 50&  197.9 &   64.3 &   49.9 &   21.9 &   -5.4 \\
\hline
\hline
\end{tabular}
\end{table*}

\end{document}